\documentclass[sn-basic]{sn-jnl}

\usepackage{graphicx}%
\usepackage{multirow}%
\usepackage{amsmath,amssymb,amsfonts}%
\usepackage{amsthm}%
\usepackage{mathrsfs}%
\usepackage[title]{appendix}%
\usepackage{xcolor}%
\usepackage{textcomp}%
\usepackage{manyfoot}%
\usepackage{booktabs}%
\usepackage{algorithm}%
\usepackage{algorithmicx}%
\usepackage{algpseudocode}%
\usepackage{listings}%
\usepackage{txfonts}

\usepackage[percent]{overpic}
\usepackage{anyfontsize}
\usepackage{xspace}
\usepackage{lipsum}
\usepackage{longtable}
\usepackage{comment}
\usepackage[normalem]{ulem}
\usepackage{subcaption}
\usepackage{lscape}
\usepackage{placeins}
\usepackage{lastpage}
\usepackage{hyperref}
\usepackage{tablefootnote}

\theoremstyle{thmstyleone}%
\theoremstyle{thmstyletwo}%

\theoremstyle{thmstylethree}%

\newcommand{\fermi}{{\em Fermi}\xspace}
\newcommand{\kw}{{\em Konus}-{Wind}\xspace}

\newcommand{\swift}{{\em Swift}\xspace}

\begin{document}

\title[Article Title]{Early Optical Follow-up of Gamma-Ray Bursts: The Critical Role of Robotic Telescopes}


\author*[1,2]{\fnm{Rahul} \sur{Gupta}}\email{rahulbhu.c157@gmail.com, rahul.gupta@nasa.gov}
\author[1]{\fnm{Judith} \sur{Racusin}}
\author[3,4]{\fnm{Gor} \sur{Oganesyan}}
\author[5,6]{\fnm{Ramandeep} \sur{Gill}}
\author[7]{\fnm{Sergey} \sur{Karpov}}
\author[8]{\fnm{Samantha R.} \sur{Oates}}
\author[9]{\fnm{Maria} \sur{Gritsevich}}
\author[9,10]{\fnm{A. J.} \sur{Castro-Tirado}}
\author[11]{\fnm{Vladimir} \sur{Lipunov}}
\author[12,13]{\fnm{Benjamin P.} \sur{Gompertz}}
\author[14]{\fnm{Maria Giovanna} \sur{Dainotti}}
\author[1]{\fnm{S. Bradley} \sur{Cenko}}
\author[15,16,17]{\fnm{Bing} \sur{Zhang}}
\author[18,19,20]{\fnm{Toktarkhan} \sur{Komesh}}
\author[21]{\fnm{Martin} \sur{Jelínek}}
\author[22]{\fnm{Stéphane} \sur{Basa}}
\author[23]{\fnm{Takanori} \sur{Sakamoto}}
\author[24]{\fnm{Carl W.} \sur{Akerlof}}
\author[25]{\fnm{Antonio Martin} \sur{Carrillo}}
\author[8]{\fnm{Sam} \sur{Shilling}}
\author[1,26,27]{\fnm{Ryan} \sur{Seeb}}
\author[28]{\fnm{Shashi} \sur{B. Pandey}}
\author[29]{\fnm{Varun} \sur{Bhalerao}}
\author[9]{\fnm{M. D.} \sur{Caballero-Garc{\'i}a}}
\author[11]{\fnm{Ivan} \sur{Panchenko}}
\author[11]{\fnm{Pavel} \sur{Balanutsa}}
\author[11]{\fnm{Artem} \sur{Kuznetsov}}
\author[11]{\fnm{Kirill} \sur{Zhirkov}}
\author[11]{\fnm{Natalia} \sur{Tyurina}}

\affil*[1]{Astrophysics Science Division, NASA Goddard Space Flight Center, Mail Code 661, Greenbelt, MD 20771, USA}
\affil[2]{NASA Postdoctoral Program Fellow}
\affil[3]{Gran Sasso Science Institute, Viale F. Crispi 7, I-67100 L'Aquila (AQ), Italy}
\affil[4]{INFN -- Laboratori Nazionali del Gran Sasso, I-67100 L'Aquila (AQ), Italy}
\affil[5]{Instituto de Radioastronomía y Astrof\'isica, Universidad Nacional Aut\'onoma de M\'exico, Antigua Carretera a P\'atzcuaro \# 8701, Ex-Hda. San Jos\'e de la Huerta, Morelia, Michoac\'an, M\'exico C.P. 58089, M\'exico}
\affil[6]{Astrophysics Research Center of the Open university (ARCO), The Open University of Israel, P.O Box 808, Ra'anana 43537, Israel}
\affil[7]{Institute of Physics of the Czech Academy of Sciences, Prague, CZ}
\affil[8]{School of Physics and Astronomy, Lancaster University, Lancaster, LA1 4YB, UK}
\affil[9]{Instituto de Astrofisica de Andaluc\'{i}a (IAA-CSIC), Glorieta de la Astronom\'{i}a s/n, E-18008, Granada, Spain}
\affil[10]{Unidad Asociada Departamento de Ingenier\'{i}a de Sistemas y Autom\'{a}tica, E.T.S. de Ingenieros Industriales, Universidad de M\'{a}laga, E-29071, Spain}
\affil[11]{Lomonosov Moscow State University, 119234, Moscow, Universitetskiy prospect, 13, 37, Russia}
\affil[12]{School of Physics and Astronomy, University of Birmingham, Edgbaston, Birmingham, B15 2TT, UK}
\affil[13]{Institute for Gravitational Wave Astronomy, University of Birmingham, Edgbaston, Birmingham, B15 2TT, UK}
\affil[14]{Division of Science, National Astronomical Observatory of Japan, 2-21-1 Osawa, Mitaka, Tokyo}
\affil[15]{The Hong Kong Institute for Astronomy and Astrophysics, The University of Hong Kong, Pokfulam Road, Hong Kong, People's Republic of China}
\affil[16]{Department of Physics, The University of Hong Kong, Pokfulam Road, Hong Kong, People's Republic of China}
\affil[17]{The Nevada Center for Astrophysics and Department of Physics and Astronomy, University of Nevada, Las Vegas, NV, 89154, USA}
\affil[18]{Energetic Cosmos Laboratory, Nazarbayev University, Astana 010000, Kazakhstan}
\affil[19]{State Key Laboratory of Radio Astronomy and Technology, Xinjiang Astronomical Observatory, CAS, 150 Science 1Street, Urumqi, Xinjiang, 830011, China}
\affil[20]{Institute of Experimental and Theoretical Physics, Al-Farabi Kazakh National University, Almaty 050040, Kazakhstan}
\affil[21]{Astronomical Institute of the Czech Academy of Sciences (ASU-CAS), Fri\v{c}ova 298, 251~65 Ond\v{r}ejov, Czech Republic}
\affil[22]{Aix Marseille Univ., CNRS, CNES, LAM, Marseille, France}
\affil[23]{Department of Physical Sciences, Aoyama Gakuin University, 5-10-1 Fuchinobe, Chuo-ku, Sagamihara, Kanagawa 252-5258, Japan}
\affil[24]{University of Michigan, Randall Laboratory of Physics, 450 Church St., Ann Arbor, MI, 48109-1040}
\affil[25]{School of Physics and Centre for Space Research, University College Dublin, Belfield, Dublin 4, Ireland}
\affil[26]{Department of Physics, Brown University, Providence, RI 02912, USA}
\affil[27]{Southeastern Universities Research Association, NASA/GSFC, CRESST II CoOp, Washington, DC 20005}
\affil[28]{Aryabhatta Research Institute of Observational Sciences (ARIES), Manora Peak, Nainital-263002, India}
\affil[29]{Department of Physics, Indian Institute of Technology Bombay, Powai, 400 076, India}

\abstract{Gamma-ray bursts (GRBs) are the most luminous electromagnetic explosions in the Universe, and offer unique laboratories for studying relativistic jets, compact-object formation, particle acceleration, and the high-redshift Universe. The early optical emission of GRBs, particularly within seconds to minutes after the burst, carries crucial information about the central engine, jet magnetization, bulk Lorentz factor, and circumburst environment. We present a comprehensive review of the early optical phenomenology of GRBs and the essential role played by ground-based robotic optical telescopes to observe the fleeting early-time phenomena through rapid, automated responses to real-time GRB alerts and high-cadence photometry. We examine the key early optical features of GRBs, including prompt optical emission coincident with the $\gamma$-ray phase, bright reverse shock optical flashes, the onset of external forward shock afterglow, and superimposed optical flares, plateaus, and discuss the diagnostic power of each in constraining jet physics. We discuss the physical mechanisms underlying these phenomena and their implications for GRB physics (e.g., estimating the initial Lorentz factor $\Gamma_0$, magnetization, and the density profile). Early optical observations have constrained the initial bulk Lorentz factor $\Gamma_0 \sim 100$--$1000$, weak-to-moderate ejecta magnetization for events with prominent reverse shocks, the circumburst density profile, and the geometry of the magnetic field in the ejecta through polarimetry. We also provide the technical capabilities and landmark contributions of major robotic facilities, and discuss future prospects in the era of SVOM, Einstein Probe, Rubin/LSST, ULTRASAT, TeV observatories, and multi-messenger alerts. The continued synergy between space-based high-energy missions and ground-based robotic optical/near-infrared telescopes will remain essential for advancing our understanding of relativistic jets, the central engine, and the role of GRBs in the transient Universe.}

\keywords{gamma-ray bursts: general, methods: observational}



\maketitle

\section{Introduction}

Gamma-ray bursts (GRBs) are brief, intense flashes of high-energy radiation and represent the most luminous electromagnetic explosions in the Universe. Since their discovery by the Vela satellites \citep{1973ApJ...182L..85K}, their origin remained uncertain for several decades, with both Galactic and cosmological scenarios being actively debated. A major observational advance came from the accumulation of large GRB samples by the Burst and Transient Source Experiment (BATSE), a wide-field $\gamma$-ray monitoring instrument designed to detect and localize transient high-energy events across the sky, aboard the \textit{Compton Gamma-Ray Observatory}. BATSE had shown that GRBs are distributed nearly isotropically on the sky, strongly disfavoring an origin in the Galactic disk and supporting either a cosmological population or an extended Galactic-halo population \citep{1996ApJ...459...40B}. The decisive breakthrough came on 28 February 1997, when the \textit{BeppoSAX} satellite detected a fading X-ray source coincident with GRB\,970228 \citep{1997Natur.387..783C}. Optical follow-up within hours revealed a fading point source: the first GRB optical afterglow \citep{1997Natur.386..686V}, located in a faint, distant galaxy. The first afterglow redshift was measured for GRB 970508 using the absorption and emission features in spectra of the optical afterglow \citep{1998ApJ...495L..99R}. Spectroscopy of the host galaxy of GRB\,970228 established the redshift, $z = 0.695$ \citep{2001ApJ...554..678B}, placing the event firmly at cosmological distances.

Early studies had already noted evidence for unusually short-duration bursts and possible substructure within the GRB population \citep[e.g.,][]{1981Ap&SS..80....3M,1984Natur.308..434N,1992AIPC..265..304D}. The definitive observational demonstration of a bimodal duration distribution was provided by the BATSE sample, which showed that GRBs separate phenomenologically into short and long-duration classes, with a division near $T_{\rm 90}$$\simeq2$~s \citep{1993ApJ...413L.101K}. This classification was originally empirical, based primarily on prompt $\gamma$-ray duration and spectral hardness. The physical interpretation of this bimodality developed in parallel and subsequently became the standard two-progenitor framework. In this picture, long-duration GRBs are generally associated with the core-collapse of massive stars, supported observationally by their occurrence in star-forming regions and by spectroscopic associations with broad-lined Type Ic supernovae \citep{1993ApJ...405..273W,2003Natur.423..847H,2003ApJ...591L..17S}. Short-duration GRBs, by contrast, were theoretically linked to compact-object mergers, particularly neutron-star--neutron-star or neutron-star--black-hole systems \citep{1989Natur.340..126E,Li87,Li95}. This interpretation gained strong observational support from the localization of short GRBs in both early- and late-type host galaxies, their typically lower circumburst densities, and the absence of bright supernovae in nearby well-observed cases \citep{2005Natur.437..851G,2014ARA&A..52...43B}. The association of GW170817 with GRB~170817A finally confirmed that at least some short GRBs are produced by binary neutron-star mergers \citep{2017PhRvL.119p1101A,2017ApJ...848L..13A}. At the same time, recent events with apparently long durations but merger-like properties, such as GRB~211211A and GRB~230307A \citep{2022Natur.612..223R,2022Natur.612..228T,2024Natur.626..737L}, demonstrate that duration alone is not a complete physical classifier, and that multi-wavelength context is essential for progenitor inference \citep{2025arXiv250800142G}.

Both major progenitor channels are thought to produce a compact central engine, either a rapidly accreting black hole or a highly magnetized neutron star, capable of launching an ultra-relativistic jet with bulk Lorentz factor $\Gamma \sim 100-1000$ \citep{2004RvMP...76.1143P}. The isotropic-equivalent energy release, $E_{\rm iso}\sim10^{50}$--$10^{55}$ erg, over timescales from milliseconds to hundreds of seconds, makes GRBs powerful probes of relativistic outflows, compact-object formation, and massive-star death \citep{2023ApJ...946L..31B, 2026MNRAS.546ag060A}.

The prompt $\gamma$-ray emission is widely attributed to internal dissipation, either internal shocks between shells of different Lorentz factors \citep{Rees1994} or magnetic reconnection in a Poynting-flux-dominated outflow \citep{2003ApJ...597..998L, 2011ApJ...726...90Z, 2024ApJ...972..166G} at radii $R \sim 10^{13}$--$10^{15}$\,cm. When the ejecta eventually decelerate against the circumburst medium, external shocks produce the long-lived multiwavelength afterglow emission \citep{MeszarosRees1997, 2002ApJ...570L..61G}. The synchrotron afterglow spectrum in the standard external-shock model is characterized by three critical break frequencies \citep{1998ApJ...497L..17S}: the synchrotron self-absorption frequency $\nu_a$, defined by an optical depth $\tau_\nu \simeq 1$ and set primarily by the shocked-particle density, magnetic field strength, and source size (for example, in the usual slow-cooling ISM case with $\nu_a<\nu_m<\nu_c$, $\nu_a \propto E_{\rm K}^{1/5} n^{3/5}\epsilon_e^{-1}\epsilon_B^{1/5}$); the characteristic synchrotron frequency of minimal energy power-law electrons, with Lorentz factor $\gamma_m\propto\Gamma$, is defined as $\nu_m\propto \Gamma\gamma_m^2B$, where it depends on the bulk Lorentz factor of the flow, $\Gamma$, and the comoving magnetic field, $B\propto\Gamma^4n^{1/2}$. If the external medium is radially stratified with $n\propto R^{-k}$, $\Gamma\propto R^{-(3-k)/2}$ after the blast wave decelerates and follows a self-similar evolution. As a result, $\nu_m\propto R^{-3(4-k)/2}\propto t^{-3/2}$ for a spherical flow, where we have used the scaling $R\propto\Gamma^2t\propto t^{1/(4-k)}$; and the cooling frequency $\nu_c\propto\Gamma\gamma_c^2B\propto\Gamma^{-4}n^{-3/2}t^{-2}\propto t^{(3k-4)/2(4-k)}$ for $\gamma_c\propto\Gamma^{-1}B^{-2}t^{-1}$, above which radiative losses become important on the dynamical timescale. The ordering of these frequencies determines the spectral regime and hence the spectral index $\beta$ in $F_\nu \propto \nu^{-\beta}$. The observed temporal decay index $\alpha$ in $F_\nu \propto t^{-\alpha}$ is then determined by the time evolution of these break frequencies and the peak flux, which depends on the blast-wave dynamics, external density profile, energy injection, and jet geometry \citep{1998ApJ...497L..17S, 2002ApJ...570L..61G, 2002ApJ...571..779P}.

While space-based high-energy missions provide rapid triggers for prompt $\gamma$-ray/hard X-ray bursts, with localizations ranging from a few degrees to sub-arcminute accuracy, early UV/optical follow-up is provided by both space and ground-based facilities. In particular, \swift/UVOT can begin UV/optical observations within tens of seconds to a few minutes after a GRB trigger, providing a vital space-based view of the early afterglow \citep[e.g.,][]{2009ApJ...690..163R,2019MNRAS.488.2855P,2009MNRAS.395..490O,Oates2023}. Ground-based robotic telescopes provide a complementary window through rapid optical/NIR coverage, wider-field tiling of large error regions, high-cadence monitoring, polarimetry, and deeper follow-up from seconds to hours after the burst. These early emissions offer diagnostics for GRB physics, including constraints on jet dynamics, composition, energy dissipation, and the surrounding environments \citep[e.g.,][]{NakarPiran2004, 2003ApJ...586L...5F, 2007Sci...315.1822M}. According to the standard relativistic fireball model \citep{MeszarosRees1997, 1999ApJ...519L..17S}, early optical emission can arise from several mechanisms: (i) internal shocks within the relativistic ejecta (responsible for prompt optical emission), (ii) the reverse shock that propagates back into the ejecta as it decelerates against the circumburst medium, and (iii) the long-lived forward shock that accelerates electrons in the swept-up external medium \citep{1999ApJ...520..641S, 2000ApJ...545..807K, 2005ApJ...628..315Z}. Additionally, late energy injections from either the ejecta having a radially stratified structure \citep{Rees-Meszaros-98,Sari-Meszaros-00,Granot+03,Schroeder24,Anderson25} or prolonged central engine activity \citep{LL98,Dai-Lu-98,Kumar-Piran-00,Sari-Meszaros-00,Zhang-Meszaros-02} can produce ``refreshed'' shocks, leading to bright optical flares or rebrightenings. The temporal evolution, spectral properties, and correlations in early optical light curves thus probe key parameters, including the initial bulk Lorentz factor $\Gamma_0$, the circumburst density profile (constant ISM or stratified wind), the magnetization parameter of the outflow, and the duration of central engine activity \citep{Li2009}.

Critically, the earliest optical observations (within seconds to minutes after the GRB trigger) capture transient phenomena such as prompt optical emission, reverse shock emission, the onset of the afterglow, and early optical flares. These features fade rapidly and are difficult to study with slower, traditional telescope response mechanisms \citep{2005Natur.435..178V, 2009ApJ...691..723B}. This requirement has driven the development of dedicated robotic telescope systems. The advent of robotic optical telescopes and networks has revolutionized early GRB follow-up observations. These autonomous systems respond to satellite alerts within seconds, slew rapidly to the GRB position, and begin imaging while the burst is still ongoing or immediately thereafter \citep{2003PASP..115..132A}, enabled by real-time data processing pipelines (see Section \ref{data analysis}). This capability has revealed a diverse phenomenology of early optical emission \citep{2023RNAAS...7...56F}, including prompt optical flashes coincident with gamma rays (e.g., the naked-eye GRB 080319B; \citealt{2008Natur.455..183R,Beskin2010}), bright reverse shock peaks (e.g., GRB 990123 observed with ROTSE; \citealt{1999Natur.398..400A}), complex temporal variability, early-time polarization signatures indicating ordered magnetic fields in reverse shocks \citep{2009Natur.462..767S, 2013Natur.504..119M}, systematic early light-curve samples from networks such as ROTSE-III and TAROT \citep{2009ApJ...702..489R, 2009AJ....137.4100K}, and clear detections of the onset of forward shock afterglow emission \citep{2007Molinari}.

In addition to detecting early prompt and early optical afterglow components, ground-based optical/NIR observations are essential for obtaining spectroscopic or photometric redshifts, providing arcsecond/sub-arcsecond localizations for robust host-galaxy association and environmental characterization.

Here, we investigate how rapid robotic optical observations have transformed our understanding of the earliest phases of GRBs. We synthesize the observed early optical phenomenology, its physical interpretation, and the constraints that seconds-to-hours optical follow-up places on prompt-emission physics, reverse shocks, forward-shock onset, ejecta magnetization, circumburst environments, and central-engine activity.

This review is organized as follows. Section~\ref{sec:features} describes the diverse early optical features observed in GRBs and their physical interpretations. Section~\ref{sec:science} summarizes the key science returns enabled by rapid robotic optical follow-up of GRBs. Section~\ref{sec:facilities} provides a literature review of major robotic facilities and networks, highlighting their technical capabilities and landmark contributions. We discuss summary, conclusion, and future capabilities of robotic telescopes in the era of rapid multi-messenger and time-domain astronomy in Sections~\ref{sec:conclusion} and ~\ref{sec:future}.

\begin{figure*}[ht]
    \centering
    \includegraphics[width=0.95\textwidth]{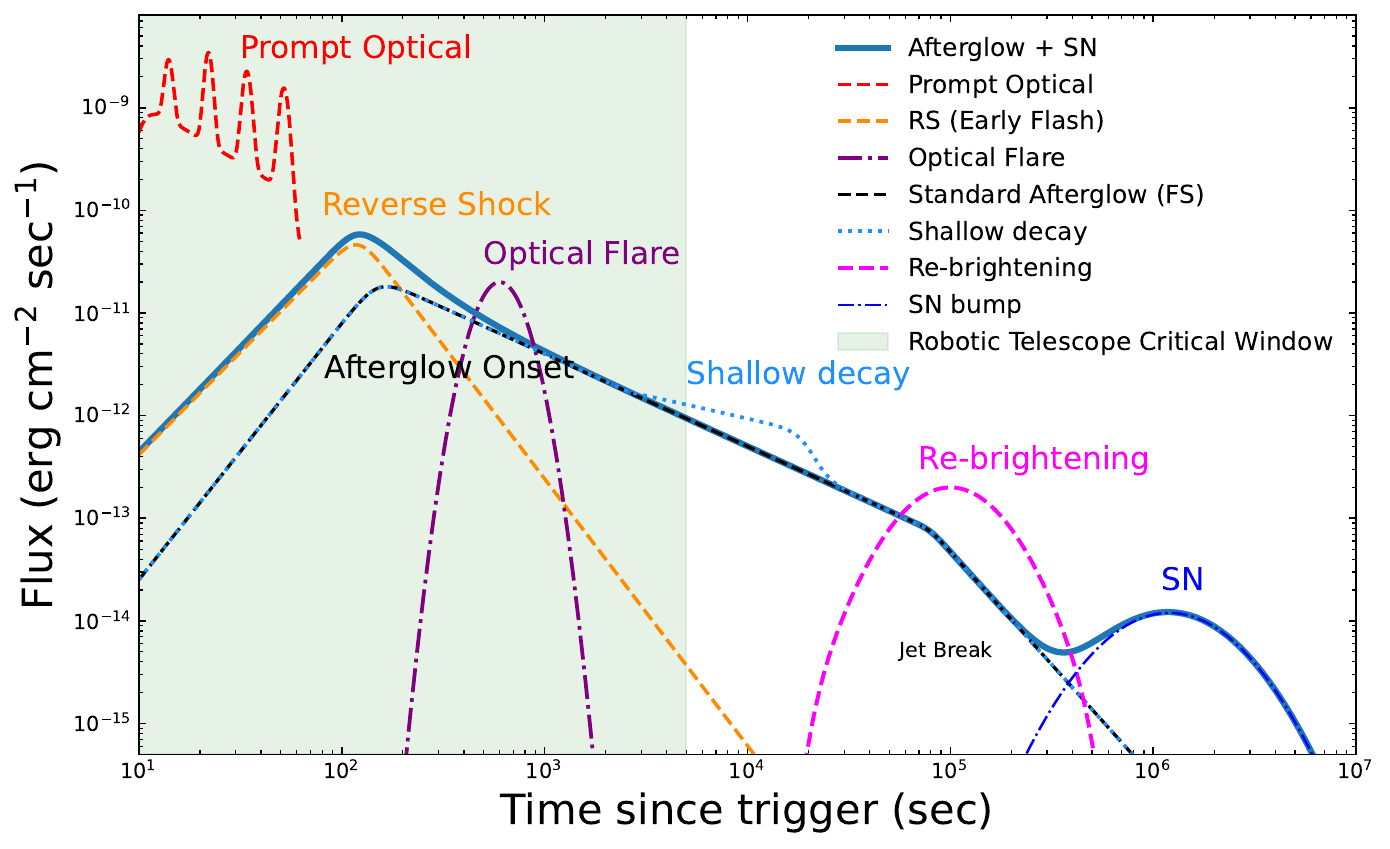}
    \caption{Schematic optical light curve of GRBs and its relevance to robotic follow-up: A schematic optical light curve constructed to illustrate the key features seen in early optical studies. The solid blue curve represents the baseline afterglow evolution, combining forward shock emission (including an illustrative jet-break steepening at $t \sim 10^5$\,s) with a late-time supernova (SN) bump. Additional components are overplotted to highlight distinct phases: Prompt optical emission (bright and rapidly variable, correlated with $\gamma$-rays), an early reverse-shock optical flash at $t \sim 10^2$\,s, an optical flare at $t \sim 10^3$\,s, a shallow-decay/plateau phase  ($t \sim 10^3$~--~$10^4$\,s), and a later re-brightening episode (e.g., refreshed shocks or density structure). The shaded region shows the ``critical window'' ($t \sim 10$~--~$5\times10^{3}$\,s) in which low-latency, high-cadence robotic observations provide the strongest leverage for separating these components and constraining outflow dynamics (e.g., $\Gamma_0$), magnetization, and the circumburst environment.}
    \label{fig:synthetic_optical_lc}
\end{figure*}

\section{Early Optical Features and Physical Interpretation}
\label{sec:features}

\begin{figure*}[!ht]
    \centering
    \includegraphics[width=1.05\textwidth, angle=0]{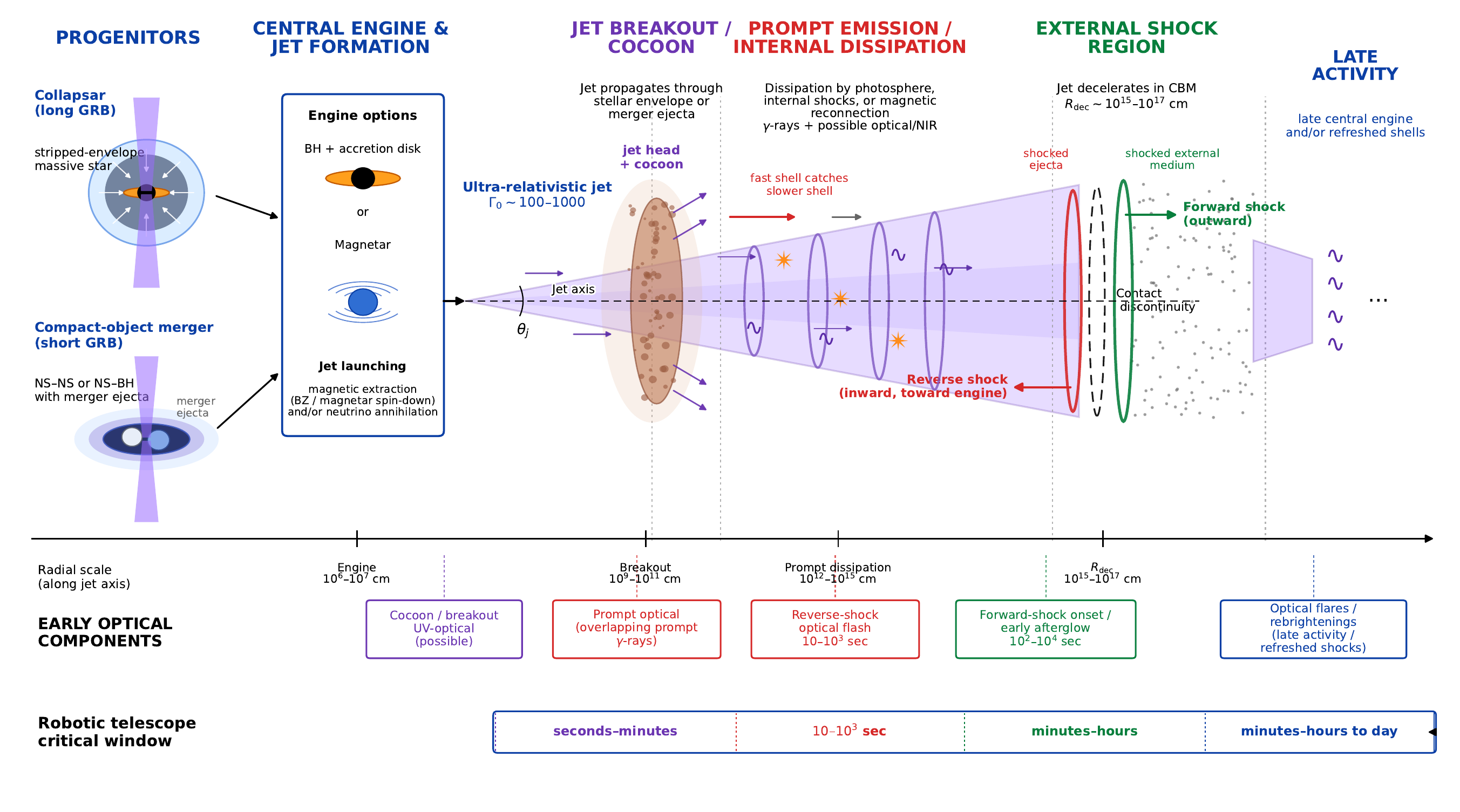}
    \caption{Schematic physical picture of early optical emission in GRBs. Long GRBs originate from collapsars, whereas short GRBs originate from compact-object mergers. Both channels can produce either a black-hole accretion-disk system or, in some cases, a rapidly rotating magnetar. After jet formation and acceleration, the ultra-relativistic outflow propagates through the stellar envelope or merger ejecta, inflating a cocoon that may produce thermal UV/optical emission. Prompt $\gamma$-ray emission and possible prompt optical/NIR emission arise from photospheric radiation, internal shocks, or magnetic reconnection at larger radii. At the deceleration radius, the ejecta interact with the circumburst medium, forming a forward shock, contact discontinuity, and reverse shock; the reverse shock can produce a rapidly fading optical flash, while the forward shock produces the onset of the standard afterglow. Later central-engine activity, refreshed shocks, or angular/radial jet structure can generate optical flares and rebrightenings. The robotic telescopes are therefore essential to detect the prompt optical, reverse shock, early forward-shock, and flare/rebrightening phases.}
    \label{fig:synthetic_early_optical_lc}
\end{figure*}

Early optical light curves of GRBs often exhibit multiple components, with their relative contributions varying significantly from one event to another \citep{2003ApJ...595..950Z, Melandri2008,2014ApJ...785...84J,Oates2023}. Dense, multi-band photometric observations during the first minutes to hours after the burst are therefore crucial, as they frequently provide the only practical route to distinguishing emission associated with the prompt phase (or internal processes) from that arising from external forward shock (afterglow) emission. A schematic representation of a typical GRB optical light curve \citep{2012ApJ...758...27L} is shown in Figure \ref{fig:synthetic_optical_lc}. In this section, we describe the key observed features and their standard interpretations (see Figure \ref{fig:synthetic_early_optical_lc}).

\subsection{Prompt Optical Emission}
\label{sec:prompt_optical}

\begin{figure}[!ht]
    \centering
    \includegraphics[width=0.7\textwidth]{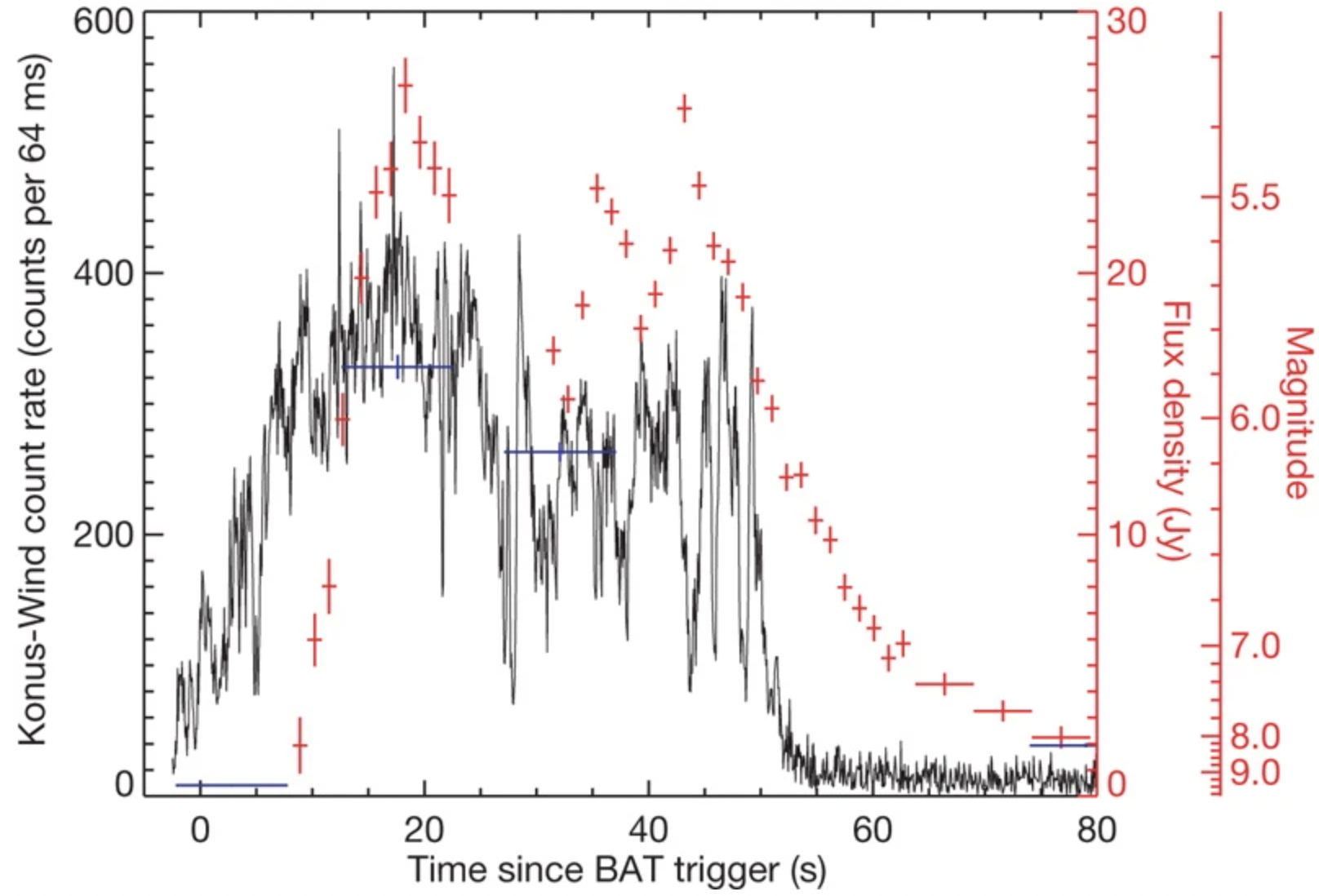}
\caption{Prompt $\gamma$-ray and optical emission comparison for the exceptionally bright naked eye burst (GRB\,080319B). The background-subtracted \kw $\gamma$-ray light curve (18--1160~keV; black), with contemporaneous optical measurements (red) from Pi of the Sky and TORTORA, is overplotted. The optical signal emerges within seconds of the high-energy onset and exhibits broad structure, including multiple sub-peaks reaching a maximum brightness of $\sim$5.3~mag in the TORTORA data. Adapted from \citet{2008Natur.455..183R}.}
\label{fig:racusin2008_prompt_optical}
\end{figure}

Prompt optical emission refers to optical/near-IR photons detected during, or directly overlapping with the high-energy prompt emission phase of a GRB (i.e., within seconds to minutes of the trigger), which provides the most direct window into the innermost dissipation region of the GRB jet. The field was inaugurated by the spectacular discovery of a bright optical flash ($\sim 9$ magnitude) for GRB~990123 using the \textit{Robotic Optical Transient Search Experiment} ROTSE-I telescope, contemporaneous with the GRB prompt phase, and starting $\sim$22~s after burst onset, established that optical photons can emerge essentially ``in real time'' with the prompt emission and demonstrated the diagnostic power of rapid, automated response \citep{1999Natur.398..400A}. The contemporaneous BATSE $\gamma$-ray light curve of GRB~990123 exhibited two main pulses \citep{1999ApJ...524...82B}, with the optical flash temporally coincident with the second pulse \citep{1999Natur.398..400A}, however, its rapidly fading optical flash is widely interpreted as being dominated by reverse shock synchrotron emission rather than by the same internal component that produced the high-energy emission. A second landmark event was GRB\,080319B, the ``naked-eye burst'', where prompt optical emission reached $\sim 5.3$ mag and was recorded with high cadence by multiple wide-field systems, including Pi of the Sky and TORTORA (see Figure \ref{fig:racusin2008_prompt_optical}), enabling stringent constraints on the emitting region(s) and spectral components \citep{2008Natur.455..183R}. The optical emission tracked the $\gamma$-ray variability over multiple sub-pulses on timescales as short as a few seconds, placing the emission site at $R \lesssim 10^{14}$\,cm -- far inside the external-shock deceleration radius. Detailed broadband modeling revealed that neither the prompt optical nor the $\gamma$-ray emission alone could be explained by a simple one-zone synchrotron model, requiring either an unusually efficient Compton component or a structured multi-zone dissipation region \citep{2008Natur.455..183R, 2009ApJ...691..723B}.

Prompt optical observations are especially valuable because they can probe emission components closely linked to the central engine and jet magnetization. GRB~160625B provides an important example: rapid observations by the MASTER Global Robotic Net detected prompt optical polarization, indicating an ordered magnetic-field component in the relativistic outflow \citep{2017Natur.547..425T}. Detailed optical photometry also revealed quasi-periodic structures in the intrinsic prompt optical emission \citep{2023ApJ...943..181L}. Together with contemporaneous multiwavelength data, these features motivated a three-stage central-engine interpretation, in which the quasi-periodic variability may be related to the forced precession of a rapidly rotating, self-gravitating superdense object, or ``spinar'' \citep{Li2007}, before its final collapse to a black hole.

In addition to these extreme cases, prompt optical emission correlated with $\gamma$-ray variability has been reported in a few more bursts: A widely cited example is GRB~041219A, for which rapid optical measurements showed variability tracking the prompt high-energy light curve, indicating a possible common origin for the optical and $\gamma$-ray emission \citep{2005Natur.435..178V}. Complementary near-IR detections contemporaneous with the later stages of the prompt phase of GRB~041219A further emphasized that low-energy prompt counterparts can extend into the IR \citep{2005Natur.435..181B}. Other examples include GRB\,050820A \citep{2006Natur.442..172V, Cenko2006GRB050820A}, GRB\,061121 \citep{2007ApJ...663.1125P}, and GRB\,110205A \citep{Zheng+12, Gendre+12} where the optical signal contains both internal and external-shock components. More recently, the Transiting Exoplanet Survey Satellite (TESS) provided a serendipitous detection of prompt optical emission from GRB~230307A. The TESS light curve showed a bright optical component coincident with the \fermi prompt emission, reaching $T_{\rm mag}=14.49$ when averaged over a 200~s full-frame image, followed by a fainter afterglow-like component that evolved over $\sim0.5$~d \citep{2023RNAAS...7...56F}. Because the prompt optical signal was detected in only one TESS exposure and was absent in the next image, the true optical peak was likely brighter than the time-averaged signal measurement, making GRB~230307A an important recent example of prompt optical emission captured by a wide-field optical survey.

The physical origin of prompt optical emission remains debated. Unlike the ``classical'' optical afterglow, generally attributed to synchrotron radiation from the external forward shock, prompt optical emission can originate from multiple radiative sites and mechanisms. In internal-dissipation scenarios \citep{2004MNRAS.354.1031F, 2014ApJ...789..145H}, prompt optical emission can in principle arise from the same region that produces the prompt $\gamma$-rays. In this case, the optical flux would correspond to the low-frequency extension of the prompt non-thermal spectrum. This interpretation requires that the characteristic synchrotron frequency associated with the emitting electrons, $\nu_m \propto \Gamma B \gamma_m^2$, lies well above the optical band, as is typically expected if the prompt spectral peak is located in the X-ray/$\gamma$-ray regime. However, whether this condition is satisfied depends on the poorly constrained microphysical parameters, including $\gamma_m$, the comoving magnetic field $B$, the bulk Lorentz factor $\Gamma$, and possible synchrotron self-absorption or cooling breaks between the optical and high-energy bands. In this context, optical and $\gamma$-ray variability are expected to be tightly correlated, as observed in GRB\,041219A \citep{2005Natur.435..178V} and GRB\,080319B \citep{2008Natur.455..183R}.

An alternative model proposed by \cite{2008Natur.455..183R}, and \cite{2009MNRAS.395..472K} suggests that synchrotron emission produces the prompt optical photons at $R \sim 10^{13}$~--~$10^{14}$\,cm, while inverse Compton (IC) upscattering of those same photons generates the observed $\gamma$-rays. The key prediction is a specific relationship between the optical and $\gamma$-ray fluxes, $F_\gamma / F_{\rm opt} \sim Y_{\rm IC}$ (the Compton $y$-parameter), and a temporal offset reflecting the seed photon population evolution. This model naturally accounts for the extreme brightness of GRB\,080319B \citep{2009MNRAS.395..472K}. Prompt optical emission can also arise from distinct dissipation zones, e.g., a faster inner region and a slower outer region producing quasi-thermal photospheric emission \citep{Meszaros2000, Peer2007, Guiriec+16} or from magnetic reconnection events at varying radii within a striped-wind outflow \citep{2011ApJ...726...90Z}. Discriminating between these scenarios requires broadband spectral energy distributions spanning optical through $\gamma$-ray energies.
 
The key observational diagnostics are: (a) the temporal correlation between optical and $\gamma$-ray light curves and spectral lag relative to $\gamma$-rays; (b) the broadband SED across optical, X-ray, and $\gamma$-ray energies; (c) the implied emission radius from variability and opacity arguments; and (d) early optical polarization, which constrains the degree of magnetic order in the emission region (see reviews by \citealt{2004RvMP...76.1143P, 2015PhR...561....1K, 2015AdAst2015E..22P}).

\subsection{Reverse Shock Signatures}
\label{Sec:RS}

When a relativistic shell of finite width expands into the circumburst medium, two shocks are expected to form: a forward shock that propagates into the ambient environment and a reverse shock that travels back into the ejecta \citep{1995ApJ...455L.143S,1999ApJ...520..641S,Kobayashi+99,Kobayashi-Sari-00}, both separated by a contact discontinuity. The shock dynamics are governed by the shell's kinetic energy $E_{\rm k,iso}$, initial Lorentz factor $\Gamma_0$, initial radial width $\Delta_0$, and the density profile of the external medium $\rho=m_pn=m_pn_0(R/R_0)^{-k}=AR^{-k}$, where $R_0$ is the radius at which the density is normalized to $n_0$ and $m_p$ is the proton mass. A combination of three of these four quantities allows us to define a critical Lorentz factor \citep{Kobayashi-Zhang-03,Granot-Ramirez-Ruiz-10}, 

\begin{eqnarray}
    \Gamma_{\rm cr} &\equiv& \left(\frac{l}{\Delta_0}\right)^{(3-k)/2(4-k)} \\
    &\simeq& 150\,(1+z)^{3/8}E_{\rm k,iso,52}^{1/8}n_{0,0}^{-1/8}t_{\rm grb,2}^{-3/8} \quad (k=0)\\
    &\simeq& 93\,(1+z)^{1/4}A_{*,-1}^{-1/4}E_{\rm k,iso,52}^{1/4}t_{\rm grb,2}^{-1/4} \quad (k=2)
\end{eqnarray}
where $\Delta_0\sim ct_{\rm grb}/(1+z)$ is equivalent to the light crossing distance over the prompt GRB duration $t_{\rm grb}$, $l\equiv(E_{\rm k,iso}/Ac^2)^{1/(3-k)}$ is the Sedov length (ignoring order unity factors), and $A=m_pn_0R_0^k$. Here, the density normalization for a wind-like external medium \citep{Chevalier-Li-00,Panaitescu-Kumar-00} is
\begin{equation}
    A = \frac{\dot M}{4\pi v} = 5\times10^{11}\,A_*\left(\frac{\dot M}{10^{-5} M_\odot\,{\rm yr^{-1}}}\right)\left(\frac{v}{10^3\,{\rm km\,s^{-1}}}\right)^{-1}\,{\rm g\over cm}\,,
\end{equation}
where $\dot M$ is the wind mass-loss rate of the progenitor star and $v$ is the wind velocity. 

Based on $\Gamma_{\rm cr}$, two distinct dynamical regimes are realized: (a) when $\Gamma_0>\Gamma_{\rm cr}$ the shell follows the \textit{thick-shell} dynamics in which the reverse shock becomes relativistic before crossing the shell, and (b) when $\Gamma_0<\Gamma_{\rm cr}$ the shell is in the \textit{thin-shell} regime in which the reverse shock remains Newtonian and only becomes mildly relativistic just before crossing the shell. In the thin-shell regime, the reverse shock is weak and unable to slow down the ejecta shell when it crosses it at $R_\times<R_{\rm dec}$. Instead, the shell decelerates upon sweeping up a mass of $\sim M_0/\Gamma_0=E_{\rm k,iso}/\Gamma_0^2c^2$ at the deceleration radius. This deceleration radius, $R_{\rm dec}$, is given by: 
\begin{eqnarray}
    R_{\rm dec} &=& \left[\frac{(3-k)E_{\rm k,iso}}{4\pi Ac^2\Gamma_0^2}\right]^{1\over(3-k)} \\
    &\simeq& 5.4\times10^{16}\,E_{\rm k,iso,52}^{1/3}n_{0,0}^{-1/3}\Gamma_{0,2}^{-2/3}\,{\rm cm} \quad (k=0) \\
    &\simeq& 1.8\times10^{15}\,A_{*,-1}^{-1}E_{\rm k,iso,52}\Gamma_{0,2}^{-2}\,{\rm cm} \quad (k=2)\,.
\end{eqnarray}
Only when $\Gamma_0=\Gamma_{\rm cr}$ do the two radii coincide and the reverse shock emission peaks at $t_\times\approx t_{\rm dec}$, where
\begin{eqnarray}
\label{eq:tdec}
    t_{\rm dec} &=& (1+z)\frac{R_{\rm dec}}{2\Gamma_0^2c}=(1+z)\left[\frac{(3-k)E_{\rm k,iso}}{2^{5-k}\pi A c^{5-k}\Gamma_0^{2(4-k)}}\right]^{1\over(3-k)} \\
    &\simeq& 90\,(1+z)E_{\rm k,iso,52}^{1/3}n_{0,0}^{-1/3}\Gamma_{0,2}^{-8/3}\,{\rm s} \quad (k=0) \\
    &\simeq& 3\,(1+z)A_{*,-1}^{-1}E_{\rm k,iso,52}\Gamma_{0,2}^{-4}\,{\rm s} \quad (k=2)\,.
\end{eqnarray}

Alternatively, in the thick shell case, the reverse shock is relativistic and is able to decelerate the ejecta upon crossing it over the shell's light crossing timescale of $t_\times\sim(1+z)\Delta_0/c\sim t_{\rm grb}$, where $t_{\rm grb}$ is the prompt GRB duration. This is also the timescale when the emission from the reverse shock peaks. More generally, the reverse shock crossing time can be expressed as $t_\times\sim(\Gamma_\times/\Gamma_{\rm cr})^{-2(4-k)/(3-k)}t_{\rm grb}$ where the Lorentz factor (LF) of the contact discontinuity at the shock crossing time is $\Gamma_\times\sim \min[\Gamma_0,\Gamma_{\rm cr}]$ \citep{1999ApJ...520..641S,2000ApJ...545..807K}.

Radiation from the reverse shock typically peaks on $\sim10$~–~$10^{3}$\,s timescales \citep{2000ApJ...545..807K,2003ApJ...597..455K} and is characterized by a fast-rise, steep-decay light curve profile, with the peak frequency $\nu_m^{\rm RS}$ typically in the optical/near-infrared (NIR) band, significantly lower than the forward shock peak frequency $\nu_m^{\rm FS}$. The peak frequency, $\nu_m\propto\Gamma\gamma_m^2B$, scales with the bulk LF of the shocked material, $\Gamma$, the LF of the minimal energy electrons, $\gamma_m\propto\langle\gamma\rangle\propto\epsilon_e(\bar\Gamma-1)\propto\bar\Gamma$ when $\bar\Gamma\gg1$ and where $\langle\gamma\rangle$ is the mean particle energy per unit rest mass, the LF of the shocked material (downstream) in the rest frame of the unshocked material (upstream), $\bar\Gamma$, and the magnetic field in the shocked material, $B$. The shocked material behind the forward and reverse shocks is expected to be in pressure equilibrium and also move at the same bulk LF. Pressure equilibrium also guarantees similar energy densities and, as a result, similar magnetic fields. Therefore, the two peak frequencies differ due to the lower $\langle\gamma\rangle$ in the shocked ejecta, which is lower due to the lower $\bar\Gamma\sim\Gamma_0/\Gamma$ between the shocked and unshocked ejecta \citep{1999ApJ...520..641S}, assuming the shock-microphysical parameters are similar in both shocked regions. The temporal decay following the peak is steep, with $F_\nu \propto t^{-\alpha}$ where $\alpha \sim 2$~--~3 for adiabatic evolution, though radiative losses can steepen the decline further \citep{2000ApJ...545..807K, 2003ApJ...597..455K}. Because of its relatively steep decay, the reverse shock optical flash is typically detectable only for the first $t\sim10^3-10^4$\,s after the burst, depending on its intrinsic brightness and the depth of the follow-up observations. Observations of such flashes provide direct constraints on the initial Lorentz factor $\Gamma_0$, the density profile ($n\propto r^{-k}$), and the magnetization of the outflow \citep{2014ApJ...785...84J}. 

Theoretical studies show that the detectability and radiative efficiency of reverse shocks depend critically on the magnetization parameter $\sigma \equiv B^2/(4\pi \rho c^2)$ of the ejecta, where $B$ is the comoving magnetic field strength and $\rho$ is the proper mass density \citep{2003ApJ...595..950Z}. In the baryonic regime ($\sigma \ll 1$), the reverse shock efficiently converts kinetic energy into internal energy, producing luminous optical emission with comparable energy output to the forward shock \citep{1999ApJ...520..641S}. However, for moderately magnetized ejecta ($\sigma \sim 0.1$--1), magnetic pressure support weakens the reverse shock compression, reducing its radiative efficiency and potentially shifting the peak to later times \citep{2004MNRAS.354.1031F}. In the highly magnetized Poynting-flux dominated regime ($\sigma \gg 1$), the reverse shock may be suppressed entirely, leading to weak or absent optical flashes \citep{2005ApJ...628..315Z}. The transition between these regimes occurs around $\sigma_c \sim (\Gamma_0/\Gamma_{\rm sh})^2$, where $\Gamma_{\rm sh}$ is the shocked-shell Lorentz factor \citep{2009A&A...494..879M}. Observationally, unambiguous reverse shock signatures are rare and are found only in a small, selection-biased subset of GRBs with sufficiently early and well-sampled optical light curves. The presence or absence of a detectable reverse shock likely reflects both observational selection effects and intrinsic ejecta properties, including the degree of magnetization \citep{2014ApJ...785...84J, 2021MNRAS.505.4086G}.

\begin{figure}[ht]
    \centering
    \includegraphics[width=0.5\textwidth]{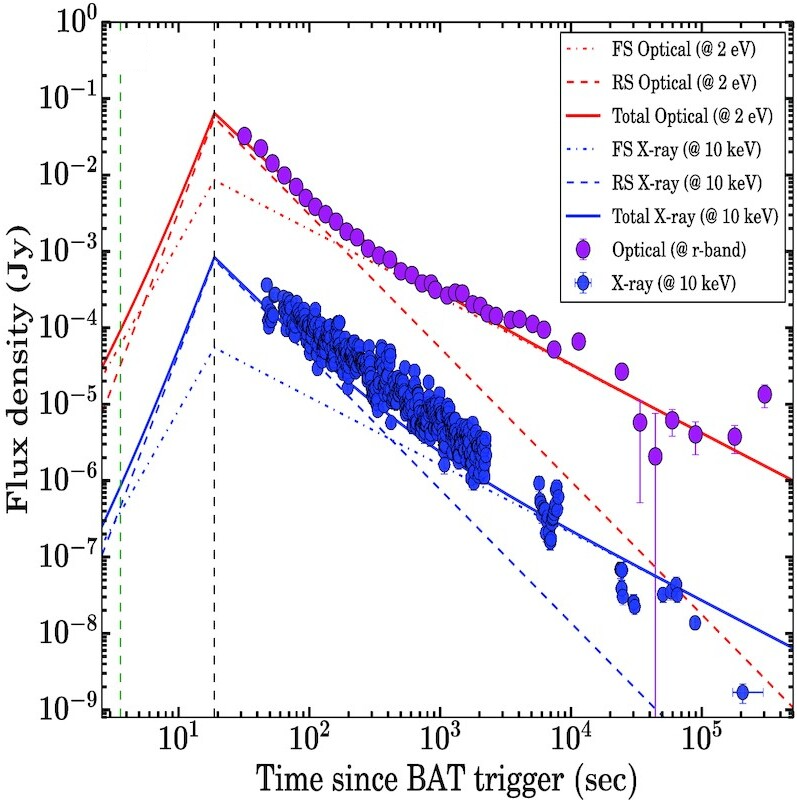}
\caption{The afterglow modeling of GRB 140102A using a combined thin shell reverse and forward shock model. The early observed optical flux is dominated by the RS component. The vertical dashed markers denote key temporal features: the green dashed line marks the end of the prompt duration ($T_{\rm 90}$), and the black dashed line indicates the inferred deceleration time for GRB 140102A. Adapted from \citet{2021MNRAS.505.4086G}.}
\label{fig:grb140102A}
\end{figure}

The prototype event is GRB~990123, for which ROTSE detected an optical flash reaching $V\sim9$ mag during the prompt $\gamma$-ray phase, followed by a steep decay with $\alpha \simeq 2$ \citep{1999Natur.398..400A}. This optical flash was far brighter than the extrapolated FS contribution and is well explained as synchrotron radiation from a reverse shock crossing the ejecta \citep{SariPiran1999,1999MNRAS.306L..39M,SoderbergRamirezRuiz2002}. In this interpretation, the RS characteristic frequency was below or near the optical band at the deceleration epoch, while the cooling frequency was above the optical band, possibly extending to the X-ray/soft-$\gamma$-ray regime \citep{SariPiran1999,SoderbergRamirezRuiz2002}. A similar early steep-to-shallow optical evolution was observed in GRB~021211. Observations using the Katzman Automatic Imaging Telescope began at $t\approx105$~s and showed a break at $t\sim10$~--~12~min, with the early component decaying as $\alpha\approx1.6$~--~1.8 before a shallower FS-dominated phase emerged; this behavior is commonly interpreted as the transition from RS-dominated to FS-dominated optical emission \citep{2003ApJ...586L...5F,2003A&A...402L...9W,2004MNRAS.353..511P}. Several {\it Swift}-era events strengthened the case that optical RS flashes are not unique to GRB~990123. GRB~061126 showed a rapidly fading early optical component that is broadly consistent with RS synchrotron emission, although detailed modeling requires non-trivial ejecta/circumburst parameters \citep{Perley2008,2008ApJ...687..443G}. GRB~120711A also displayed a powerful early optical flash peaking at $R\sim11.5$ mag at $t\sim126$~s, with the optical light curve well described by the superposition of RS and FS components \citep{2014A&A...567A..84M}. GRB~140102A showed a distinct early optical RS component in the thin-shell regime for a constant-density medium (see Figure \ref{fig:grb140102A}), with a decay $F_\nu\propto t^{-1.72}$ and a moderate magnetization parameter, $R_{\rm B}$ ($\epsilon_{B,\rm RS}/\epsilon_{B,\rm FS}$) $\approx18$, suggesting a baryonic jet with dynamically relevant magnetic fields \citep{2021MNRAS.505.4086G}.

More recent broadband studies have shown that RS emission can dominate not only the earliest optical emission but also the radio/mm and even high-energy components in favorable cases. For GRB~130427A, combined radio, millimetre, UV/optical/NIR, and X-ray modeling required distinct RS and FS components, with the RS dominating the radio/mm emission and the UV/optical/NIR emission at $t\lesssim0.1$~d \citep{2013ApJ...776..119L}. Finally, GRB~180720B has provided one of the most compelling recent cases in which early optical synchrotron emission and GeV $\gamma$-rays can be explained by synchrotron and inverse-Compton radiation from the RS propagating into weakly magnetized ejecta; the associated optical polarization evolution further indicates a transition from RS- to FS-dominated emission \citep{Arimoto+24}. 

Polarimetry provides powerful complementary diagnostics of shock microphysics and magnetic field topology. Early optical polarimetry requires specialized instrumentation on rapid-response telescopes but yields uniquely decisive constraints. GRB~090102 provided one of the clearest polarimetric signatures: its early optical light curve showed a steep-to-shallow transition characteristic of RS-to-FS evolution, and optical ring-polarimeter (RINGO) measured a high optical polarization of $\sim10$ per cent at $t\sim160$~s, suggesting large-scale magnetic fields in the ejecta \citep{2009Natur.462..767S}. GRB~120308A showed even higher polarization ($\Pi = 28 \pm 4$\%) with minimal evolution in the polarization position angle ($\theta_{\rm PA}$), suggesting large-scale toroidal or helical fields advected from the central engine \citep{2013Natur.504..119M}. In GRB\,180720B, the reverse-shock dominated optical light curve had a peak linear polarization of $\Pi\sim5\%$, and it showed rapid oscillations in the polarization degree as the emission switched from being reverse to forward shock dominated, where the more typical $\Pi\sim2\%$ was measured during the later forward shock dominated phase. More interestingly, the polarization angle jumped by $\sim90^\circ$ between the reverse and forward shock dominated emissions, indicating a similar change in the orientation of the magnetic field in the two distinct emission regions \citep{Arimoto+24}. These measurements constrain the magnetic field coherence length and anisotropy in both the ejecta and the shocked medium behind the two shocks \citep{2020MNRAS.491.3343G}. Such inferences then allow us to understand the magnetic field amplification mechanism, e.g., turbulent dynamo processes in the shock transition layer can generate small-scale tangled fields, while primordial fields compressed by the reverse shock can maintain coherence \citep{2001ApJ...563L..15G}.

The spectral evolution of reverse shock emission provides additional constraints on physical parameters. The synchrotron self-absorption frequency $\nu_a^{\rm RS}$ is typically in the radio band, below optical frequencies, placing the optical regime above $\nu_a^{\rm RS}$ but often below the cooling frequency $\nu_c^{\rm RS}$, corresponding to the slow-cooling regime \citep{2023NatAs...7..986B}. The spectral slope at frequencies $\nu_m^{\rm RS}<\nu<\nu_c^{\rm RS}$ is thus $F_\nu \propto \nu^{-\beta}$ with $\beta = (p-1)/2 \approx 0.5$~--~1.0 for electron power-law index $p \approx 2$~--~3. However, in cases where inverse Compton scattering is important, the cooling frequency can be pushed to lower energies, transitioning to the fast-cooling regime ($\nu_c^{\rm RS} < \nu_{\rm opt}$) and steepening the spectrum to $\beta = p/2$ \citep{2001ApJ...548..787S}. Multi-wavelength observations are essential for locating these break frequencies and for separating reverse and forward shock contributions: GRB~990123 showed evidence for $\nu_m^{\rm RS} < \nu_{\rm opt} < \nu_c^{\rm RS}$ during the peak, with X-ray observations constraining $\nu_c^{\rm RS}$ to be in the keV range.

The absence of reverse shock signatures can be attributed to several factors: (1) high magnetization ($\sigma \gtrsim 1$) suppressing reverse shock formation, (2) low circumburst densities ($n \lesssim 0.1$~cm$^{-3}$), which increase the deceleration time and therefore delay both the peak of the forward shock and the reverse shock crossing time beyond typical early observational windows, (3) very high initial Lorentz factors ($\Gamma_0 \gtrsim 500$), which reduce the deceleration time and shift the forward-shock onset and any associated reverse shock peak to very early times ($t_{\rm dec}\lesssim 10$~s), before optical follow-up begins, or (4) thin-shell conditions ($t_\times > t_{\rm grb}$) where the reverse shock is weak and short-lived \citep{2007AdSpR..40.1186Z}. Furthermore, the correlation between reverse shock brightness and prompt  $\gamma$-ray properties, with more energetic bursts showing stronger optical flashes, supports models where both emissions originate from related shock dissipation processes in stratified outflows \citep{2018ApJ...859...70F}. Thus, reverse shock signatures serve as critical probes of the physical conditions in relativistic outflows, constraining jet composition, ejecta magnetization, shell thickness, and energy dissipation mechanisms from the central engine to the external shock phase \citep{2005ApJ...628..315Z,2014ApJ...785...84J}.

\subsection{Optical Flares}
\label{subsec:optical_flares}

Early-time optical afterglow light curves often deviate from the smooth power-law decay, instead exhibiting transient rebrightenings (flares) superimposed on the underlying afterglow emission. These optical flares, occurring on timescales from minutes to hours post-burst, represent a common phenomenon in GRB afterglows. They are characterized by localized flux enhancements with comparatively short variability timescales. The temporal compactness of these features is typically quantified by the dimensionless ratio $\Delta t/t_{\rm pk}\ll 1$, where $t_{\rm pk}$ is the flare peak time and $\Delta t$ its characteristic width \citep{2005ApJ...631..429I, 2007MNRAS.380.1744N}. The parameter $\Delta t/t_{\rm pk}$, combined with multi-band spectral evolution and simultaneous X-ray monitoring, provides robust diagnostics for discriminating between flares originating from late internal dissipation versus external shock processes \citep{2007ApJ...671.1903C}. Interpreting these optical flares requires distinguishing intrinsic flux variability from chromatic evolution induced by the passage of spectral break frequencies through the observing bands, as well as establishing temporal correlations with X-ray flares \citep{2006ApJ...642..389N, 2006ApJ...642..354Z}.
 
\begin{figure}[ht]
    \centering
    \includegraphics[width=0.55\textwidth]{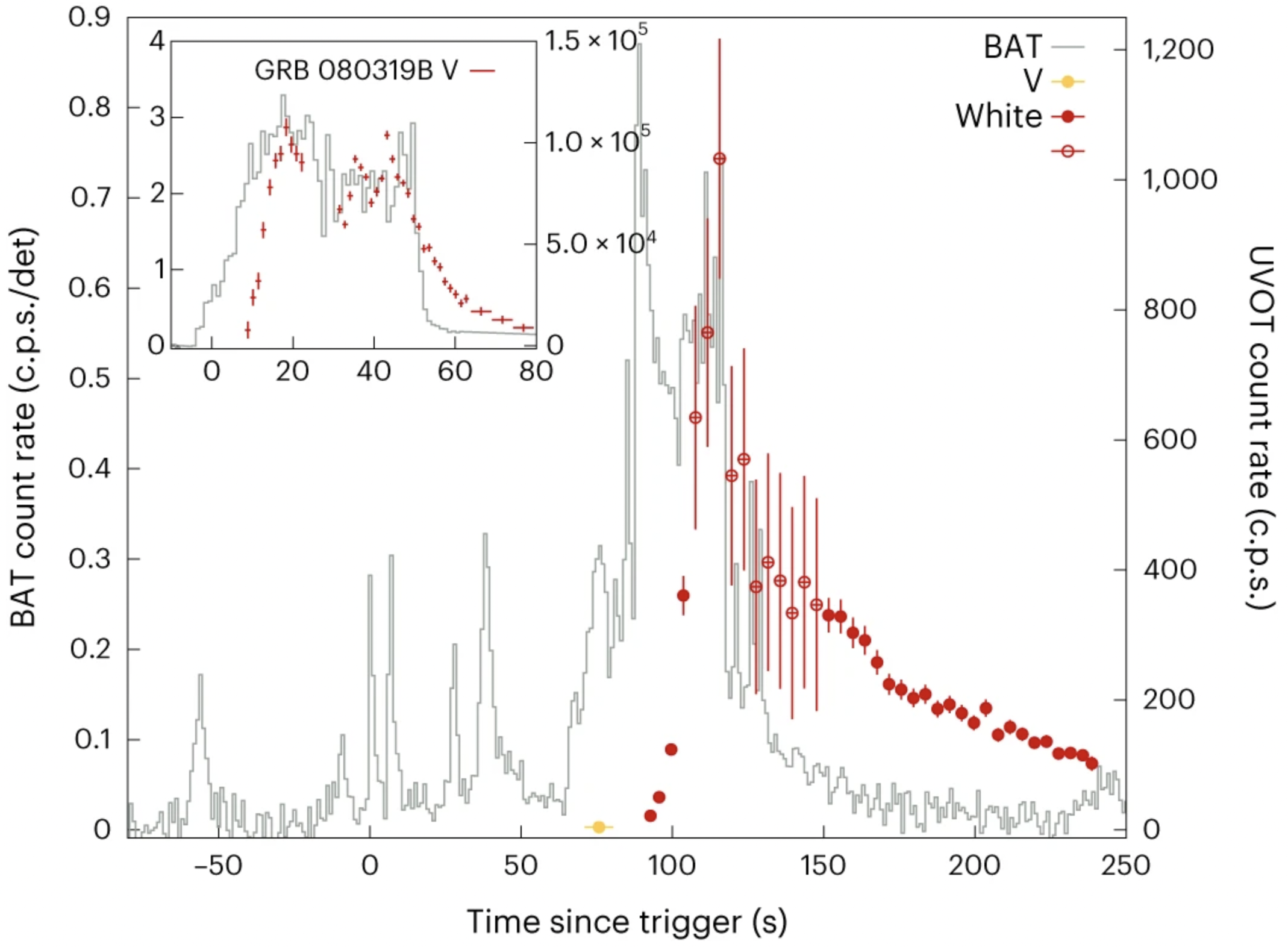}
\caption{Optical--UV flare of GRB~220101A compared with the prompt $\gamma$-ray
activity: The \swift/BAT prompt emission light curve, shown over approximately $T_0-70$ to $T_0+250$~s, exhibits rapid prompt $\gamma$-ray variability, whereas the contemporaneous \swift/UVOT light curve in the $V$ and white bands, obtained mainly during $\sim$ $T_0+80$ to $T_0+230$~s, evolves more smoothly and does not closely track the $\gamma$-ray temporal structure. This behavior differs from the well-known case of GRB~080319B, shown in the inset for reference (axes follow the same conventions as the main panel), where the optical emission more closely followed the high-energy activity. Instead, following \citet{2023NatAs...7.1108J}, we refer to this component as an optical--UV flare because its temporal behavior does not simply trace the $\gamma$-ray activity and likely has a distinct physical origin, possibly refreshed shocks from late ejecta catching up with the earlier decelerating outflow. Adapted from \citet{2023NatAs...7.1108J}.}
\label{fig:grb220101A}
\end{figure}

Optical flares exhibit morphological diversity; some are symmetric, with a fast rise and decay, while others show asymmetric evolution. Systematic searches using \swift/UVOT and coordinated ground-based observations have established that UV/optical flaring represents a common phenomenon, though detectability depends critically on observational cadence and photometric depth. The availability of homogeneous UV/optical observations, particularly from \swift/UVOT, has enabled systematic flare identification and statistical characterization \citep{2013ApJ...774....2S, 2025MNRAS.543.2404R}. \citet{2017ApJ...844...79Y} compiled a sample of 119 UV/optical flares from the UVOT catalog, including 77 with spectroscopic redshift measurements, and identified tight empirical correlations among flare timescales: rise time scales with decay time, and total duration correlates with peak time. Furthermore, they demonstrated that the frequency distributions of multiple flare parameters, including duration, rise time, decay time, peak time, and waiting time, are consistent with power-law forms and exhibit striking similarity to the corresponding distributions observed in X-ray flares. These empirical similarities support the hypothesis that a substantial fraction of optical flares share a common physical origin with X-ray flares, implicating prolonged and intermittent central-engine activity \citep{2017ApJ...844...79Y}.

Occasionally, optical/UV flares reach extraordinary luminosities that provide stringent constraints on the emission site and radiative mechanism. A notable example is the exceptionally energetic UV/optical flare observed in GRB~220101A (see Figure \ref{fig:grb220101A}), which reached an absolute AB magnitude of approximately -39.4 \citep{2023NatAs...7.1108J}. The UV/optical emission overlaps the prompt $\gamma$-ray phase, but its temporal evolution does not simply trace the $\gamma$-ray activity. This behavior distinguishes GRB~220101A from classical prompt optical events such as GRB~080319B and suggests that the optical--UV flare likely arose from a different component, possibly refreshed shocks produced by late-ejected energetic material catching up with the earlier decelerating outflow \citep{2023NatAs...7.1108J}.

While optical flares are sometimes contemporaneous with X-ray flares, this correspondence is not universal; optical flares can appear without a detectable X-ray counterpart, and vice versa. The statistical similarities between optical and X-ray flare properties and parameter distributions reported by \citet{2017ApJ...844...79Y} motivate a multi-mechanism framework for interpreting the diverse phenomenology of the flares. Several physical scenarios have emerged from observational and theoretical considerations: (i) \emph{Late internal dissipation:} Renewed central-engine activity can drive internal energy dissipation at late times through internal shocks \citep{2006ApJ...642..354Z} or magnetic reconnection in a Poynting-flux-dominated outflow \citep{2011MNRAS.413.2031M}. This mechanism naturally produces temporally narrow flares with $\Delta t/t_{\rm pk}\ll 1$ and characteristically steep rise and decay indices. The expected emission is inherently chromatic, reflecting high-latitude curvature effects and evolving microphysical parameters within the dissipation region. (ii) \emph{Refreshed shocks and energy injection:} Slower ejecta shells catching up to the decelerating blast wave can inject additional energy into the external forward shock. Refreshed shocks are expected to occur after the blast wave has begun to decelerate and therefore may appear at somewhat later epochs, depending on the Lorentz-factor distribution of the ejecta. This process typically yields broader temporal profiles ($\Delta t/t_{\rm pk}\sim 1$) and more nearly achromatic rebrightenings across the optical-to-X-ray spectrum \citep{Rees-Meszaros-98, 1999ApJ...520..641S}. (iii) \emph{Circumburst density structure:} Inhomogeneities in the circumburst medium, arising from stellar-wind variability or asymmetries in the progenitor environment, can in principle modulate the external-shock emissivity as the blast wave encounters over-dense clumps, shells, or wind-termination shocks \citep{2002A&A...396L...5L,2002ApJ...565L..87D}. Such features are also expected to become relevant after the blast wave has propagated into the external medium. However, this mechanism is strongly limited by relativistic equal-arrival-time effects, the finite radial extent of the emitting shock front, and geometric smoothing. Analytic arguments show that sharp external-density fluctuations are strongly smoothed in the observed light curve \citep{2005ApJ...631..429I,2007MNRAS.380.1744N}, while relativistic hydrodynamic simulations of blast waves crossing sudden density increases or decreases find that such transitions produce smooth changes in the afterglow normalization or decay slope rather than distinct flares \citep{2013ApJ...773....2G}. Density structure may therefore contribute to broad, approximately achromatic modulations, but is unlikely to explain the rapid, high-amplitude optical flares that are more naturally associated with late internal dissipation or refreshed shocks. A related scenario involves \emph{two-component jet structures}, wherein jets composed of multiple outflow components with different bulk Lorentz factors produce complex light curves with apparent flares as slower, wider components decelerate into the observer's line of sight \citep{2003ApJ...591.1086G, 2008Natur.455..183R}. 

Discriminating between these physical channels observationally requires three key capabilities: (a) high-cadence robotic monitoring capable of resolving the rapid variability timescale $\Delta t/t_{\rm pk}$ on minutes to hours scales, (b) rapid multi-band photometry to quantify spectral evolution and chromaticity, thereby constraining the evolution of synchrotron break frequencies through the observing bands, and (c) contemporaneous X-ray coverage (e.g., with \swift/XRT or \emph{Einstein Probe}/FXT) to establish temporal and spectral correlations between optical and X-ray emission components and to enable broadband spectral energy distribution (SED) modeling. Simultaneous multi-wavelength observations during flare episodes provide the maximum diagnostic power: chromatic flares, with discordant temporal behavior in optical and X-ray bands, suggest emission from distinct components or high-latitude contributions from late internal dissipation, whereas achromatic flares implicate hydrodynamic processes or energy injection into a common emitting region \citep{2010ApJ...720.1513K, 2011MNRAS.410.1064M}.

\subsection{Forward Shock and Onset of the Afterglow}

The forward shock, propagating outward into the circumburst medium at relativistic velocities, produces long-lived afterglow emission via synchrotron radiation from shock-accelerated electrons. The onset and early evolution of this emission encodes fundamental information about the blast wave dynamics, including the deceleration radius, the external density profile (constant or wind-like medium), and the total kinetic energy content of the relativistic outflow \citep{1998ApJ...497L..17S, 2006RPPh...69.2259M}. Consequently, early multi-wavelength observations during the first minutes to hours post-trigger provide a unique diagnostic window into the physical conditions at the shock front and the structure of the circumburst environment.

In the standard fireball model, the relativistic ejecta initially undergoes a coasting phase during which the bulk Lorentz factor remains approximately constant at $\Gamma_0$. As the outflow sweeps up the circumburst medium, deceleration begins once the swept-up external mass becomes dynamically important, i.e., when
\[
m_{\rm ext}(R_{\rm dec}) \sim \frac{M_0}{\Gamma_0}
\simeq \frac{E_{\rm k,iso}}{\Gamma_0^2 c^2},
\]
where $M_0 \simeq E_{\rm k,iso}/(\Gamma_0 c^2)$ is the initial baryonic rest mass of the ejecta. This condition defines the deceleration radius $R_{\rm dec}$ (see Eq.~(5), \citealt{1992MNRAS.258P..41R, 1995ApJ...455L.143S}), and the corresponding observer-frame deceleration time is approximately $t_{\rm dec} \simeq (1+z)R_{\rm dec}/(2c\Gamma_0^2)$. For canonical parameters ($E_{\rm k,iso} \sim 10^{53}$ erg, $n_0 \sim 1$ cm$^{-3}$, $\Gamma_0 \sim 100$), the deceleration radius is of order $10^{16}$--$10^{17}$ cm, corresponding to observer frame timescales of hundreds to thousands of seconds. The timing of the early afterglow peak thus provides strong constraints on $\Gamma_0$, given reasonable estimates of $E_{\rm k}$ and $n_0$.

The observed characteristics of the afterglow onset depend sensitively on both viewing geometry and the deceleration dynamics. For observers within the jet opening angle $\theta_{\rm jet}$, the forward shock emission is visible from the onset of the external shock afterglow, because the relativistically beamed emission is directed toward the observer. In contrast, for off-axis observers with viewing angle $\theta_{\rm obs}>\theta_{\rm jet}+\Gamma_0^{-1}$, the early forward-shock emission is strongly suppressed, since the relativistic beaming cone of angular size $\sim \Gamma_0^{-1}$ does not initially include the line of sight. As deceleration proceeds and $\Gamma$ decreases, the half-opening angle of the beaming cone ($\sim 1/\Gamma$) expands, eventually encompassing off-axis sightlines and producing a characteristic rising light curve \citep{1999ApJ...520..641S, 2002ApJ...570L..61G}. 

\begin{figure}[ht]
    \centering
    \includegraphics[width=0.6\textwidth]{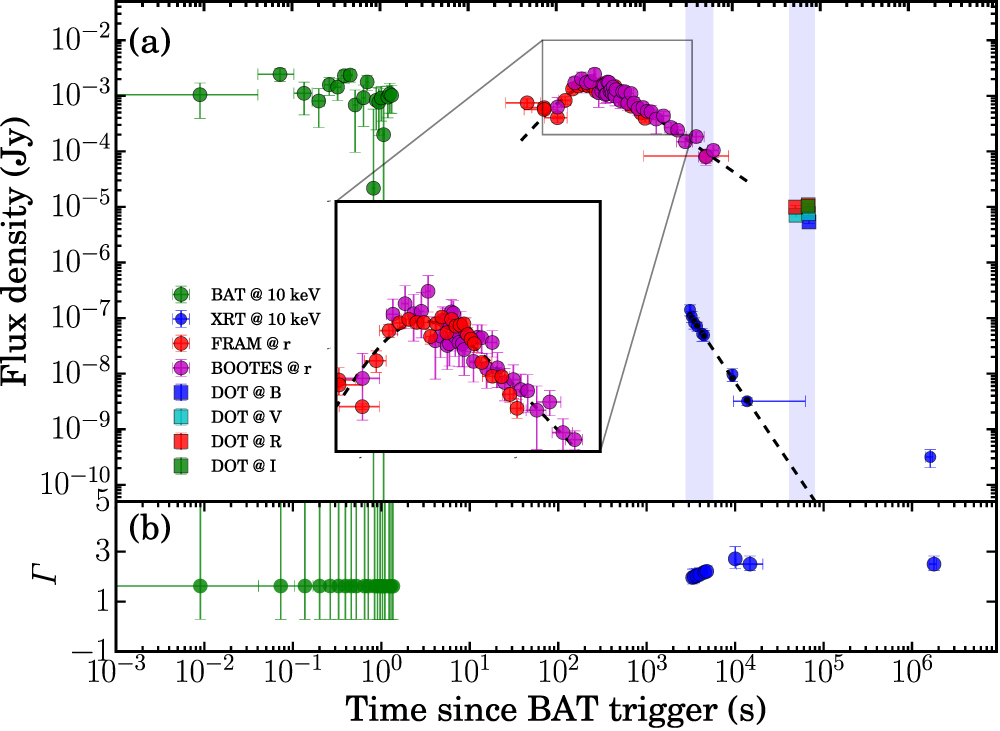}
\caption{Multiwavelength temporal and spectral evolution of GRB~201015A with the onset of afterglow. {(a)} Flux-density light curves in the high-energy and optical bands, including \swift/BAT ($\sim$10~keV; green), \swift/XRT ($\sim$10~keV; blue), and optical emission in the $r$ band (red/magenta). The inset highlights the early-time optical bump at higher temporal resolution. (b) Time evolution of the photon index derived from spectral fits in the \swift/BAT and \swift/XRT energy ranges, illustrating changes in the prompt-to-afterglow spectral behavior. Adapted from \citet{2023ApJ...942...34R}.}
\label{fig:grb201015A}
\end{figure}

\begin{figure}[t]
    \centering
    \includegraphics[width=0.6\columnwidth]{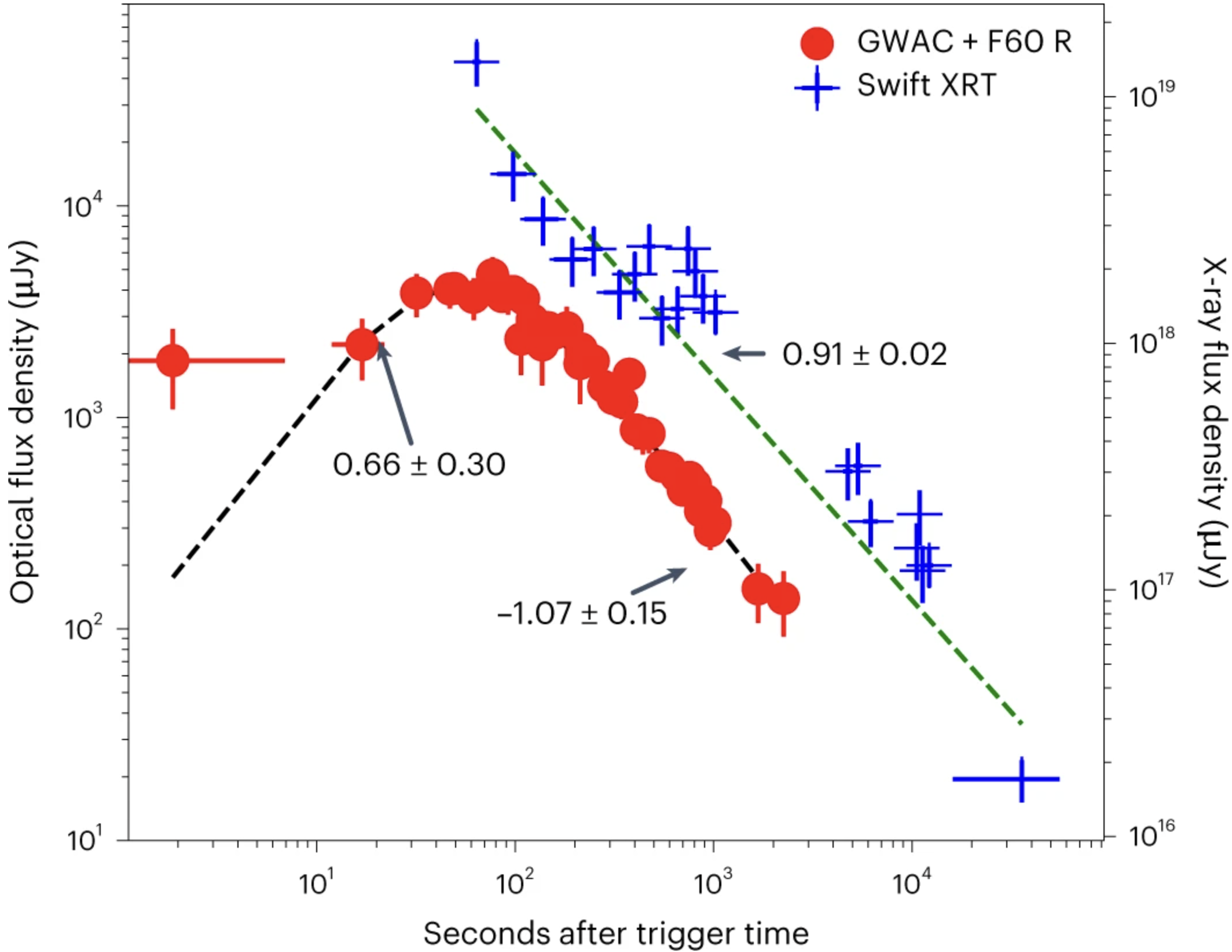}
    \caption{Optical and X-ray light curves of GRB~201223A and their modeling. The optical data (from GWAC and GWAC-F60A) show an initial shallow phase, a rise to a peak at $\sim 52$~s, and a subsequent decay, while the \swift/XRT light curve exhibits broadly similar early-time behavior. This figure illustrates a clear example of the prompt-to-afterglow transition and supports an interpretation in which the later optical evolution is dominated by the onset of the forward shock. Adapted from \citet{2023NatAs...7..724X}.}
    \label{fig:grb201223a_transition}
\end{figure}

High-cadence early optical monitoring has revealed diverse onset behaviors that challenge simplified theoretical models. GRB 060418, observed by the REM telescope beginning at $\sim$ 70 seconds after the \swift/BAT trigger, exhibited a smooth optical rise consistent with forward shock emission propagating into a uniform-density medium with $n_0 \approx 0.1$ cm$^{-3}$ \citep{2007Molinari, 2009MNRAS.395..490O}. Both the temporal and spectral evolution during this phase were well described by standard synchrotron models. Similarly, other bursts, including very high energy detected GRB 201015A (see Figure \ref{fig:grb201015A}), exhibited an onset of afterglow \citep{2023ApJ...942...34R}. Another particularly useful recent example is GRB~201223A (see Figure \ref{fig:grb201223a_transition}), for which GWAC and the follow-up F60A telescope captured the optical evolution from the prompt phase into the early afterglow. After the initial prompt optical detection, the light curve showed a short shallow phase, then a rise to a peak at $t_{\rm p}\sim 52$~s, followed by a power-law decay. This smooth optical rise and decay are consistent with the onset of the external forward shock, making GRB~201223A an important case in which wide-field robotic monitoring directly traced the prompt-to-afterglow transition and helped separate the early prompt optical component from the subsequent afterglow emission \citep{2023NatAs...7..724X}.

The broadband SED during the afterglow onset phase provides powerful constraints on the underlying synchrotron model parameters. In the standard framework, the SED is characterized by three critical break frequencies: the self-absorption frequency $\nu_a$, below which the emitting region becomes optically thick; the characteristic synchrotron frequency $\nu_m$ of freshly accelerated electrons with the minimum Lorentz factor at the shock; and the cooling frequency $\nu_c$, above which radiative losses on the dynamical timescale become efficient \citep{1998ApJ...497L..17S, 2002ApJ...570L..61G, 2014PASA...31....8G}. The relative ordering of these frequencies evolves with time and determines the observed spectral slopes in different wavebands. Simultaneous early observations spanning optical through X-ray wavelengths enable direct measurement of these characteristic frequencies and yield robust estimates of the micro-physical shock parameters, particularly the fractions of dissipated energy channeled into relativistic electrons ($\epsilon_e$) and magnetic fields ($\epsilon_B$) \citep{2002ApJ...571..779P}. Deviations from canonical power-law spectral indices can furthermore reveal additional physical processes, including inverse Compton scattering, Klein-Nishina suppression of high-energy photon production, or pair-loading effects in highly magnetized jets  \citep{2009ApJ...703..675N, 2010ApJ...712.1232W}.

\subsection{Optical plateau phases and the resulting correlations}
\label{sec:optical_plateaus}

Another important feature observed in early optical afterglows is the presence of shallow-decay or plateau phases, typically occurring on timescales of $\sim10^{2}$--$10^{4}$~s after the GRB trigger, although in some events the shallow phase can extend to later times. Observationally, an optical plateau is characterized by a temporal decay index significantly shallower than the standard forward-shock expectation, often with $\alpha_{\rm pl}\lesssim0.5$, followed by a transition to a steeper normal afterglow decay with $\alpha\sim1$--$1.5$ \citep{PanaitescuVestrand2011, 2012ApJ...758...27L}. These plateaus may be either achromatic, when the optical and X-ray bands show similar temporal evolution and break times, or chromatic, when the optical and X-ray light curves evolve differently \citep{2006MNRAS.369.2059P, 2007ApJ...670..565L}. The achromatic case is generally interpreted as evidence that the optical and X-ray emission arises from the same external forward shock, whose decay has been flattened by continued energy injection into the blast wave. Such energy injection may be produced by prolonged central-engine activity, refreshed shocks from slower ejecta shells catching up with the decelerating blast wave, a broad distribution of ejecta Lorentz factors, or spin-down power from a rapidly rotating magnetar \citep{Dai-Lu-98,2006ApJ...642..354Z,Rowlinson2013}. In contrast, chromatic optical plateaus suggest that a single simple forward-shock component is insufficient, and may instead require multiple emission zones, late internal dissipation, spectral-break evolution between the optical and X-ray bands, evolving microphysical parameters, or angular structure in the jet \citep{2007ApJ...658L..75G, 2022Galax..10...93S, PanaitescuVestrand2011}.

\begin{figure}[!ht]
    \centering
    \includegraphics[width=0.7\columnwidth]{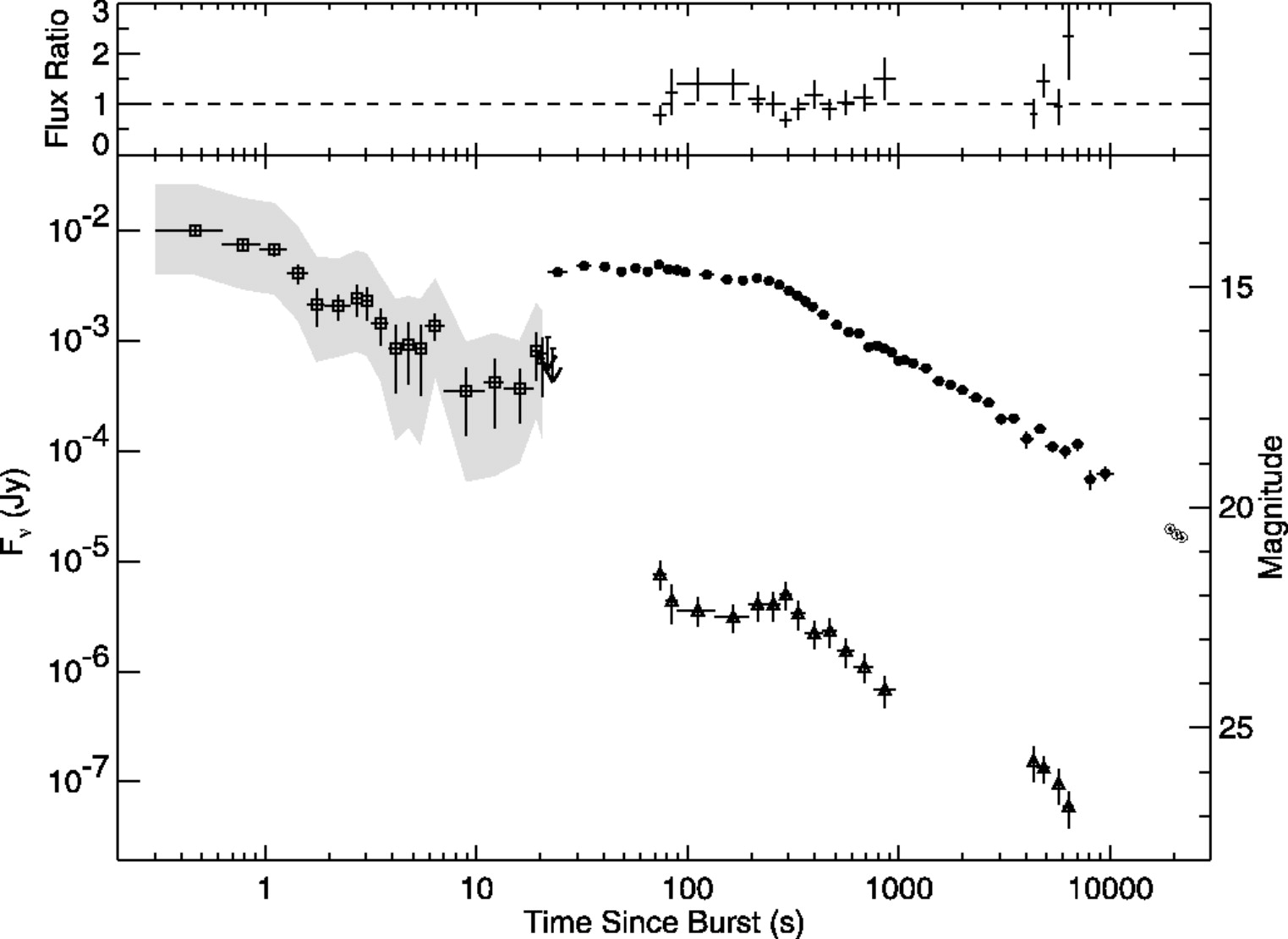}
    \caption{Early optical and X-ray plateau in GRB~050801, illustrating the importance of rapid robotic optical follow-up. ROTSE-III observations began only $\sim22$~s after the burst and revealed an approximately flat optical light curve during the first $\sim250$~s, followed by a transition to a normal afterglow decay. The contemporaneous \swift/XRT light curve showed a similar shallow phase and break time, while the nearly constant optical-to-X-ray flux ratio indicated approximately achromatic evolution. This behavior supports an external-shock origin with continued energy injection rather than prompt optical emission or a dominant reverse-shock component. Adapted from \citet{2006ApJ...638L...5R}.}
    \label{fig:grb050801_plateau}
\end{figure}

A particularly clear example of an early optical plateau enabled by robotic follow-up is GRB~050801. ROTSE-III began observing this burst only $\sim22$~s after the trigger and revealed an approximately flat optical light curve during the first $\sim250$~s, followed by a transition to a normal power-law decay. The contemporaneous \swift/XRT light curve showed a similar shallow phase and break time, while the optical-to-X-ray flux ratio remained nearly constant, indicating an achromatic evolution and supporting an external-shock origin with continued energy injection rather than a purely prompt or reverse-shock origin (Figure~\ref{fig:grb050801_plateau}; \citealt{2006ApJ...638L...5R,2007MNRAS.377.1638D}). This event illustrates why rapid-response optical/UV facilities are essential for early afterglow studies. \swift/UVOT played a key role in constraining the early plateau and its transition to the normal decay phase, demonstrating its importance for capturing such behavior. Robotic ground-based telescopes complement UVOT by extending the temporal coverage, increasing observing cadence, and providing additional filters and sensitivity during the rapidly evolving first few tens to hundreds of seconds.

Other well-observed events demonstrate that optical plateaus are phenomenologically diverse. GRB~060729 exhibited an unusually long-lived shallow decay phase, often discussed in the context of sustained energy injection or magnetar-powered evolution \citep{2007ApJ...662..443G}. GRB~080710 showed a broad, chromatic optical/NIR rise followed by a shallow decay and later steepening, indicating that a simple single-component
afterglow model is insufficient \citep{2009A&A...508..593K}. While \citet{2009A&A...508..593K} discussed an off-axis jet interpretation, recent detailed modeling by \citet{2024JHEAp..41....1O} found that the observed chromatic peaks cannot be reproduced by an off-axis GRB model. GRB~091029 displayed chromatic optical/NIR and X-ray evolution, suggesting that multiple emission components or spectral evolution were important during the early afterglow \citep{2012A&A...546A.101F}. The optical plateau at $\sim 10$ ks can be interpreted within a two-component jet framework, similar to that proposed for GRB~191221B \citep{191221B}. In this scenario, an initially fast, narrow jet component ($\Gamma_0 \sim 400$, $\theta_{\rm j}\sim1.4^\circ$) dominates the early optical decay, while a slower and wider component ($\Gamma_0 \sim 25$, $\theta_{\rm j}\sim2.8^\circ$) becomes dominant at later times and produces the optical plateau. The lack of a corresponding X-ray plateau can be explained if the wide component has a steeper spectrum, resulting in a less significant flux contribution in the X-ray band than in the optical. These examples show that an optical plateau is a phenomenological signature rather than a unique physical mechanism, and that broadband temporal and spectral coverage is required to distinguish between external-shock energy injection, central-engine activity, refreshed shocks, two-component jet framework, and geometric effects.

\begin{figure*}[!ht]
    \centering
    \includegraphics[width=0.850\textwidth]{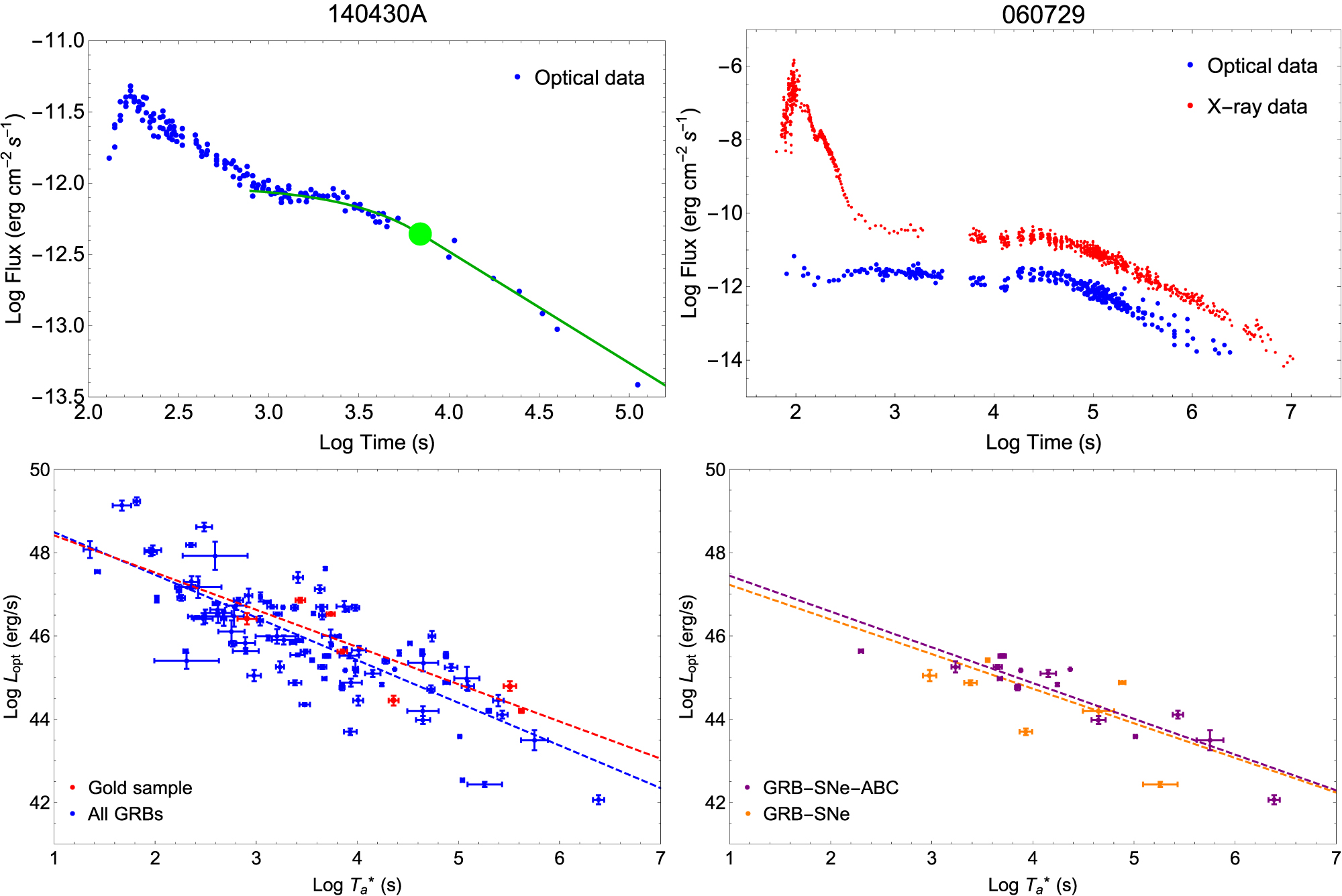}
    \caption{Upper left: the \cite{2007ApJ...662.1093W} fit for a well-sampled optical plateau shown as a green line, with the green dot representing the best fit. Upper right: another example of a well-sampled light curve with the coincident observation of the X-ray plateau. Lower panels: the relation for the Gold and the total sample (left) and for the GRB-SNe total and the GRB-SN (A, B, C) (right). The best-fit lines are calculated using a linear model fit in log scale and plotted in matching colors as dashed lines. Adapted from \citet{Dainotti2020}.}
    \label{fig:correlations}
\end{figure*}

Recent statistical studies have further shown that optical plateau luminosities and durations may follow correlations analogous to those identified in the X-ray band \citep{Dainotti2008,Dainotti2010,dainotti17a,Dainotti_2020}. In particular, the optical plateau luminosity is anti-correlated with the rest-frame plateau duration, with brighter plateaus tending to end earlier \citep{Dainotti2020,Dainotti2023,Dainotti2022ApJS..261...25D}, see fig \ref{fig:correlations}. This correlation is tighter when we consider the gold sample of GRBs, for which an almost flat plateau is found with a good coverage of data, and a tighter correlation is also found when we consider GRBs associated with SNe Ib/c, whose association is confirmed spectroscopically, or a clear optical bump resembling the spectroscopic association of other spectroscopically associated GRBs. An extension of the 2D relation by adding the peak of the prompt luminosities has led to the 3D Dainotti relation in optical, see  \citep{Dainotti2022ApJS..261...25D}, which has been further used as a cosmological tool \citep{Dainotti2022} similarly to how the 3D Dainotti X-ray correlation \citep{dainotti2022MNRAS.tmp.2639D} has been used as a cosmological tool. Similar relations have been reported between plateau properties and prompt/high-energy observables, extending the well-known X-ray Dainotti relation to the optical regime \citep{Dainotti2008,Dainotti2010,Dainotti2022}. These correlations suggest that optical plateaus may encode information about the GRB energy reservoir, jet evolution, and the lifetime of central-engine activity, and they may eventually provide useful cosmological probes once selection effects, cadence biases, redshift incompleteness, and intrinsic scatter are properly accounted for.

The detection and characterization of optical plateaus, therefore, provide important constraints on jet dynamics, ejecta stratification, energy injection, and central-engine longevity. Robotic telescopes play a central role because they can respond within seconds, obtain dense early-time light curves, and identify whether the shallow phase begins immediately after the prompt emission or emerges after early flares or reverse shock components. When combined with simultaneous X-ray observations and multi-band optical/NIR data, robotic optical monitoring provides one of the most direct ways to determine whether plateau phases are achromatic signatures of blast-wave energy injection or chromatic signatures of multiple emission components.

\subsection{Smooth Optical Self-Similar Emission of GRBs}

A subset of GRBs with very early optical coverage exhibits a remarkably smooth, single-peaked optical evolution: the flux rises gradually, reaches a broad maximum, and then decays without strong flares or discontinuities. This phenomenological behavior has been termed Smooth Optical Self-Similar Emission, hereafter SOSSE emission \citep{Lipunov2017SOSSE}. The defining feature of this class is that, after normalization by the peak flux and peak time, the early optical light curves can be described by a common dimensionless profile.

\begin{figure}[!ht]
    \centering
     \includegraphics[width=0.8\columnwidth]{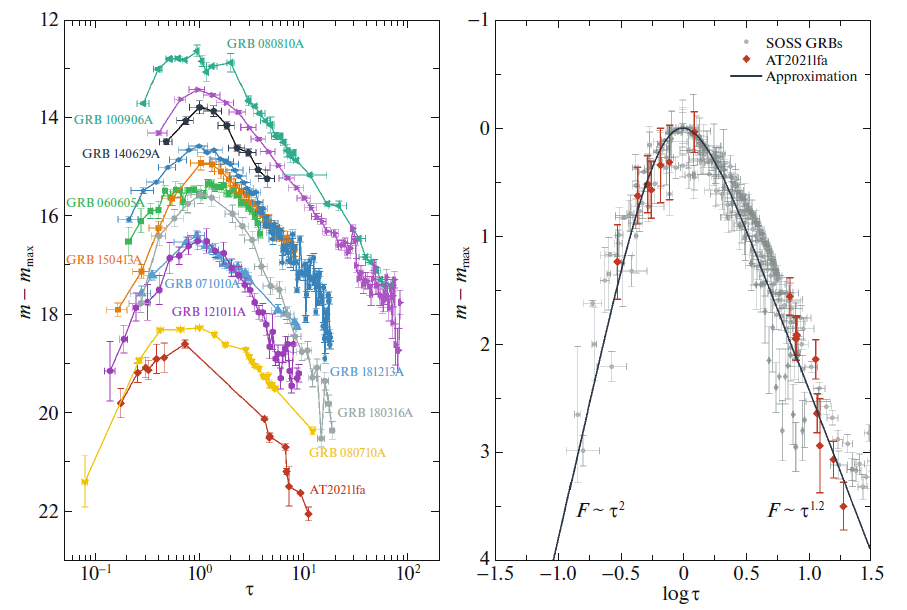}
    \caption[width=\columnwidth]{Smooth optical self-similar emission in early GRB afterglows. Left: observed optical light curves of the selected smooth, single-peaked GRB sample, shown as apparent magnitude versus time since the high-energy trigger. Right: the same light curves after normalization by the peak flux and peak time, compared with the empirical SOSSE profile given by Eq.~\eqref{eq:sosse}. The red points illustrate the application of the SOSSE scaling to the orphan afterglow candidate AT2021lfa/ZTF21aayokph \citep{Lipunov2022AT2021lfa}, for which the explosion epoch can be recovered from the optical light curve.}
    \label{fig:sosse}
\end{figure}

The selection of SOSSE candidates requires well-sampled optical observations beginning during the prompt or very early afterglow phase, ideally close to or before the end of the prompt $\gamma$-ray emission. We further require a smooth, single optical maximum, with no strong flaring, rebrightening, or discontinuous transition between the rising and decaying phases. Although the selected bursts span more than two orders of magnitude in peak time and optical brightness, their normalized light curves occupy a narrow region in the scaled plane shown in Figure~\ref{fig:sosse}.

For each burst, the optical flux density is normalized to its peak value, $F_{\nu,\max}$, and the time is normalized by the delay between the adopted burst onset time, $t_0$, and the optical peak time, $t_{\rm pk}$:
\begin{equation}
    m-m_{\rm pk} =
    -2.5\log_{10}\left(\frac{F_\nu}{F_{\nu,\max}}\right),
    \qquad
    \tau =
    \frac{t-t_0}{t_{\rm pk}-t_0}.
\end{equation}
In most cases, we take $t_0$ to be the high-energy trigger time. Since both the numerator and denominator in $\tau$ are observed-frame time intervals, the normalization largely removes the common cosmological time-dilation factor. However, systematic uncertainty can remain if the high-energy trigger was caused by a precursor or occurred after the true onset of the event.

The normalized SOSSE profile can be represented by the empirical function \citep{Lipunov2017SOSSE}
\begin{equation}
    \frac{F_\nu(\tau)}{F_{\nu,\max}} =
    \left[
    \frac{\beta \tau^{\beta-1}}
    {1+(\beta-1)\tau^\beta}
    \right]^\alpha ,
    \label{eq:sosse}
\end{equation}
where $\alpha \simeq 1.2$ and $\beta \simeq 2.7$. This form peaks at
$\tau=1$ and asymptotically gives a fast early rise,
$F_\nu \propto \tau^{\alpha(\beta-1)} \approx \tau^2$, followed by a
standard afterglow-like decay, $F_\nu \propto \tau^{-\alpha} \approx
\tau^{-1.2}$.

The physical origin of SOSSE is naturally connected to the onset of the external shock as the relativistic ejecta are decelerated by the circumburst medium \citep{BlandfordMcKee1976,MeszarosRees1997,1998ApJ...497L..17S}. In this interpretation, the smooth optical peak marks the transition from the rising phase of the forward shock to the self-similar deceleration regime. A reverse shock contribution may also be important at early times and can modify the detailed shape of the optical peak \citep{SariPiran1999,2000ApJ...545..807K}. Therefore, SOSSE should be regarded as a phenomenological description of a subset of smooth early optical afterglows, rather than as a unique physical model.

The same scaling can also be used to estimate the explosion epoch of optical transients without a detected high-energy trigger. For the orphan afterglow candidate AT2021lfa/ZTF21aayokph, the SOSSE framework was used to constrain the burst time to within $\lesssim 14$ min \citep{Lipunov2022AT2021lfa}, as shown by the red points in Figure~\ref{fig:sosse}.

\section{Physical constraints from rapid optical observations}
\label{sec:science}

Robotic telescopes that capture optical emission within seconds to hours of GRB triggers have opened unprecedented windows into jet physics, central engine properties, and radiative processes. Section~\ref{sec:features} summarized the theoretical expectations for the main early optical components: prompt optical emission, reverse-shock flashes, forward-shock onset, optical flares, plateaus, and rebrightenings. In this section, we focus instead on the physical constraints enabled by rapid robotic optical follow-up (see Table \ref{tab:obs_diagnostics}). The central question is not only whether an early optical counterpart is detected, but what can be inferred from its timing, luminosity, color evolution, and temporal correlation with high-energy emission and polarization. These observables provide constraints on the emitting region, the initial Lorentz factor, ejecta magnetization, circumburst density profile, dust extinction, the duration of central-engine activity, and the transition phases between prompt and afterglow regimes \citep{2013EAS....61..203P}.

In practice, early optical light curves often contain several overlapping contributions. A single early detection is therefore rarely sufficient for a unique physical interpretation. The strongest constraints come from well-sampled, multi-band light curves obtained within seconds to minutes after the trigger, together with simultaneous $\gamma$-ray, X-ray, radio/mm, and, where possible, polarimetric data.

\begin{table*}
\centering
\caption{Main observational diagnostics from rapid optical follow-up and
the physical constraints they provide.}
\label{tab:obs_diagnostics}
\begin{tabular}{p{0.24\textwidth}p{0.30\textwidth}p{0.34\textwidth}}
\hline
Observable & Diagnostic use & Main physical constraints \\
\hline
Temporal overlap with prompt $\gamma$-rays &
Tests whether optical emission is produced by internal dissipation or by
the external shock. &
Emission radius, prompt-radiation mechanism, optical-to-$\gamma$-ray
spectral connection, internal-shock or magnetic-reconnection origin. \\

Fast optical rise and steep decay &
Identifies reverse-shock dominated emission when followed by a shallower
forward-shock component. &
Initial Lorentz factor, shell thickness, ejecta magnetization, ratio of
magnetic-field energy in reverse and forward shocks. \\

Smooth achromatic onset peak &
Identifies the deceleration time of the external forward shock. &
Initial bulk Lorentz factor $\Gamma_0$, deceleration radius, kinetic
energy, and circumburst density normalization. \\

Optical--X-ray chromaticity &
Separates single-component external-shock evolution from multiple
emission zones. &
Energy injection, late internal activity, refreshed shocks, spectral
break evolution, and central-engine lifetime. \\

Multi-band optical/NIR SED &
Measures the spectral slope and reddening at early times. &
Host extinction $A_V$, extinction-curve shape, high-redshift dropout,
intrinsic synchrotron spectrum, dust in the circumburst environment. \\

Early optical polarization &
Tests whether the magnetic field is ordered or shock-generated. &
Large-scale magnetic fields in the ejecta, reverse-shock magnetization,
jet structure, transition from reverse- to forward-shock dominance. \\
\hline
\end{tabular}
\end{table*}

The first observational task is to determine which physical component dominates the early optical light curve. The features described in Section \ref{sec:features} rarely appear in isolation, and identifying which one dominates a given early optical light curve requires comparing its temporal and spectral behavior with the prompt $\gamma$-ray and early X-ray emission (see Table \ref{tab:obs_diagnostics}). This component separation is one of the most important contributions of robotic telescopes. Without observations during the first few hundred seconds, reverse shocks, afterglow onsets, and prompt optical components can be missed or blended into a single poorly constrained early point. High-cadence observations are therefore essential for determining whether the early emission probes the central engine, the ejecta, or the external medium.

\subsection{Optical radiation as a probe of the prompt emission mechanism}
\label{prompt_optical_models}

The physical origin of the MeV prompt emission remains one of the central open questions in GRB physics. The rapid variability observed in prompt emission light curves, from milliseconds to sub-second timescales, strongly indicates that the radiation is produced by internal dissipation within ultra-relativistic jets \citep{Rees1994}. However, the nature of the dissipation mechanism and the radiative process responsible for shaping the observed keV--MeV spectra remains debated.

A historically important starting point is the pair fireball model, in which a compact, optically thick plasma is launched from the vicinity of the central engine and accelerates under its own radiation pressure \citep{Paczynski1986,Goodman1986,Shemi1990}. In this simple model, the fireball releases quasi-thermal radiation at the photosphere, with luminosities of order $10^{50}$~--~$10^{52}\,\mathrm{erg\,s^{-1}}$. This simple model captures the typical luminosities, photon energies (MeV), and relativistic expansion of GRB jets, but it fails to reproduce the observed prompt emission spectra. Most GRB photon spectra are much broader than a blackbody and are usually described by the Band function, namely two power laws, $N(E)\propto E^\alpha$ and $N(E)\propto E^\beta$, smoothly connected around the peak of the $\nu F_\nu$ (or $E^2N(E)$) spectrum at $E_{\rm p}\sim100\,\mathrm{keV}$~--~$1\,\mathrm{MeV}$ \citep{Band1993}. The typical values, $\alpha\sim-1$ and $\beta\sim-2.3$ \citep{Gruber2014}, point to a non-thermal or strongly modified photospheric origin.

A natural extension of the pair fireball scenario is obtained when the outflow carries baryons. In this case, part of the initial radiation energy is converted into kinetic energy \citep{Shemi1990}. Variability in the central engine then produces shells with different Lorentz factors, and collisions between faster and slower shells can dissipate kinetic energy through internal shocks above the Thomson photosphere \citep{Rees1994}. In this class of models, the temporal separation between prompt pulses reflects the variability of the central engine, while the duration of an individual pulse is limited by the angular timescale, $t_{\rm ang}\sim R_\gamma/(2c\Gamma^2)$, where $R_\gamma$ is the prompt-emission radius and $\Gamma$ is the bulk Lorentz factor of the emitting region.

\begin{figure*}[!ht]
    \centering
    \includegraphics[width=0.8\textwidth]{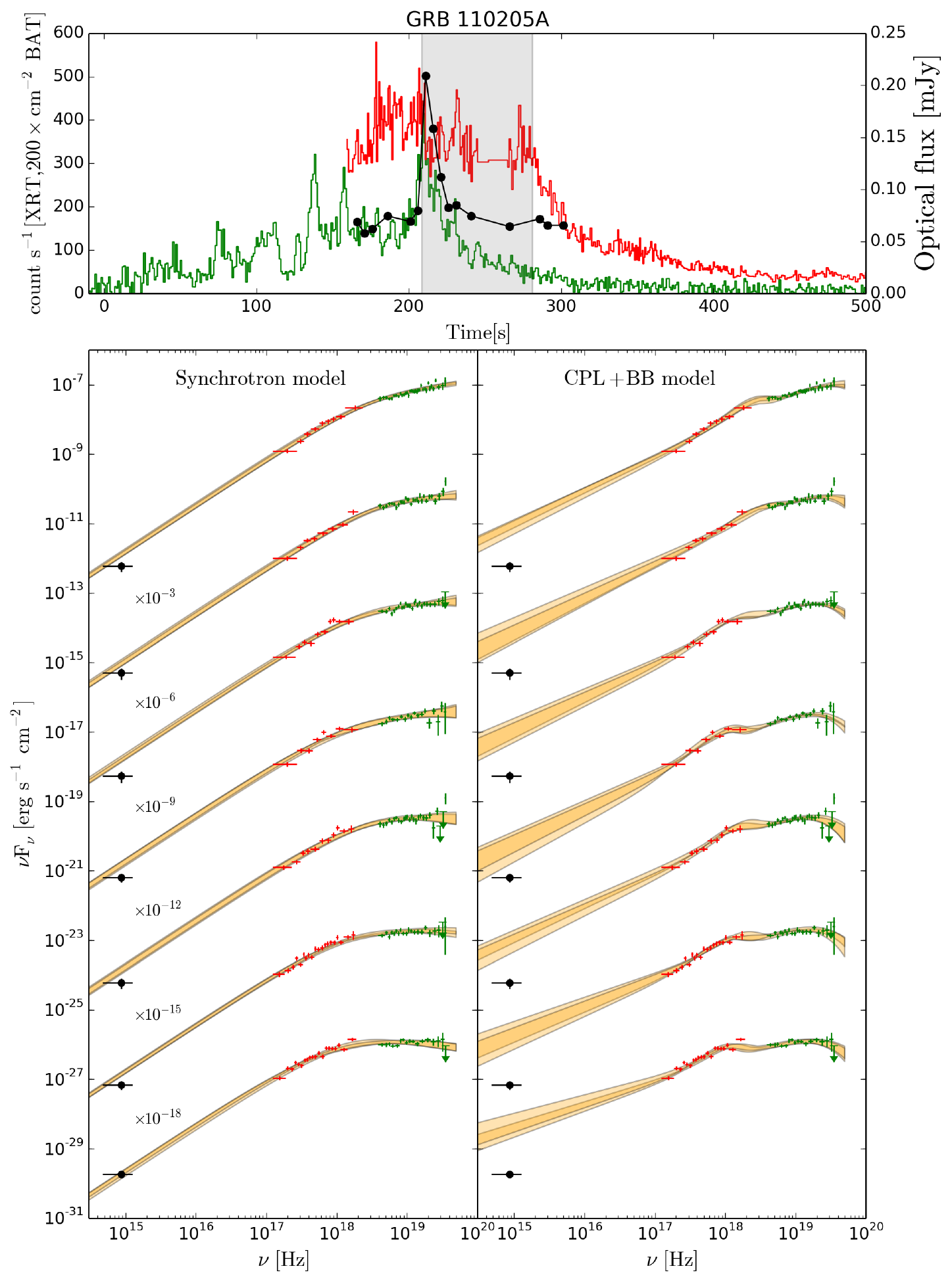}
    \caption{Example of broadband time-resolved prompt emission spectra including the optical data, shown for GRB~110205A. The optical data are compared with the extrapolation of the high-energy prompt emission model, illustrating the cases in which the optical emission is consistent with the same synchrotron component that produces the X-ray and $\gamma$-ray prompt emission and inconsistent with the empirical, two-component model. Adapted from \citet{Oganesyan2019}.}
    \label{fig:110205A}
\end{figure*}

In optically thin internal dissipation models, synchrotron radiation from shock-accelerated or reconnection-accelerated electrons is the most natural emission mechanism \citep{2022MNRAS.511.1694G, 2024ApJ...972..166G}. A power-law energy distribution of relativistic electrons can account for the high-energy spectral slope, $\beta\simeq-2.3$, in a straightforward way. The difficulty arises at lower energies. If the electrons cool efficiently within the dynamical time, as expected in a radiatively efficient GRB source, the synchrotron spectrum should be in the fast-cooling regime and should have $\alpha=-3/2$ below the spectral peak. This is softer than the typical observed value, $\alpha\sim-1$, and much softer than the hardest spectra that approach the synchrotron limit $\alpha=-2/3$ \citep{Preece1998}. However, when prompt X-ray/optical coverage is available, as in GRB~110205A (see Figure~\ref{fig:110205A}), the spectrum can be described by a broken synchrotron-like shape with slopes consistent with fast cooling \citep{Zheng+12,Oganesyan2019}. The problem is particularly severe because low radiative efficiency would make the energy budget of the jet even more demanding, especially in internal-shock scenarios where the dissipation efficiency is already modest \citep{Beniamini2016}. Thus, the low-energy spectral index problem is a generic challenge for optically thin synchrotron models, independently of whether the jet energy is mainly kinetic or magnetic. However, fitting the exact synchrotron spectrum from cooling electrons to observations, rather than the phenomenological Band function, has led to a better match at lower energies \citep{Oganesyan2019,Burgess+20}.

This tension motivated a second broad class of models in which dissipation takes place below, or close to, the photosphere. In these models, the observed spectrum is shaped by Comptonization of lower-energy photons by hot electrons at optical depths larger than unity \citep{GhiselliniCelotti1999,Rees2005,Beloborodov2010}. In baryonic jets, sub-photospheric dissipation may occur through radiation-mediated shocks, which can produce a broadened, non-thermal-looking spectrum while maintaining high radiative efficiency (see \citet{Levinson2020} for the review). Recent radiation-mediated shock calculations show that such models can produce time-resolved spectra with low-energy slopes evolving from $\alpha\sim-0.5$ to $\alpha\sim-1$, high-energy slopes $\beta\sim-2$ to $-2.4$, and peak energies in the observed GRB range \citep{Alamaa2026}. In magnetized jets, magnetic dissipation and non-thermal particle injection can generate synchrotron photons below the photosphere: these photons are then Comptonized while the radiation is advected outward and eventually released near transparency \citep{Thompson-Gill-14,Gill-Thompson-14,Vurm2016,Gill+20}.

Prompt-emission models can therefore be broadly divided into two families: optically thin models, in which the radiation is produced above the photosphere, and photospheric models, in which the spectrum is shaped by sub-photospheric dissipation that may continue across the photosphere and radiative transfer. Both families can be tuned to reproduce the observed $10\,\mathrm{keV}$~--~$10\,\mathrm{MeV}$ spectra, the energy range most commonly accessible to space-based GRB detectors \citep{Band1993}. However, they can predict very different amounts of optical radiation. This makes prompt optical emission a powerful diagnostic of the radiation mechanism \citep{Oganesyan2019}.

\begin{figure*}[!ht]
    \centering
    \includegraphics[width=1.1\textwidth, angle=0]{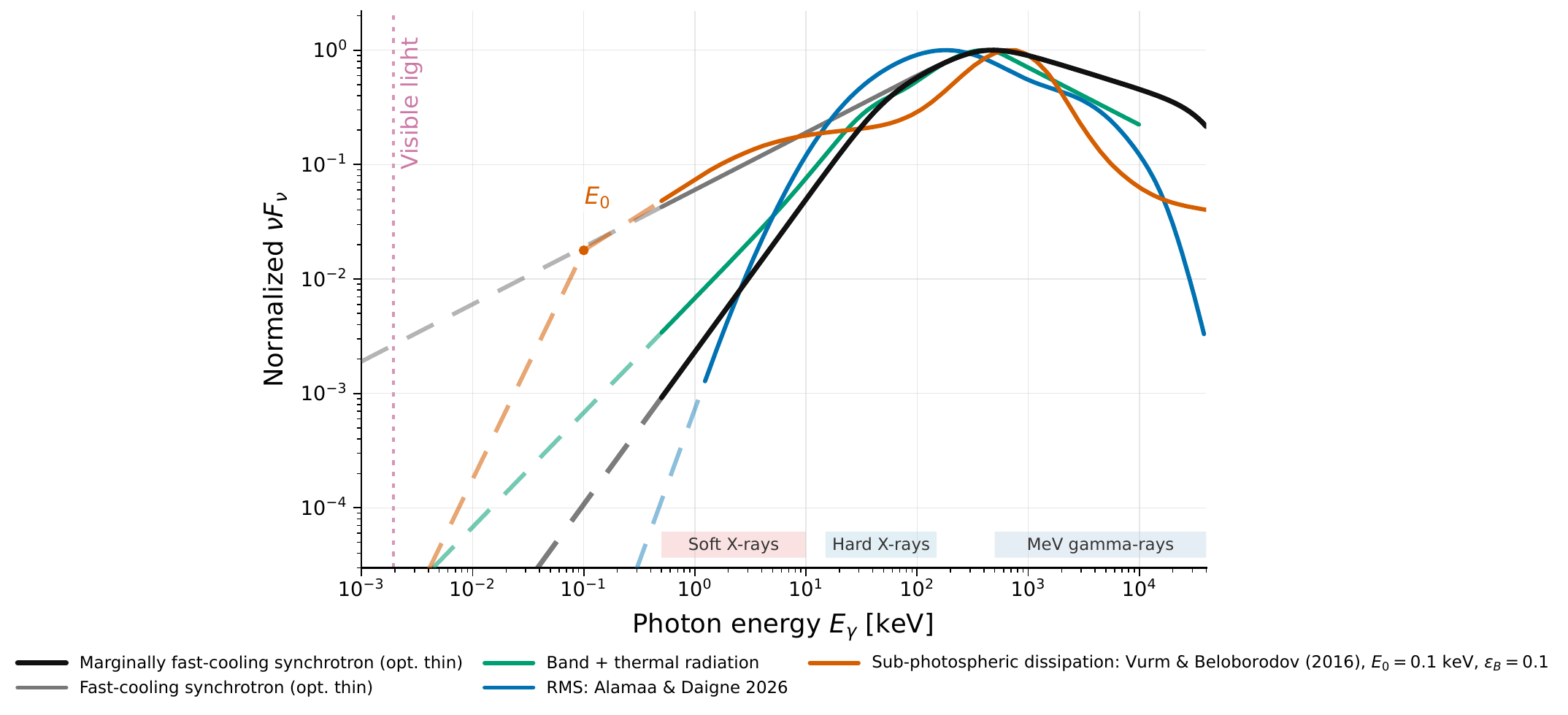}
    \caption{Expected broadband prompt emission spectra from representative models. The comparison includes optically thin synchrotron spectra in fast-cooling and marginally fast-cooling regimes, empirical Band+blackbody spectra, unmagnetized radiation-mediated shock spectra, and the magnetized sub-photospheric model. The figure illustrates that models giving similar keV--MeV spectra can differ strongly at optical frequencies, making prompt optical measurements a sensitive diagnostic of the emission mechanism. Sub-photospheric dissipation models are adapted from \citet{Alamaa2026} and \citet{Vurm2016}. In sub-photospheric dissipation models with magnetized jets, the seed synchrotron radiation is expected to be suppressed below an energy $E_0$ because of synchrotron self-absorption \citep{Thompson-Gill-14,Gill+20}.
    In this illustrative sketch, we adopt an arbitrary value $E_0 = 0.1~{\rm keV}$. In optically thin models, by contrast, the low-energy spectrum can extend down to the optical band if the emission radius is sufficiently large. This behavior is often relevant when optical radiation tracking the MeV emission is observed \citep{Oganesyan2019}.}
    \label{fig:pe_models}
\end{figure*}

Figure~\ref{fig:pe_models} compares the expected optical emission from representative prompt emission models. In unmagnetized radiation-mediated shock models, adapted from \citet{Alamaa2026}, the optical flux is expected to be weak because the radiation field is dominated by quasi-thermal seed photons that are Comptonized into the keV~--~MeV band. In magnetized sub-photospheric models, adapted from \citet{Vurm2016}, the optical output can be much larger because synchrotron emission from hot electrons supplies a population of low-energy photons below the photosphere. The precise optical flux depends sensitively on the jet magnetization, the radial distribution of dissipation, and the efficiency of non-thermal particle injection.

Figure~\ref{fig:pe_models} shows an empirical Band-plus-blackbody model, often used to represent a dominant non-thermal prompt component together with a sub-dominant photospheric thermal component \citep{Ghirlanda2003,Ryde2005,Guiriec2011,Axelsson2012}. Such two-component models can reproduce X-ray and MeV spectra, but optical observations provide an important additional constraint: if the non-thermal component is extrapolated as a simple power law to low energies, it may overpredict the optical flux \citep{Oganesyan2019}. The figure also shows optically thin synchrotron spectra in the fast-cooling and marginally fast-cooling regimes. A completely cooled electron population cannot reproduce the typical MeV spectrum, whereas marginally fast-cooling synchrotron models can naturally accommodate an additional low-energy break, usually interpreted as the cooling frequency \citep{Oganesyan2017,Ravasio2018,Ravasio2019}.

The importance of this low-energy break has become clear from broadband prompt observations, including soft X-rays. Several studies have demonstrated that \swift/XRT+BAT (0.5 - 150 keV) and Fermi/GBM (8 keV - 40 MeV) data have shown that prompt spectra often require three power-law segments rather than a single Band function \citep{Oganesyan2018,Ravasio2019}. The lowest-energy segment is typically consistent with the synchrotron slope $\alpha\simeq-2/3$, the intermediate segment with $\alpha\simeq-3/2$, and the high-energy segment with $\beta<-2$. This is naturally interpreted as synchrotron emission in the moderately fast-cooling regime, with the cooling break located at $E_{\rm break}\sim2$~--~$200\,\mathrm{keV}$ \citep{Oganesyan2017,Ravasio2019}. Extending the spectrum further down to optical frequencies provides an even stronger test of this interpretation \citep{Oganesyan2019}. 

High-cadence optical observations during the prompt emission phase are therefore crucial. If the optical emission is temporally correlated with the MeV pulses and lies on the low-energy extrapolation of the prompt emission model, it strongly supports a common internal origin. This behavior is observed in several events for which the optical and X-ray/$\gamma$-ray light curves track each other, and broadband fitting shows that synchrotron radiation can provide a good description of the optical-to-$\gamma$-ray data (Figure~\ref{fig:110205A} as an example for the time-resolved spectra of GRB~110205A). However, an important counterexample, GRB~080319B showed an extraordinarily bright prompt optical flash with optical flux far above a simple extrapolation of the MeV spectrum \citep{Beskin2010}. This required either multiple emission zones or an additional inverse-Compton component \citep{2009ApJ...691..723B, 2008Natur.455..183R}.

A larger sample of GRBs with very early optical observations with high time resolution is therefore essential. Such data can distinguish between optically thin synchrotron emission, photospheric models, magnetized sub-photospheric dissipation, and multi-component scenarios.  Robotic non-detections are also important. Deep upper limits obtained during the prompt phase constrain the optical-to-$\gamma$-ray spectral slope and can rule out bright optical flashes similar to GRB~990123 or GRB~080319B for many events.

\subsection{Prompt optical flashes from pair-loaded external shocks}
\label{pair-loading}

In the standard GRB phenomenology, a fraction of the jet energy is dissipated internally and released as prompt keV -- MeV radiation, while the remaining kinetic energy is transferred to the circumburst medium through the external blast wave. In the simplest afterglow picture, the external medium is assumed to be cold, static, and unaffected before it is swept up by the forward shock. This assumption, however, can break down during the prompt emission phase. The prompt MeV radiation front propagates ahead of the jet and substantially modifies the immediate circumburst medium by loading it with electron--positron pairs and by accelerating it outward \citep{Thompson2000,Meszaros2001,Beloborodov2002}.

The physical origin of this pair precursor is well understood. Prompt emission photons overtake the jet and stream through the external medium. Although the medium is optically thin and only a small fraction of X-ray photons are scattered, the number of prompt emission photons per ambient electron is enormous. The newly created pairs increase the scattering opacity and, in turn, scatter additional X-ray prompt emission photons. This feedback leads to a runaway pair-loading process \citep{Thompson2000,Meszaros2001,Beloborodov2002}. The external medium is therefore enriched by a lepton multiplicity $Z_\pm \equiv n_\ell/n_e$, where $n_\ell$ is the total lepton density after pair creation and $n_e$ is the original electron density of the circumburst medium. At the same time, Compton scattering and pair creation deposit momentum into the medium, pre-accelerating it to a Lorentz factor $\gamma_{\rm pre}$ \citep{Madau2000,Beloborodov2002}.

The blast wave that forms in this pair-enriched and pre-accelerated medium differs qualitatively from the standard Blandford--McKee solution \citep{Beloborodov2005,Beloborodov2014}. First, the relative Lorentz factor between the blast wave and the upstream medium is reduced by pre-acceleration. This decrease delays the onset of the usual post-deceleration regime. Then, when $Z_\pm\gg 1$, the number of leptons greatly exceeds the number of ions. The shock energy is then shared among a much larger number of particles, and the typical Lorentz factor of the shocked leptons is regulated by both pair loading and pre-acceleration, approximately as $\gamma_{\rm inj}\simeq \Gamma_{\rm rel}(\gamma_{\rm th}+\epsilon_e\mu_e m_p/Z_\pm m_e)$, where $\Gamma_{\rm rel}\simeq \Gamma/[\gamma_{\rm pre}(1+\beta_{\rm pre})]$ is the relative Lorentz factor between the blast wave and the upstream medium, $\Gamma$ is the blast-wave Lorentz factor, $\gamma_{\rm th}$ describes pre-heating of the upstream plasma, $\mu_e m_p$ is the ion mass per ambient electron, and $\epsilon_e$ is the fraction of ion energy transferred to leptons \citep{Beloborodov2005,Beloborodov2014,Hascoet2015}. 

The freshly shocked pairs cool rapidly. While the prompt radiation is still crossing the blast wave, the dominant cooling channel is external inverse Compton scattering of prompt MeV photons. This produces a bright GeV flash \citep{Beloborodov2014,Vurm2014,Hascoet2015}. The same thermal pair population can also emit synchrotron radiation if magnetic fields are present behind the forward shock, giving rise to a prompt optical flash. Thus, in this model, the GeV and optical flashes are two radiative signatures of the same pair-loaded external shock: the GeV component is produced by the external inverse Compton emission, while the optical component is produced by the synchrotron emission.

The transient nature of these flashes is controlled by the rapid radial decline of $Z_\pm$. As the blast wave expands beyond the strongly pair-loaded region, the lepton multiplicity drops steeply, the shock approaches the standard external-shock regime, and the emission evolves toward the usual afterglow behavior. This naturally produces a fast optical flare during the prompt emission phase. The optical flash is expected to be smoother than the MeV prompt light curve and will not track individual MeV pulses.

\begin{figure*}[!ht]
\centering
\includegraphics[width=0.51\textwidth]{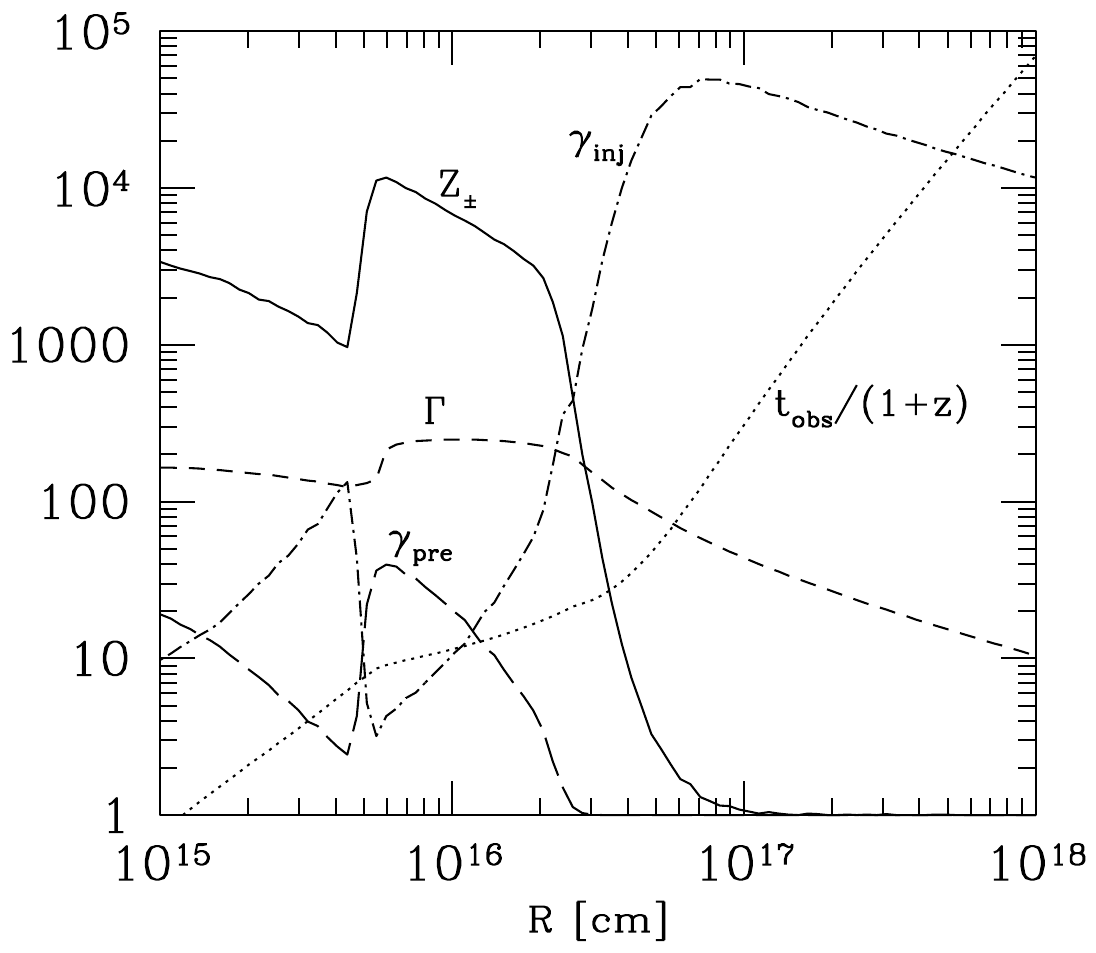}
\includegraphics[width=0.48\textwidth]{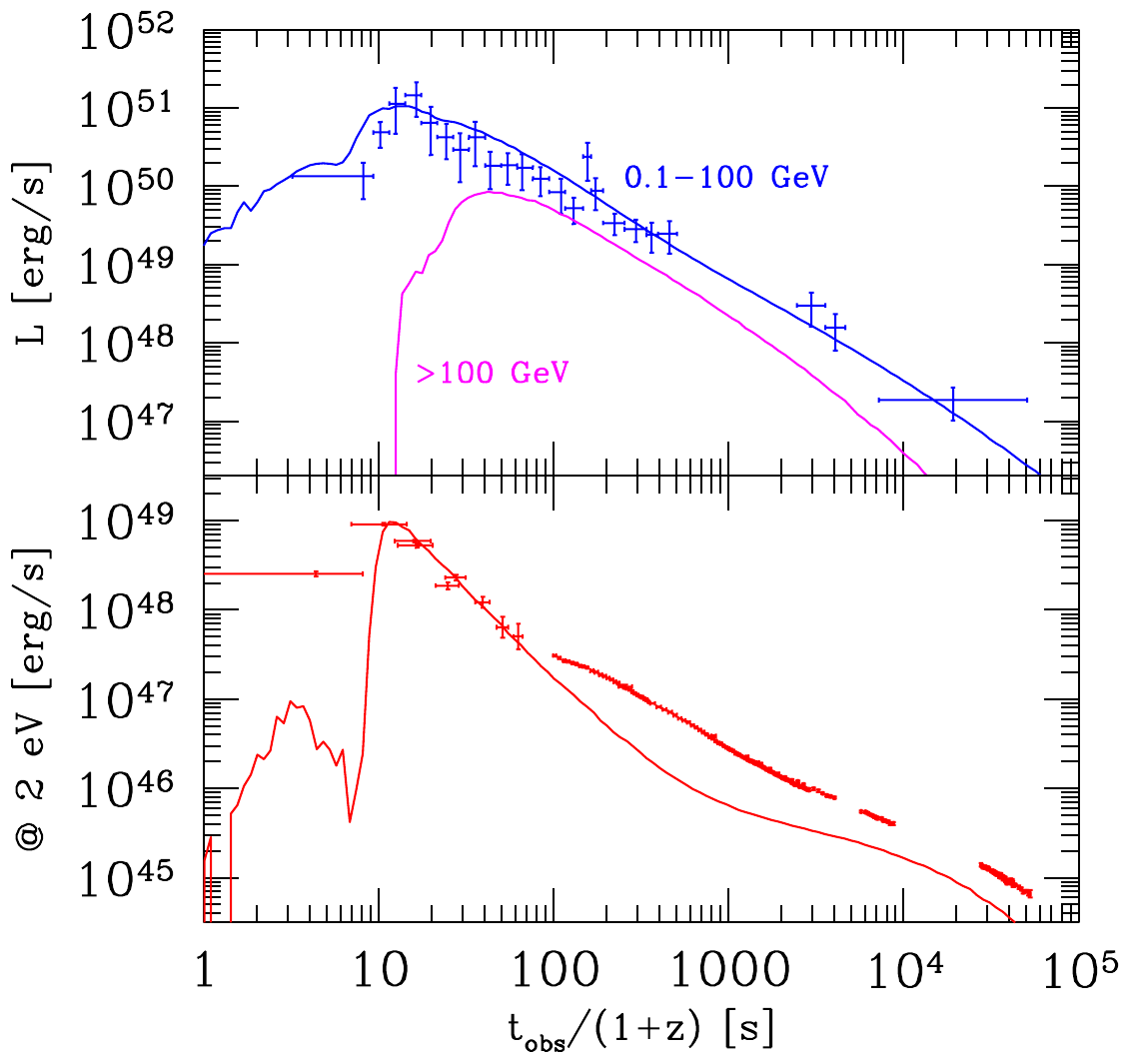}
\caption{Optical and GeV emission from a pair-loaded external blast wave. Left: radial evolution of the upstream pre-acceleration Lorentz factor $\gamma_{\rm pre}$, pair multiplicity $Z_\pm$, blast-wave Lorentz factor $\Gamma$, and injection Lorentz factor $\gamma_{\rm inj}$ of the shocked leptons. The sharp decrease of $Z_\pm$ marks the end of the pair profile. Right: application to GRB~130427A, in which the GeV flash is produced by inverse Compton cooling of shocked pairs by the prompt MeV radiation and the optical flash by synchrotron radiation from the same thermal pairs. Adapted from \citet{Vurm2014}.}
\label{fig:pair-loading}
\end{figure*}

The left panel of Figure~\ref{fig:pair-loading} illustrates this effect using the pair-loading profile calculated for a long-duration GRB in a wind-like circumburst medium \citep{Vurm2014}. The figure shows the radial evolution of the pre-acceleration Lorentz factor $\gamma_{\rm pre}$, the lepton multiplicity $Z_\pm$, the blast-wave Lorentz factor $\Gamma$, and the injection Lorentz factor $\gamma_{\rm inj}$ of the shocked leptons. The sharp decline of $Z_\pm$ marks the transition from the pair-dominated phase to the ordinary afterglow regime.

The right panel of Figure~\ref{fig:pair-loading} shows an application to GRB~130427A, where the observed GeV and optical flashes can be reproduced by emission from a pair-loaded blast wave propagating in a wind medium. In this interpretation, the GeV flash is produced by inverse Compton cooling of the thermal pairs by the prompt MeV photons, whereas the optical flash is synchrotron emission from the same shocked pairs. The observed optical luminosity then provides a direct probe of the post-shock magnetic field, while the timing and luminosity of the GeV flash constrain the density of the external wind. 

This mechanism is most efficient in wind-like medium, $\rho=A R^{-2}$, as expected around massive-star progenitors \citep{Chevalier1999}. In such a medium, the blast wave reaches the pair-loaded region while the lepton multiplicity is still large, making the pair-enriched phase observationally important. Therefore, the detection of a bright, rapidly decaying optical flash during the prompt emission phase, especially if accompanied by a GeV flash and not correlated pulse-by-pulse with the MeV light curve, can provide a powerful diagnostic of both the circumburst density profile and the microphysics of the external shock.

\subsection{Early optical emission as a probe of shock-acceleration physics}
\label{thermal electrons}
In the standard GRB afterglow framework, the broadband emission from the external forward shock is usually modeled by assuming that the swept-up circumburst electrons are accelerated into a single non-thermal power-law distribution. This assumption is convenient, but it is not guaranteed by the microphysics of relativistic collisionless shocks. Particle-in-cell simulations have shown that the downstream particle distribution is composed of a relativistic thermal component, approximately Maxwellian in shape, together with a non-thermal high-energy tail \citep{S08,Sironi2011,Sironi2013}. In other words, shock heating appears to be efficient, but these early simulations showed that acceleration into a power-law tail is not necessarily complete. More recent long-term simulations \citep{Groselj+24} do in fact find a growing population of \textit{suprathermal} (i.e., above the peak of the thermal distribution) electrons that smoothly connect the thermal distribution to the high-energy non-thermal electrons. The energy of the suprathermal particles grows over time at the expense of the thermal distribution, and therefore, it remains to be seen whether longer running simulations will find a predominantly power-law particle energy distribution downstream of relativistic collisionless shocks propagating into an almost unmagnetized external medium.

This has direct observational consequences for early afterglow emission. If a fraction of the post-shock electron energy remains in a thermal pool, the characteristic Lorentz factor of these electrons is expected to be of order $\gamma_{\rm th}\sim \epsilon_e \Gamma m_p/m_e$, up to factors of order unity that depend on the exact shape of the distribution and on how the electron energy is divided between the thermal and non-thermal components \citep{GS2009,Warren_dainotti}. The observed synchrotron frequency associated with the thermal electrons is then $\nu_{\rm th}\sim \Gamma \gamma_{\rm th}^2 eB'/(2\pi m_ec)$, where $B'$ is the comoving magnetic field downstream of the shock. For a Blandford--McKee blast wave in a homogeneous external medium, $\Gamma\propto t^{-3/8}$ and $B'\propto \Gamma$, which gives $\nu_{\rm th}\propto t^{-3/2}$. Thus, the thermal synchrotron peak is expected to sweep rapidly.

\begin{figure*}[!ht]
    \centering
    \includegraphics[width=1.100\textwidth]{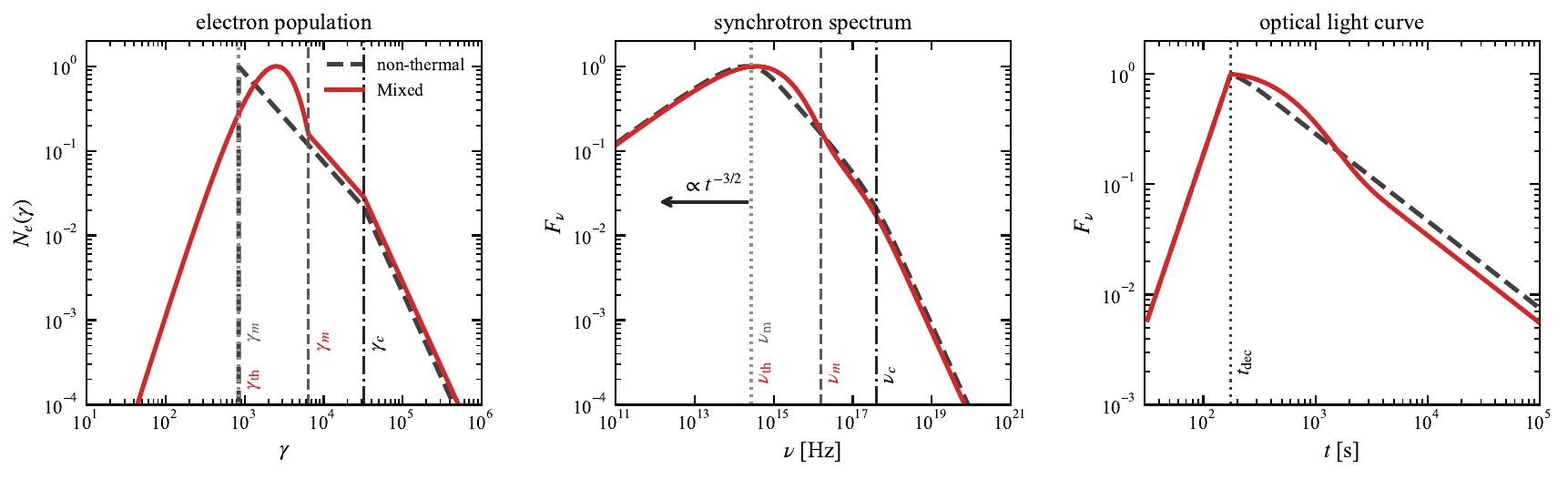}
    \caption{Schematic thermal-electron signature in an early GRB afterglow. Left: comparison between a standard power-law electron distribution and a hybrid Maxwellian plus power-law distribution. Middle: corresponding synchrotron spectra and temporal evolution of the characteristic thermal, injection, and cooling frequencies. Right: expected optical light curve, where the passage of the thermal synchrotron frequency through the optical band produces a broad flare before the ordinary non-thermal forward-shock afterglow dominates.}
    \label{fig:thermal_electrons}
\end{figure*}

Detecting this passage is challenging because the position and amplitude of the thermal peak depend on uncertain shock-microphysical parameters, including the fraction of energy in the non-thermal tail, the electron heating efficiency, and the magnetic-field amplification \citep{GS2009,Warren_dainotti}. However, recent very-high-energy detections of GRB afterglows have strengthened the case for low magnetic equipartition parameters, typically $\epsilon_B\sim10^{-4}$~--~$10^{-3}$, in models where the GeV -- TeV radiation is produced by inverse Compton emission (see \citet{MN2022} for the review). For typical blast-wave Lorentz factors near the deceleration time, such weak magnetic fields naturally place the thermal synchrotron frequency close to the optical band \citep{Jelinek2026}. Early optical observations can therefore provide a direct probe of the shape of the electron distribution.

Figure~\ref{fig:thermal_electrons} sketches this idea. The left panel compares a standard pure power-law electron distribution with a hybrid distribution composed of a Maxwellian thermal component smoothly connected to a non-thermal power-law tail \citep{GS2009}. The relevant characteristic Lorentz factors are the thermal Lorentz factor $\gamma_{\rm th}$, the Lorentz factor $\gamma_m$ associated with the non-thermal tail, and the cooling Lorentz factor $\gamma_c$. The middle panel shows the corresponding synchrotron spectrum and the temporal evolution of the characteristic frequency. The right panel shows the expected optical light curve. As $\nu_{\rm th}$ crosses the optical band, the thermal population produces a broad and smooth flare, followed by the emergence of the ordinary non-thermal forward shock afterglow.

This thermal-electron signature differs from other early optical components. It is not expected to track individual MeV pulses, as prompt optical emission from internal dissipation does. It is also broader and later than the very fast optical flashes expected from pair-loaded external shocks during the prompt phase. Finally, unlike a standard reverse-shock flash from a homogeneous shell, it can naturally produce a rise--plateau--steep-decay morphology around and after the forward-shock deceleration time. The key observable is therefore not the presence of an early optical flare but rather its duration, smoothness, temporal offset from the prompt emission pulses, and transition into the late non-thermal afterglow. 

A promising example is GRB~250702F \citep{Jelinek2026}. High-cadence optical observations starting $\sim$ 28 s after the trigger revealed two distinct optical flares. The first flare, observed between $\sim30$ and $100$ s, is temporally coincident with the brightest MeV prompt emission pulses and is spectrally consistent with the low-energy extrapolation of the prompt MeV spectrum. This component is therefore naturally interpreted as prompt optical emission from internal dissipation. The second flare, extending from $\sim100$ to $\sim1400$ s, shows a rapid rise, a shallow decline, and a steep decay before transitioning to a standard afterglow. During this phase, the optical emission is decoupled from the strong X-ray flares, suggesting that it is not simply the low-energy extension of the continuing internal activity.

The second optical flare of GRB~250702F can be reproduced by a forward shock model in which the downstream electron distribution contains both a thermal Maxwellian component and a non-thermal power-law tail \citep{Jelinek2026}. In this interpretation, the steep optical decline is caused by the passage of $\nu_{\rm th}$ through the optical band. The best-fit model gives a non-thermal energy fraction $\delta \approx 0.8$, an electron power-law index $p=2.05$, a deceleration time $t_{\rm dec} \approx 175$ s, and a thermal synchrotron frequency at deceleration of $\log_{10}(\nu_{\rm th,0}/{\rm Hz}) \approx 14.4$. These parameters imply an initial bulk Lorentz factor $\Gamma_0\simeq160$ and a characteristic thermal electron Lorentz factor $\gamma_{\rm th}\sim900$. The inferred comoving magnetic field at deceleration is $B'\sim1$ G, corresponding to $\epsilon_B\simeq5\times10^{-4}$.

\begin{figure}[!ht]
    \centering
    \includegraphics[width=0.5\textwidth]{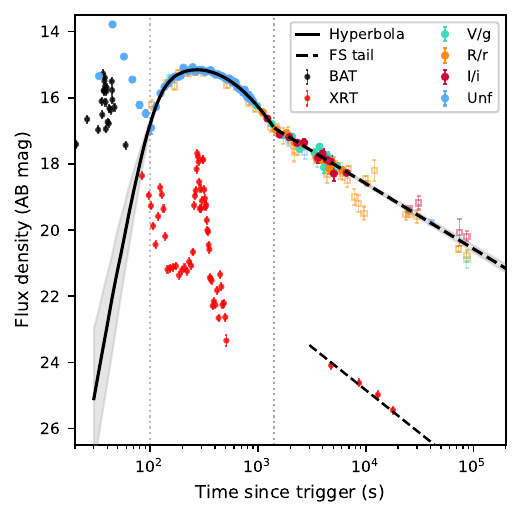}
\caption{Early optical observations of GRB~250702F. The light curve shows the prompt optical flare temporally coincident with the brightest MeV pulses and the later, broader optical flare attributed to synchrotron emission from a hybrid thermal-plus-non-thermal electron distribution in the forward shock. The transition to the late standard afterglow occurs after the thermal synchrotron peak has crossed the optical band. Adapted from \citet{Jelinek2026}.}
\label{fig:250702F}
\end{figure}

This interpretation has an important physical implication. The value $\gamma_{\rm th}\sim900$ is several times larger than the bulk Lorentz factor of the blast wave. Relativistic collisionless-shock simulations show that electrons can receive a substantial fraction of the ion energy downstream of the shock, so the characteristic electron thermal Lorentz factor can scale as $\gamma_{\rm th}\sim10^3(\epsilon_e/0.05)(\Gamma/160)$ \citep{S08,Sironi2011,Sironi2013,warren}. The GRB~250702F optical flare may therefore represent a direct observational signature of electron heating and incomplete particle acceleration in an ultra-relativistic shock.

Future high-cadence optical observations within the first few tens of minutes after the GRB trigger will be essential for testing this model. Well-sampled early optical light curves can distinguish between prompt optical emission, reverse-shock flashes, pair-loaded external-shock flashes, and the broader thermal electron feature. Such observations provide one of the few direct ways to probe how relativistic collisionless shocks divide energy between thermal electrons, non-thermal particles, and magnetic fields.

\subsection{Reverse shocks, ejecta magnetization, and shell structure}
\label{subsec:reverse_shock}

Reverse shock detections are among the most valuable observational outcomes of rapid optical follow-up because they probe the ejecta itself (in the rest frame of the contact discontinuity), not only the swept-up circumburst medium. The key observables are the early optical peak time, post-peak decay slope, reverse-to-forward shock flux ratio, optical color, and the connection between optical, radio/mm, and X-ray emission. These quantities constrain the magnetization and radial structure of the ejecta, although such constraints remain model-dependent.

A useful observational quantity is the ratio $R_B=\epsilon_{B,\rm RS}/\epsilon_{B,\rm FS}$, which compares the magnetic-field energy fraction in the reverse and forward-shock regions. Reverse-shock samples indicate that bursts with clear reverse-shock signatures can have $R_B$ significantly larger than unity, implying stronger magnetization in the ejecta than in the shocked external medium \citep{2014ApJ...785...84J}. However, $R_B$ is not identical to the global magnetization parameter $\sigma$; it is an observationally inferred microphysical ratio and depends on the assumed shock model, density profile, electron index, and spectral regime. For this reason, reverse shock detections should be described as evidence for weak to moderate ejecta magnetization, rather than as direct measurements of $\sigma$ without modeling assumptions (see Section \ref{Sec:RS}).

Detection of a reverse shock optical flash also enables complementary constraints on the initial bulk Lorentz factor $\Gamma_0$ of the ejecta. Combined with estimates of the circumburst density from late-time afterglow modeling and the isotropic equivalent energy from  $\gamma$-ray observations, the observed reverse shock peak time can therefore constrain $\Gamma_0$ to within factors of a few. It is important to note that many published estimates of $\Gamma_0$ in early-optical samples have been derived primarily from the onset of the forward shock afterglow, rather than directly from reverse shock modeling. These studies typically infer $\Gamma_0 \sim 100$~--~$1000$ \citep{2007Molinari, 2010ApJ...725.2209L}, although significant scatter exists reflecting burst-to-burst variations. The main uncertainties in deriving $\Gamma_0$ stem from degeneracies between the initial Lorentz factor, the circumburst density, and the microphysical parameters governing particle acceleration and magnetic field amplification at the shock front. Multi-wavelength observations that constrain the spectral peak of the reverse shock synchrotron component can break these degeneracies. Specifically, if the peak frequency $\nu_m^{\rm RS}$ of the reverse shock spectrum is observed to pass through the optical band, the combination of peak time and peak flux provides tighter joint constraints on both $\Gamma_0$ and the magnetic field strength in the ejecta \citep{2000ApJ...545..807K}. Simultaneous constraints from both the forward shock onset timing and the reverse shock peak timing, when both components are clearly detected, yield the most robust estimates of $\Gamma_0$ and can further reveal whether the ejecta possesses a narrow or broad distribution of Lorentz factors.

\subsection{Initial Lorentz factor from the afterglow onset}
\label{subsec:LF}

A robust way to constrain the initial bulk Lorentz factor, $\Gamma_0$, is to identify the \emph{afterglow onset} feature, a smooth rise and peak in the early forward shock light curve associated with the deceleration of the relativistic ejecta by the external medium \citep{2007ApJ...655..973K}. Using the dynamical relations introduced in Section \ref{sec:features}, the observed peak time in this case is interpreted as the deceleration time of the blast wave. The advantage of the observational approach is that the peak time can be measured directly from the light curve, while the kinetic energy and density are constrained from broadband afterglow modeling. 

To that end, it is important to identify the peak in the light curve caused by deceleration and not by the crossing of a characteristic synchrotron break frequency across the observing band. Emission coming from the different power-law segments of the synchrotron spectrum shows a different temporal rise/decay trend ($F_\nu\propto t^{-\alpha}$) pre- and post-deceleration (see, e.g., Table\,2 of \citealt{Beniamini+20}) that depends on the external medium radial density power-law index ($k$) and the radiating electron energy power-law index ($p$). In the slow-cooling regime ($\nu_m<\nu_c$), the temporal power-law index of the light curve pre-deceleration in the optically thin synchrotron emission for $p=2.5$ and $k=\{0,2\}$ is: $\alpha_{\rm pre}=k/2-3=\{-3,-2\}$ for $\nu<\nu_m<\nu_c$ (D from \citealt{Granot-Sari-02}), $\alpha_{\rm pre}=k(p+5)/4-3=\{-3,0.75\}$ for $\nu_m<\nu<\nu_c$ (G), and $\alpha_{\rm pre}=k(p+2)/4-2=\{-2,0.25\}$ for $\nu_m<\nu_c<\nu$ (H). The temporal index post-deceleration is: $\alpha_{\rm post}=(k-2)/(4-k)=\{-0.5,0\}$ (D), $\alpha_{\rm post}=[k(3p-5)-12(p-1)]/4(4-k)\simeq\{1.13,1.63\}$ (G), and $\alpha_{\rm post}=(2-3p)/4\simeq1.38$ (H). At early times when the optical emission is produced at $\nu<\nu_m$, it will not show a peak but a break in the light curve instead at $t=t_{\rm dec}$. Later, when $\nu_m$ crosses the optical band at $t>t_{\rm dec}$, the light curve will show a peak for $k<2$, which could be mistaken as that caused by blast wave deceleration. In this particular case, having simultaneous coverage in X-rays can alleviate the problem as they are always produced at frequencies $\nu>\nu_m$ and will show a light curve peak for $k<2$ at the time of deceleration.

Once the deceleration time is correctly identified, an estimate of the initial (coasting) bulk Lorentz factor of a uniform jet viewed on-axis can be obtained using Eq.\,\ref{eq:tdec} that yields
\begin{eqnarray}
\label{eq:Gamma0}
    \Gamma_0 &\simeq& 96\,(1+z)^{3/8}E_{\rm k,iso,52}^{1/8}n_{0,0}^{-1/8}t_{\rm dec,2}^{-3/8}\quad(k=0) \\
    &\simeq& 74\,(1+z)^{1/4}A_{*,-2}^{-1/4}E_{\rm k,iso,52}^{1/4}t_{\rm dec,2}^{-1/4}\quad(k=2)\,.
\end{eqnarray}
Notice the very weak dependence on $E_{\rm k,iso}$ and $n_0$ for the ISM ($k=0$) case, which means that in many bursts when the data is not sufficient to determine these parameters more accurately, a robust estimate of $\Gamma_0$ can still be obtained. 

$\Gamma_0$ values confirm that GRB jets are ultra-relativistic and provide an important connection between prompt emission physics and the later external-shock afterglow. However, the inference of $\Gamma_0$ is subject to several observational caveats. The early optical peak must be separated from the reverse shock emission, prompt optical emission, and the passage of synchrotron break frequencies through the optical band. Chromatic peaks or peaks coincident with strong high-energy variability are therefore less secure as pure forward-shock onset signatures. The most robust cases are those with dense optical sampling, stable optical colors across the peak, and broadband consistency with the later afterglow evolution.

\subsection{Magnetic field configuration from early polarimetry}
\label{subsec:polarimetry}

The early-time optical polarization measurements during the reverse shock phase directly probe the magnetic field structure in the ejecta \citep{1999ApJ...525L..29G}. If magnetic fields in the ejecta are predominantly toroidal and globally ordered due to the jet being launched magnetically dominated by, e.g., the \citet{Blandford-Znajek-77} mechanism, polarization levels can approach $\Pi=\Pi_{\max}=(\alpha+1)/(\alpha+5/3)$, where $\alpha\equiv-d\ln F_\nu/d\ln\nu$ is the local spectral index \citep[see more detailed discussion in, e.g.,][]{Granot-Taylor-05,Gill+21}. For slow-cooling emission with frequency $\nu_m<\nu<\nu_c$, the asymptotic $\alpha=(p-1)/2$ that yields a significant local polarization of $70\%\lesssim\Pi_{\max}=(p+1)/(p+7/3)\lesssim74\%$ for $2.2\lesssim p \lesssim 2.8$. It is possible that the magnetic field is not fully ordered in the entire observed region of the jet that has an angular size of $1/\Gamma$, in which case the net polarization will be diluted by misaligned polarization vectors. Such a situation is realized if the magnetic field is ordered within $N$ mutually incoherent patches of angular size $\theta_B\ll1/\Gamma$, in which case the net polarization is $\sim\Pi_{\max}/\sqrt{N}$ \citep{Gruzinov-Waxman-99}. Another way to dilute the polarization produced by ordered fields in the ejecta is by having some contribution to the flux from small-scale random magnetic fields that naturally arise at shocks propagating into weakly magnetized media. If such small-scale and tangled field are produced at the reverse shock, they will contribute zero net polarization in a uniform jet viewed on axis. Since they will contribute some fraction of the total flux, it will dilute the net polarization from ordered fields \citep[e.g.][]{Granot-Konigl-03}. A similar situation is obtained when a fraction of the total flux comes from the weakly polarized overlapping emission from the forward shock. When the ejecta is weakly magnetized or unmagnetized, the local polarization can be substantial from the small-scale shock-produced tangled magnetic field. However, it will entirely cancel out to zero in an axisymmetric and uniform jet viewed on-axis. In such a case, the only way to produce net non-zero polarization is if the jet is misaligned and the observer's line-of-sight passes near the sharp edge of a uniform jet \citep{1999ApJ...525L..29G} or it has angular structure that gives unequal weights to emission coming from within the observed region \citep{Rossi+04,Gill-Granot-18,Gill-Granot-20}. In both scenarios, imperfect cancellation of the polarization vectors will yield net non-zero polarization.

The GRB 090102, GRB 120308A, and GRB 180720B polarization measurements (see Section \ref{Sec:RS}) illustrate this theoretical picture directly: their high, slowly evolving polarization is most consistent with ordered fields superposed on smaller-scale turbulence generated at the shock front \citep{2009Natur.462..767S, 2013Natur.504..119M, Arimoto+24}. A similar pattern was reported for GRB 190114C by MASTER \citep{Jordana-Mitjans+20}. The temporal evolution of polarization degree and position angle as the angular size of the observable part of the jet grows during jet deceleration provides additional constraints on the configuration of the magnetic field in the ejecta. However, early polarimetry remains technically challenging, requiring specialized instrumentation on rapid-response telescopes. Future systematic polarimetric campaigns can distinguish between magnetic field configurations predicted by different jet launching models, including models with significant Poynting flux at the base.

\subsection{Circumburst density profile and shock microphysics}

After the earliest reverse-shock or prompt optical components fade, the
optical emission usually becomes dominated by the forward shock. At this
stage, the temporal decay index $\alpha$ and spectral index $\beta$
provide observational tests of the external density profile and the
synchrotron spectral regime. Comparing optical and X-ray closure
relations can distinguish between an ISM-like medium and a wind-like medium, although in practice, the interpretation is often affected by
energy injection, evolving microphysical parameters, or additional
emission components \citep[e.g.][]{Angulo-Valdez+26,Akl+26,Gill+26}.

Early optical data are especially valuable because they connect the deceleration phase to the later afterglow. When the same forward-shock component can be followed from the onset through hours or days, the density profile, kinetic energy, and microphysical parameters $\epsilon_e$ and $\epsilon_B$ are more tightly constrained than from late-time data alone. Broadband observations extending to radio/mm wavelengths further help locate the characteristic synchrotron frequencies and break degeneracies between density, energy, and magnetic field strength.

Nevertheless, observational constraints on the circumburst environment are rarely unique from optical data alone. Similar optical decay slopes can arise from different combinations of density profile, spectral regime, energy injection, and jet geometry. Reliable circumburst-density constraints therefore require broadband afterglow modeling, ideally combining optical/NIR, X-ray, and radio/mm coverage to locate the synchrotron break frequencies, while accounting for host-galaxy extinction and possible late-time host contamination \citep{1998ApJ...497L..17S,Granot-Sari-02,2002ApJ...571..779P,
2003ApJ...590..992F,2012ApJ...746..156C,2015ApJ...815..102F,2011A&A...534A.108K}. 

\subsection{Probing dust properties and host galaxy extinction}
\label{subsec:dust_extinction}

Rapid multi-band optical and NIR photometry initiated within minutes of a GRB trigger enables construction of the instantaneous optical--NIR spectral energy distribution before substantial temporal evolution and evolution of spectral break frequencies complicate the analysis. When combined with Galactic foreground extinction corrections derived from dust maps and a spectroscopic or photometric redshift constraint, these early SEDs provide one of the most direct methods for measuring line-of-sight extinction and isolating the host-galaxy contribution that directly probes the progenitor environment \citep{2003ApJ...585..775P, 2006ApJ...641..993K}. The observed attenuation generally comprises contributions from Milky Way dust, absorption in intervening systems, and extinction intrinsic to the host galaxy \citep{2010MNRAS.401.2773S}. Broad wavelength coverage is therefore essential to break degeneracies among the intrinsic spectral slope, curvature induced by synchrotron break frequencies, and reddening from dust.

A primary motivation for obtaining very early multi-color observations is the potential for time-dependent dust processing in the circumburst medium. The intense UV and X-ray radiation field associated with the prompt emission and early afterglow can, in principle, sublimate, or photodesorb dust grains at small radii, leading to evolving extinction on minute-to-hour timescales \citep{2000ApJ...537..796W}. Empirically, however, definitive evidence for the temporal evolution of the extinction has proven elusive. Most multi-epoch SED studies find that the V-band extinction $A_V$ remains consistent with a constant value from the earliest observations through days or weeks post-burst \citep{2010MNRAS.406.2473P}. This stability may indicate that the bulk of the dust resides at sufficiently large distances where the radiation field intensity is inadequate for efficient destruction, that grain sublimation is less efficient than theoretical predictions suggest, or that modest temporal variations are masked by photometric uncertainties and systematic modeling effects. While a small number of events have shown tentative evidence for early-time color evolution consistent with decreasing extinction or dust destruction (e.g. GRB~120119A; \citealt{Morgan2014}), robust confirmation remains challenging because intrinsic spectral evolution, reverse-shock/forward-shock
transitions, and calibration offsets can produce similar color changes
\citep[e.g.][]{Perley2008,Starling2008}. Therefore, establishing genuine time-variable extinction requires high signal-to-noise, preferably simultaneous multi-color photometry obtained at multiple epochs within the first few hours, ideally with coordinated optical--NIR coverage, so that the evolving SED can be separated from intrinsic afterglow spectral evolution and systematic cross-calibration effects.

The wavelength dependence of extinction provides additional constraints on dust grain properties in GRB host galaxies. Optical and NIR SED analyses often favor extinction curves that exhibit steeper wavelength dependence and lack the prominent 2175\,\AA\ absorption feature characteristic of Milky Way dust, instead resembling Small Magellanic Cloud (SMC)-like extinction laws \citep{2006ApJ...641..993K, 2008ApJ...685..376K, 2010MNRAS.406.2473P, 2013MNRAS.432.1231C}. This result suggests smaller average grain sizes or different grain compositions compared to local Galactic dust. However, a subset of bursts do exhibit Milky Way-like extinction curves, indicating significant diversity in dust properties across host environments. The derived total-to-selective extinction ratio $R_V \equiv A_V/E(B-V)$ spans a broad range, reflecting variations in grain size distributions and dust processing histories. The total sightlines host-galaxy extinction likewise varies widely, from negligible reddening to heavily obscured with $A_V \gtrsim 5$ mag that constitute the ``dark GRB'' population \citep{2004ApJ...617L..21J, 2009AJ....138.1690P, 2012A&A...537A..15S}. Because such heavily extinguished events are systematically underrepresented in optical-selected samples, rapid NIR follow-up is critical for recovering a more complete census of the GRB population and for quantifying selection biases. In the NIR, dust attenuation is substantially reduced (typically $A_K \approx 0.1 A_V$ for standard extinction curves), often enabling detections when optical observations yield only upper limits.

At moderate to high redshift, the interpretation of early-time SEDs requires care because intergalactic neutral-hydrogen absorption can affect the observed bands whenever they sample rest-frame wavelengths blueward of Ly$\alpha$ ($1216$\,\AA).  Thus, the importance of this effect depends not only on redshift but also on the wavelength coverage of the SED: at $z\gtrsim 2$--3, Ly$\alpha$-forest absorption can already suppress flux in the bluer observer-frame optical/UV filters, while at $z\gtrsim 5$--6 \citep{2026arXiv260531132P} the break is shifted through much of the optical window. At still higher redshift, the increasing opacity of the intergalactic medium, including strong Ly$\alpha$ absorption and possible Gunn--Peterson absorption, can strongly attenuate or effectively extinguish the flux blueward of the redshifted Ly$\alpha$ wavelength. If not modeled, this suppression can mimic strong host-galaxy dust reddening or lead to an overestimate of the intrinsic optical darkness. Therefore, for GRBs whose SEDs include bands probing rest-frame UV wavelengths below Ly$\alpha$, intergalactic absorption must be included explicitly before attributing the observed spectral curvature or optical faintness to dust extinction. Consequently, early NIR photometry becomes essential not only for identifying high-redshift candidates through Lyman-$\alpha$ dropout signatures, but also for disentangling the relative contributions of dust extinction and neutral hydrogen absorption to the observed continuum shape. Furthermore, NIR photometry enables robust photometric redshift estimation when immediate spectroscopic observations are unavailable, a capability that is particularly valuable given the rapid fading of GRB afterglows.

\subsection{Optical transients from compact object mergers}
\label{subsec:kilonovae}

Following gravitational wave detections of binary neutron star or neutron star-black hole mergers, optical surveys can identify kilonova emission powered by radioactive decay of $r$-process elements synthesized in the neutron-rich ejecta \citep{2017LRR....20....3M}. The kilonova associated with GW170817 exhibited a blue optical component peaking around one day post-merger, followed by redder emission emerging at later times \citep{2017ApJ...848L..13A,  MasterGW, 2017ApJ...848L..18N, 2017ApJ...848L..27T, 2017Natur.551...67P, 2017ApJ...851L..21V, 2017ApJ...848L..19C}. The early blue component arises from lanthanide-poor ejecta with high opacity in the red and infrared, while the red component traces lanthanide-rich material with high opacity in the blue and optical. Rapid optical follow-up within the first hours is critical for detecting the blue component before it fades \citep{2017ApJ...851L..21V, 2017ApJ...848L..18N}. For events where a short GRB is observed on-axis, the afterglow emission may dominate over the kilonova in the first hours \citep{2022Natur.612..223R, 2013Natur.500..547T, 2024Natur.626..737L, 2019ApJ...883...48L}, complicating identification. Off-axis GRB jets produce afterglows that rise over days to weeks as the jet decelerates and its beaming cone expands toward the observer, creating a distinct temporal signature \citep{2017ApJ...848L..20M, 2018NatAs...2..751L}. Robotic networks capable of rapidly tiling large gravitational wave localization regions are essential for maximizing kilonova discoveries \citep{2020MNRAS.497..726G}.

\subsection{Searching for orphan afterglows and off-axis jets}
\label{subsec:orphan_afterglows}

Standard GRB jet models predict a population of ``orphan'' afterglows, i.e., afterglow-like transients detected without an associated prompt $\gamma$-ray trigger. In the classical off-axis jet scenario, an observer initially located outside the prompt-emission beam or jet opening angle can detect the afterglow only at later times, when the jet has decelerated, and the relativistic beaming cone has widened, and/or when lateral spreading makes the emission visible \citep{1997ApJ...487L...1R}. However, orphan afterglows are not unique to this off-axis deceleration/spreading picture. They may also arise from baryon-loaded ``dirty'' or failed fireballs, in which the Lorentz factor is too low to produce
a bright prompt GRB, but is still sufficient to power an afterglow \citep{2003ApJ...591.1097R}, or from different beaming angles for the prompt $\gamma$-ray emission and the lower-energy afterglow, producing so-called on-axis orphan afterglows \citep{2003NewA....8..141N}. Therefore, rate estimates based solely on jet geometry should be regarded as model-dependent. For a typical jet opening angle of $\theta_{\rm jet}\sim5$--$10^\circ$, the geometric number of sightlines outside the prompt-emission cone can exceed that of on-axis GRBs by factors of order $\sim100$--$400$, but the actually detectable
orphan afterglow rate depends sensitively on survey depth and cadence, jet structure, ambient density, Lorentz-factor distribution, and the relative beaming of the prompt and afterglow emission \citep{2010MNRAS.403..300V}.

Systematic searches with wide-field optical surveys have now identified a growing sample of orphan afterglows ($\sim$ 10) without detected prompt $\gamma$-ray counterparts \citep{2025ApJ...985..124L,2025MNRAS.537.2362P,2022ApJ...938...85H,
Lipunov2022AT2021lfa,2021ApJ...918...63A,2020ApJ...905...98H}. Some of them may be ordinary on-axis GRB afterglows for which the prompt emission was missed by high-energy instruments, whereas others may represent intrinsically $\gamma$-ray-weak or mildly/off-axis relativistic explosions. The strongest constraints on an accompanying prompt emission currently come from cases such as AT2019pim and AT2021lfa, for which deep high-energy limits and early optical coverage are available \citep{2025MNRAS.537.2362P,2022ApJ...938...85H,Lipunov2022AT2021lfa}. AT2021lfa / MASTER OT J123248.62–012924.5 transient \citep{Lipunov2022AT2021lfa}  with X-ray afterglow, detected by \swift, but without direct $\gamma$-ray detection, provided an explanation within the framework of the ``dirty fireball'' model with an underdeveloped jet and a low Lorentz factor.

For a genuinely off-axis orphan afterglow from a top-hat jet, the expected observational signature is a rising afterglow light curve that peaks only after the relativistic beaming cone has widened into the observer's line of sight. In this case, for observers with viewing angles $\theta_{\rm obs}>\theta_{\rm jet}$, the peak occurs at a time $t_{\rm pk}>t_{\rm jet}$, i.e. later than the jet-break time. At $t>t_{\rm jet}$ the outflow spreads sideways with $\Gamma(t)=\theta_{\rm jet}^{-1}(t/t_{\rm jet})^{-1/2}$. The light curve peaks when the observer sees the jet core that occurs when $\Gamma(t)=\theta_{\rm obs}^{-1}$, which corresponds to a peak time of $t_{\rm pk}=(\theta_{\rm obs}/\theta_{\rm jet})^2t_{\rm jet}\propto\theta_{\rm obs}^2$ \citep{1999ApJ...525..737R,2002ApJ...570L..61G} typically ranging from hours to days depending on the viewing geometry and jet energetics. At $t>t_{\rm pk}$, the light curve evolution is the same as the post-jet-break light curve seen by an on-axis observer. Unlike on-axis afterglows, which peak at or near the burst trigger time, orphan afterglows exhibit characteristic delayed rises that make them identifiable through their unusual temporal evolution \citep{2015A&A...578A..71G}.

Structured jet models \citep{2002MNRAS.332..945R,2003ApJ...591.1086G,Kumar-Granot-03,Rossi+04,Beniamini+20,Beniamini+22}, where the energy and Lorentz factor decrease smoothly with angle from the jet axis rather than terminating sharply at $\theta_{\rm jet}$, predict modified orphan afterglow rates and light curve shapes. In such models, observers at moderate off-axis angles ($\theta_{\rm obs} \sim 2-3\theta_{\rm core}$, where $\theta_{\rm core}$ is the core opening angle) can still detect emission from the jet core as it decelerates, though at reduced brightness and with delayed peaks. The detection rate depends critically on the steepness of the angular energy profile and the survey's limiting magnitude and cadence.

GW170817/GRB~170817A provided the first unambiguous detection of an off-axis GRB jet, viewed at $\theta_{\rm obs} \sim 15-20$ degrees \citep{Troja+17,Troja+18,DAvanzo+18,2018Natur.561..355M, Gill-Granot-18, Lamb-Kobayashi-18,Lazzati+18,Resmi+18,2019A&A...628A..18S,Gill+19}. The optical and radio afterglow exhibited rising light curves peaking around 150 days post-merger, consistent with an initially off-axis relativistic jet that became visible as it decelerated \citep[e.g.][]{Margutti+18,Mooley+18,Dobie+18}. This observation confirmed that structured jet models better describe at least some short GRB jets and demonstrated that nearby off-axis events are indeed detectable with sufficiently deep and sustained follow-up observations \citep{2017ApJ...848L..20M, 2018NatCo...9.4089T}.

Modern wide-field optical surveys have already begun to open the discovery space for GRB afterglows identified independently of a high-energy trigger. In particular, PTF/iPTF \citep{2013ApJ...769..130C}, MASTER, and especially ZTF have demonstrated that fast-fading, afterglow-like transients can be identified directly from optical survey streams \citep{2015A&A...578A..71G, Lipunov2022AT2021lfa}. ZTF has now yielded $\sim$ ten orphan afterglows with no detected associated GRB, including six published examples: AT2019pim, AT2020blt, AT2021lfa, AT2021any, AT2023lcr, and AT2023sva \citep{2025MNRAS.537.2362P,2020ApJ...905...98H,2022ApJ...938...85H,2022JApA...43...11G, 2021ApJ...918...63A,2025ApJ...985..124L,2025MNRAS.538..351S}. These events show that orphan-afterglow searches are already an active observational program, although in some cases the absence of a detected prompt counterpart may reflect satellite sensitivity, sky coverage, or Earth occultation rather than an intrinsically $\gamma$-ray dark explosion. Therefore, the Vera C. Rubin Observatory's Legacy Survey of Space and Time (LSST) should be viewed not as initiating orphan-afterglow searches, but as extending them to a substantially deeper, wide-field survey regime, where its faint limiting magnitudes will improve sensitivity to more distant and/or intrinsically fainter afterglows, even though its baseline cadence will generally be less optimized for the fastest optical afterglow evolution than dedicated high-cadence searches \citep{2019PASP..131a8002B}.

Rapid robotic follow-up will therefore remain essential for measuring the early light-curve shape, color evolution, and fading rate, and for triggering deeper optical, X-ray, and radio observations. Multi-wavelength follow-up, especially in the radio and X-ray bands, is crucial for confirming a relativistic blast-wave origin and for distinguishing orphan-afterglow candidates from supernovae, tidal disruption events, stellar flares, cataclysmic variables, and other fast optical transients. A well-vetted sample of such events will provide unique constraints on jet structure, opening angles, angular energy distributions, viewing-angle effects, and beaming-corrected GRB event rates.

\subsection{Population-level constraints and selection effects}

The population of early optical afterglows is highly heterogeneous. Some bursts show prompt optical emission correlated with $\gamma$-rays, some show luminous reverse-shock flashes, some show smooth forward-shock onsets, some show optical flares or plateaus, and many show no detectable early optical counterpart. This diversity reflects real physical differences between GRB outflows, but also strong observational selection effects.

The detectability of early optical emission depends on response time, limiting magnitude, cadence, filter choice, sky brightness, Galactic extinction, host extinction, redshift, localization accuracy, and whether observations begin before or after the relevant component has peaked. Wide-field robotic systems are optimized for rapid response and prompt coverage of large localizations, whereas larger robotic telescopes provide deeper follow-up and better color information. Global networks further reduce day--night and weather gaps, improving the probability of capturing the first minutes after the trigger.

Statistical samples from ROTSE, TAROT, \swift/UVOT, GROND, and other facilities have shown that early optical afterglows span several orders of magnitude in luminosity and exhibit a broad range of temporal behavior \citep{2009ApJ...702..489R,2009AJ....137.4100K,2009MNRAS.395..490O,2009ApJ...690..163R,2011A&A...526A..30G, 2025MNRAS.543.2404R}. These samples are essential for moving beyond individual showcase events, but they must be interpreted with survey selection functions in mind. For example, a low observed fraction of reverse shock flashes may reflect intrinsic magnetization, but it may also reflect insufficient cadence or depth. Similarly, the observed rate of prompt optical emission depends strongly on whether observations start during the prompt $\gamma$-ray phase.

A robust population-level interpretation, therefore, requires reporting not only detections but also upper limits, exposure start times, cadence, filters, and limiting magnitudes. Uniform treatment of non-detections is especially important for constraining the intrinsic frequency of prompt optical emission, reverse shocks, dark bursts, and early optical flares.

\section{Robotic telescopes and technical capabilities}
\label{sec:facilities}

GRB detections and localizations from space-based observatories such as \swift \citep{2004ApJ...611.1005G}, \fermi \citep{2009ApJ...702..791M},
{\it AstroSat} \citep{2022ApJ...936...12C}, SVOM \citep{2022IJMPD..3130008A}, and Einstein Probe \citep{2022hxga.book...86Y} are rapidly disseminated to the astronomical
community through the General Coordinates Network (GCN)
\citep{Bart98,Bart08,Bart14,Bart2000,2026ApJS..283...30S}. These alerts, typically issued within seconds to minutes of the high-energy trigger, enable ground-based telescopes to respond promptly and capture the rapidly fading optical emission during the prompt and early afterglow phases. Traditional observatory operations, characterized by human-directed pointing decisions, telescope queue management, and overhead times of tens of minutes to hours, cannot achieve the rapid response ($\lesssim$ 1 minute) necessary for probing the physical conditions during shock breakout, reverse shock emission, and the onset of the forward shock.

Robotic telescopes address this observational challenge through fully autonomous operation. Upon receiving a GCN alert, these systems automatically interrupt current observations, execute rapid slews (often $>$10 degrees per second) to the GRB coordinates, perform autonomous guide star acquisition, and commence time-series imaging with minimal overhead. This automation eliminates human reaction time and optimizes observing efficiency, achieving first-light latencies as short as $\sim$10 – 60 seconds post-trigger \citep{2003PASP..115..132A}. The scientific priorities driving robotic GRB follow-up include:

\begin{itemize}
\item \textit{Low latency response}: Automated reaction to GCN notices within seconds, fast slew capabilities ($> 5^\circ$ s$^{-1}$), and minimal acquisition overhead to capture emission during the first few seconds to minutes, when reverse shock and early forward shock signatures emerge.

\item \textit{High-cadence photometry}: Sub-minute sampling during the first $\sim$ 10 minutes post-trigger to resolve rapid variability and color evolution, transitioning to logarithmically spaced observations (e.g., intervals of $\Delta t/t \sim 0.1$
) extending to days or weeks for afterglow decay monitoring \citep{2010ApJ...720.1513K}.

\item \textit{Wide dynamic range}: Capability to observe optical afterglows spanning $\sim$6 -- 20 magnitude, from naked-eye-bright bursts (V $\lesssim$ 6) to faint, high-redshift afterglows requiring deep integration (R $\gtrsim$ 22), often within the same facility through adaptive exposure control \citep{2006Natur.442..172V}.

\item \textit{Multi-band and polarimetric observations}: Simultaneous or rapid-sequence photometry in BVRI and/or Sloan filters to constrain spectral slopes ($\beta_{\rm opt}$), host galaxy extinction (A$_{V}$), and redshift; polarimetry ($\lesssim$ 1 \% precision) to diagnose magnetic field structure and GRB jet geometry.

\item \textit{Autonomous data pipelines}: Real-time image subtraction against reference catalogs (e.g., Pan-STARRS, SDSS), astrometric and photometric calibration to Gaia DR3 and PS1, transient detection algorithms, and automated re-alerting to enable rapid community follow-up.
\end{itemize}

\subsection{Major Robotic Telescope Facilities}

Below, we summarize widely used robotic telescope systems in the GRB literature (see Table \ref{tab:facilities}), emphasizing their technical specifications and key contributions to time-domain GRB science. 

\begin{table*}
\caption{Representative key robotic optical facilities for early GRB follow-up. Apertures, response times, and capabilities vary by site, instrument configuration, and epoch. Field of view (FOV) values are approximate. References provide representative facility descriptions and GRB science results. MITSuME includes multiple installations with varying specifications.}
\label{tab:facilities}
\begin{center}
\begin{tabular}{lcp{10cm}}
\hline
Facility & Aperture & Notable Capability \\
\hline
\\
ROTSE-I & 0.11\,m & $\leq$ 3 second slew time, four camera array  \\
 & & \citep{1999Natur.398..400A} \\[3pt]

ROTSE-III & 0.45\,m & Very low latency, wide-field ($1.85^\circ \times 1.85^\circ$) \\
 & & \citep{1999Natur.398..400A,2003PASP..115..132A,2009ApJ...702..489R} \\[3pt]

TAROT & 0.25\,m & Ultra-fast slew ($\sim$10~--~30\,s), uniform archive \\
 & & \citep{2008AN....329..275K, 2009AJ....137.4100K} \\[3pt]

RAPTOR & 0.4\,m & Prompt optical correlated with \\
 & & $\gamma$-rays, high-cadence multi-color \\
 & & \citep{2002SPIE.4845..126V, 2006Natur.442..172V, 2005Natur.435..178V} \\[3pt]

REM & 0.6\,m & Rapid simultaneous optical/NIR, early SEDs \\
 & & \citep{2001AN....322..275Z, 2007Molinari}  \\[3pt]

MASTER & 0.4 -- 0.9\,m & Global network, wide-field tiling, polarimetry, total FOV $9^\circ \times 9^\circ$ \\
   & & \citep{2010AdAst2010E..30L} \\[3pt]

BOOTES & 0.3 -- 0.6\,m & Multi-site network, rapid response, spectroscopy \\
  & & \citep{CastroTirado1999, 2004AN....325..679C} \\[3pt]

Palomar 60-inch & 1.524\,m & Automated interruption, broad optical filter coverage \\
  & & \citep{Cenko2006P60, Cenko2009P60Catalog} \\[3pt]

LCO & 0.4 -- 2\,m & Global network, longitudinal coverage\\
 & & ToO cadence control \\
 & & \citep{2013PASP..125.1031B} \\[3pt]

Liverpool Telescope & 2\,m & Deep multi-band, rapid, spectroscopy \\
 & & polarimetry (RINGO) \\
 & & \citep{2004SPIE.5489..679S, 2009Natur.462..767S,2013Natur.504..119M} \\[3pt]

GOTO & 0.4\,m & Rapid large-area coverage, GW follow-up capability \\
 & & \citep{2022MNRAS.511.2405S} \\[3pt]

MITSuME & 0.5\,m & Simultaneous multi-color \\
 & & ($g',r',i'$) imaging, early SED \\
 & & \citep{2008AIPC.1000..543S} \\[3pt]

NUTTelA-TAO & 0.7\,m & Simultaneous $g',r',i'$ (BSTI), Longitude gap coverage \\
 & & \citep{2022SPIE12184E..8AG} \\[3pt]

D50 & 0.5\,m & Capable of detecting faint signals, Fast response  \\
 & & \citep{Jelinek2019, Trcka2019} \\[3pt] 

FRAM & 0.25\,m &  Atmospheric monitoring, wide-field, rapid \\
 & & \citep{2021JInst..16P6027A} \\[3pt]

GROWTH-India & 0.7\,m &  Multi-band photometry, geographically optimal for sky coverage \\
 & & \citep{Kumar2022GIT, Kasliwal2019GROWTH} \\[3pt] 

COLIBRÍ & 1.3\,m & Ultra-fast ($\lesssim$20\,s access), optical+NIR, SVOM-era \\
 & & \citep{2020JAI.....950001F,2026arXiv260424259B} \\[3pt]
\hline
\end{tabular}
\end{center}
\end{table*}

\subsubsection{ROTSE-I and ROTSE-III}
The Robotic Optical Transient Search Experiment (ROTSE) was one of the first
successful demonstrations of fully autonomous optical follow-up of GRBs. The first-generation system, ROTSE-I, was a precursor to the later ROTSE-III network rather than a contemporaneous facility. ROTSE-I used a wide-field array of small telephoto lenses designed to cover the large GRB localization regions available before the \textit{Swift} era. Its most famous success was the detection of the prompt optical flash from GRB~990123, obtained while the $\gamma$-ray emission was still in progress \citep{1999Natur.398..400A}. This event provided the first simultaneous optical--$\gamma$-ray detection of a GRB and demonstrated the scientific importance of robotic response on timescales of seconds.

ROTSE-III was the next major generation of the experiment, developed after the
ROTSE-I proof of concept and optimized for rapid follow-up of more accurately
localized GRBs. In contrast to ROTSE-I, ROTSE-III consisted of 0.45~m robotic
telescopes with faster slewing, improved sensitivity, and a global distribution.
The ROTSE-III network operated during the main GRB follow-up era from 2003 to
approximately 2017 and included four nearly identical 0.45~m, f/1.85 telescopes
with 1.85$^\circ \times 1.85^\circ$ fields of view, located in Texas
(ROTSE-IIIb), Australia (ROTSE-IIIa), Namibia (ROTSE-IIIc), and Turkey
(ROTSE-IIId) \citep{2003PASP..115..132A}. The telescopes used unfiltered CCD
imaging, approximately calibrated to the $R$ band, and typically reached
5$\sigma$ limiting magnitudes of $m_{\rm lim}\sim17.5$--18.5 in 60~s
exposures. Their rapid slew capability, of order $\sim20^\circ$~s$^{-1}$,
allowed observations to begin within tens of seconds after receipt of a GCN notice.

ROTSE/ROTSE-III observations have been particularly valuable for early-time GRB
afterglow studies. They are commonly used to (i) measure optical rise and decay
indices during the first minutes after the burst, (ii) test for reverse-shock
emission when the optical light curve is temporally distinct from the
high-energy prompt emission \citep{2010ApJ...714..799P}, and (iii) provide
early upper limits that constrain the afterglow onset time when no detection is obtained. For example, ROTSE-IIIc, located at the H.E.S.S. site near
Mt.~Gamsberg in Namibia, detected optical emission from GRB~050525A beginning
only 8.7~s after the GCN notice \citep{2005GCN..3465....1R, 2005ApJ...631.1032R}. Thus, ROTSE-I established the feasibility and scientific impact of prompt robotic optical observations, while ROTSE-III transformed this concept into a coordinated
multi-site robotic follow-up network.

\subsubsection{RAPTOR}
The RAPid Telescopes for Optical Response (RAPTOR) system, developed and operated by Los Alamos National Laboratory, was designed for autonomous optical monitoring and rapid GRB follow-up. Its architecture combined wide-field patrol cameras, with fields of view of order $\sim 9^\circ \times 9^\circ$, and narrower-field follow-up telescopes, with fields of view of order $\sim 0.15^\circ \times 0.15^\circ$, enabling both wide-area monitoring and rapid pointed observations of external triggers \citep{2002SPIE.4845..126V}. In later configurations, including the RAPTOR-T system, co-aligned optical channels provided simultaneous multi-color imaging, including SDSS-like $g^{\prime}r^{\prime}i^{\prime}z^{\prime}$ coverage. The system was capable of autonomous responses on timescales of $\lesssim 10$ s, with the wide-field cameras optimized for prompt counterpart searches and the narrow-field telescopes reaching substantially deeper limits, typically $m_{\rm lim}\sim18$--19 depending on exposure time, sky conditions, and observing mode.

RAPTOR delivered several landmark early-time GRB observations. For GRB~041219A, RAPTOR detected optical emission contemporaneous with the prompt high-energy activity, with temporal behavior correlated with the $\gamma$-ray light curve, supporting an internal-dissipation origin for at least part of the optical component \citep{2005Natur.435..178V}. RAPTOR also provided critical early optical coverage of GRB~050820A, where the optical light curve showed both prompt-phase variability and a smoother early afterglow component, demonstrating how sub-minute optical observations can separate internal emission from the emerging external-shock afterglow \citep{2006Natur.442..172V}. RAPTOR-like architectures motivate current pipelines that combine alert ingestion, immediate scheduling, and near-real-time photometric products suitable for rapid community response.

\subsubsection{TAROT}
\label{subsubsec:tarot}
The TAROT network (\textit{Télescopes à Action Rapide pour les Objets Transitoires}) consists of fully robotic, wide-field instruments designed primarily for the prompt optical and early-afterglow follow-up of GRBs. The two original TAROT telescopes, TCA at Calern Observatory in France and TCH at ESO La Silla in Chile, are 0.25\,m fast-slewing telescopes with wide-field CCD cameras covering approximately $2^\circ \times 2^\circ$.
A third TAROT station, TRE, operates from Les Makes Observatory on Réunion Island with a smaller aperture but a similar rapid-response strategy. The longitudinal distribution of the network improves sky coverage and enables rapid follow-up of satellite triggers from both hemispheres. TAROT was specifically optimized to minimize latency: after receipt of a GRB notice, observations can begin within a few to tens of seconds, using pre-defined exposure sequences tailored to capture prompt optical emission and the earliest afterglow evolution \citep[e.g.,][]{2008AN....329..275K,2009AJ....137.4100K}. Observations are usually obtained in clear or unfiltered mode, broadly calibrated to the
$R$ band, with typical limiting magnitudes of $R \sim 16$--18 depending
on exposure time, sky brightness, and observing conditions.

TAROT observations have played an important role in characterizing the diversity of early optical GRB behavior, including prompt optical limits, early detections, smooth rises, optical flares, and rebrightening episodes. The homogeneous sample presented by \citet{2009AJ....137.4100K} showed that TAROT detected 20 optical afterglows between 2001 and 2008 and provided many early upper limits, making the network valuable for
constraining the incidence of bright prompt optical emission. TAROT data have also been used to study rising optical afterglows and cases where the optical and X-ray light curves require more than a single simple forward-shock component \citep{2009AIPC.1133..175G}. Although modest in aperture, TAROT is particularly effective in the bright-to-intermediate optical regime, where rapid cadence and low latency are often more important than depth for separating prompt, reverse-shock, and
forward-shock components.

\subsubsection{REM}
The Rapid Eye Mount (REM) telescope is a 60 cm fast-slewing robotic telescope located at the ESO La Silla Observatory in Chile, designed primarily for rapid optical and near-infrared follow-up of GRBs \citep{2001AN....322..275Z}. A key strength of REM is its fully autonomous response to GRB alerts and its ability to obtain nearly simultaneous optical and NIR imaging using the optical ROSS/ROSS2 camera and the REMIR near-infrared camera. This configuration provides early broad-band coverage from the optical to the NIR, approximately $\sim 0.4$--$2.3\,\mu$m, within minutes of a GRB trigger, with typical limiting magnitudes of order $R\sim18$ and $K\sim15$--16 for minute-scale exposures, depending on observing conditions and afterglow brightness. Such rapid optical--NIR coverage is particularly valuable for identifying dust-obscured ``dark'' bursts, recognizing high-redshift candidates through Ly-$\alpha$ absorption and optical dropouts, and constructing early SEDs while the afterglow is still bright and rapidly evolving. By measuring optical--NIR colors at very early epochs, REM observations help separate intrinsic extinction, intergalactic absorption, and spectral curvature, thereby reducing degeneracies that arise when only optical data are available. REM has made several landmark GRB contributions, including the detection of the smooth early optical rise of GRB~060418, which provided one of the clearest measurements of the forward-shock onset and the initial Lorentz factor \citep{2007Molinari}; rapid NIR observations of the high-redshift GRB~050904 at $z=6.29$, where Ly-$\alpha$ absorption strongly suppresses the optical flux \citep{2005A&A...443L...1T}; and optical--NIR studies of extinguished afterglows used to constrain line-of-sight dust extinction and host-galaxy environments \citep{2013MNRAS.432.1231C}.
 
\subsubsection{MASTER Global Robotic Net}
\label{subsubsec:master}

The Mobile Astronomical System of Telescope-Robots (MASTER) is a longitudinally distributed network of fully robotic wide-field optical telescopes designed for rapid transient discovery, prompt GRB, gravitational wave, and high-energy neutrino follow-up, as well as autonomous sky-survey operations \citep{2010AdAst2010E..30L, MasterGW, MasterNeutrino}. The network includes sites in Russia, South Africa, Argentina, the Canary Islands, and Mexico, providing broad longitudinal coverage and rapid access to a large fraction of the sky. The MASTER-II design was initially based on twin 40-cm, fast optical tubes, each covering a field of view of approximately $2^\circ \times 2^\circ$, which can be used either to observe the same field with different filters or polarizers or to tile adjacent regions \citep{Kornilov}. This configuration, together with automated GCN response, rapid slewing, and real-time transient detection pipelines, makes MASTER particularly well suited for early GRB observations, especially for events with relatively large localization uncertainties, such as those detected by \fermi/GBM \citep{2016MNRAS.455..712L,2026ApJ...997..246G,2022MNRAS.511.1694G,2020ARep...64..126E}.

MASTER has made a lot of important contributions to early optical GRB and gravitational waves science. The main highlights are the independent discovery of short GRB and GW source Kilonova \citep{2017ApJ...848L..12A}, prompt optical emission polarization discovery \citep{2017Natur.547..425T}, Prompt and early optical observations of GRB~100901A and GRB~100906A began while the high-energy emission was still ongoing, providing rare constraints on the relation between optical and $\gamma$-ray emission. In GRB~100901A, the optical and high-energy emission showed evidence for a common origin, whereas GRB~100906A exhibited a smoother optical peak more consistent with an external-shock onset \citep{2012MNRAS.421.1874G}. MASTER has also contributed to early optical polarimetry. Observations of GRB~150301B and GRB~150413A demonstrated the capability of the network to obtain polarimetric measurements during the earliest observable afterglow phases \citep{2016MNRAS.455.3312G}. A particularly important case is GRB~160625B, for which MASTER observations contributed to the first detection of significant and variable linear polarization during the prompt optical flash, providing direct evidence that ordered magnetic fields can be present in the emitting region during the active prompt phase \citep{2017Natur.547..425T}. Detailed MASTER optical photometry of GRB~160625B has also been used in multi-wavelength studies of its complex, multi-episode prompt emission \citep{2023ApJ...943..181L}.

The wide-field capability of MASTER is especially valuable for GRBs detected by instruments with large error regions. A recent example is GRB~230204B, an energetic long GRB detected by \fermi/GBM and MAXI, for which the optical afterglow was independently discovered by the MASTER and BOOTES robotic telescope networks at $\sim1.3$ ks after the burst trigger \citep{2026ApJ...997..246G}. MASTER observations covered the high-energy localization region and identified a bright, rapidly fading optical transient, enabling subsequent multi-wavelength follow-up and broadband afterglow modeling. This event illustrates the practical importance of wide-field robotic tiling for transforming a coarse high-energy localization into a precise optical position suitable for deeper optical, X-ray, radio, and spectroscopic observations.

\subsubsection{BOOTES}
\label{subsubsec:bootes}

The BOOTES (Burst Observer and Optical Transient Exploring System) network is the first global network of robotic observatories deployed on all continents \citep{2023NatAs...7.1136C}, and one of the longest-running efforts in robotic GRB follow-up. It began operations in Spain in 1998 (with first light in late 1997) using wide-field cameras and all-sky monitors to survey large sky areas and the error boxes of contemporary high-energy missions, including HETE, while performing real-time optical transient detection on the imaging data \citep{CastroTirado1999, 2001grba.conf..412P, 2004AN....325..679C}. With the launch of \textit{Swift} and the resulting arcsecond GRB localizations, BOOTES added a rapid pointed follow-up mode; the BOOTES-1 station, which hosts several smaller telescopes and wide-field instruments, included the 0.3-m BOOTES-1B telescope. Initially equipped with an experimental slitless wide-field spectrograph \citep{2004BaltA..13..696D}, BOOTES-1B was soon reconfigured for unfiltered direct imaging in favour of sensitivity, in which form it began routinely collecting GRB afterglows and remains active to this day, more than 20 years after first light. The GRB follow-up record accumulated over the first decade of operation was reviewed by \citet{2006AIPC..836..688J} and \citet{2016AdAst2016E..12J}.

The deployment of the 0.6-m BOOTES-2 telescope at La Mayora (M\'alaga, Spain), equipped with the Compact Low Resolution Spectrograph (COLORES), marked the beginning of the expansion of BOOTES into a worldwide network \citep{2016AdAst2016E..12J}. The network now comprises six identical 0.6-m robotic telescopes distributed across the globe: BOOTES-2 (M\'alaga, Spain), BOOTES-3 (New Zealand), BOOTES-4 (Yunnan, China), BOOTES-5 (San Pedro M\'artir, Mexico), and BOOTES-6 and BOOTES-7 (Argentina and Bloemfontein, South Africa) \citep{2023NatAs...7.1136C, 2023FrASS..10.2887H}. This longitudinal spread enables near-continuous coverage of transient alerts and coordinated follow-up across time zones.

The telescopes achieve slew times of $\lesssim 10$~s and observe through high-quality Sloan $grizy$ filters (similar to those used by Pan-STARRS) as well as clear bands, reaching limiting magnitudes of $m_{\rm lim} \sim 17$--$19$. The COLORES spectrographs can provide early spectral classification and redshift constraints for sufficiently bright transients. For GRB afterglows, BOOTES typically contributes fast detections and limits that constrain the early light curve, multi-band photometry for early spectral indices, and rapid spectroscopy of bright events that link the initial detection to later deep spectroscopy on 8--10\,m-class telescopes \citep{2021MNRAS.505.4086G}.

\subsubsection{Palomar 60-inch telescope}
\label{subsubsec:Palomar}

The Palomar 60-inch telescope (P60), a 1.52-m robotic telescope at Palomar Observatory, played an important role in the \swift-era transition from isolated GRB afterglow detections to systematic early optical follow-up. The facility was converted from classical operation to a fully robotic, queue-scheduled system in 2004, with the explicit goal of providing moderately rapid ($\lesssim 3$ min) and sustained ($R \lesssim 23$ mag) observations of GRB afterglows and other optical transients \citep{Cenko2006P60}. Although P60 was not as fast or as wide-field as instruments optimized for prompt optical flashes, its larger aperture, automated interruption of the observing queue, real-time reduction pipeline, and broad optical filter coverage made it especially effective for measuring early afterglow colors and following fading afterglows beyond the reach of smaller robotic telescopes. The scientific impact of this strategy was demonstrated by the P60--\swift early optical afterglow catalog, which presented multi-color observations of 29 long-duration \swift GRBs for which P60 observations began within one hour of the trigger; optical afterglows were recovered for $\sim 80\%$ of this prompt sample, largely owing to the use of redder filters such as $R_C$, $i^{\prime}$, and $z^{\prime}$ \citep{Cenko2009P60Catalog}. This sample provided an important, relatively homogeneous view of early \swift afterglows and showed that a large fraction of events classified as optically faint or ``dark'' have suppressed optical-to-X-ray flux ratios, with host-galaxy dust extinction likely contributing significantly to the observed darkness \citep{Cenko2009P60Catalog}. P60 also contributed to detailed multi-wavelength studies of individual bursts, most notably GRB~050820A, for which observations began 3.43 min after the \swift trigger and helped show that the early optical emission did not simply track the prompt $\gamma$-ray light curve but was instead consistent with the emergence of the external-shock afterglow and later energy injection \citep{Cenko2006GRB050820A}. Thus, the primary legacy of P60 in GRB studies is its role as a medium-aperture robotic bridge between ultra-fast small telescopes and large-aperture spectroscopy; it provided rapid, calibrated, multi-color afterglow light curves that were essential for constraining early spectral slopes, dust extinction, afterglow diversity, and the selection biases affecting \swift-era optical samples.

\subsubsection{Las Cumbres Observatory Global Telescope Network (LCO)}
\label{subsubsec:lco}

The Las Cumbres Observatory (LCO) Global Telescope Network is a longitudinally distributed, fully robotic optical facility designed to operate as a single, centrally scheduled observatory \citep{2013PASP..125.1031B}. The network comprises several aperture classes, including 2.0\,m, 1.0\,m, and 0.4\,m telescopes, with instruments and observing modes standardized as far as possible within each aperture class. This architecture is particularly well-suited to time-domain astronomy because observations can be dynamically assigned to the most appropriate available telescope according to target visibility, weather, lunar constraints, priority, and requested cadence.

For GRB follow-up, LCO provides several complementary capabilities. First, its robotic target-of-opportunity mode enables rapid optical imaging of newly localized afterglows, especially for events that remain sufficiently bright minutes to hours after the high-energy trigger. Second, the geographical distribution of the network reduces diurnal gaps and allows well-sampled multi-epoch light curves to be constructed over timescales from hours to days, which is essential for measuring early decay slopes, detecting breaks or rebrightenings, and separating prompt, reverse-shock, and forward-shock components. Third, the availability of broad optical filter coverage, including Johnson--Cousins and Sloan-like filters depending on the telescope and instrument configuration, enables color evolution and spectral-slope measurements that help distinguish chromatic internal components from achromatic external-shock evolution.

The 2.0\,m LCO nodes also provide rapid low-resolution spectroscopy with
FLOYDS, covering approximately the optical range $\lambda \simeq 3200$--$10000$\,\AA, when the afterglow is bright enough. Such spectra can provide early constraints on the continuum shape, dust reddening, broad absorption features, and, in favorable cases, the redshift. The LCO network is particularly valuable for population-level studies of GRB afterglows, where uniform observing strategies, consistent calibration, and automated data processing reduce the systematic uncertainties that often arise when combining heterogeneous data from many individual facilities.

\subsubsection{Liverpool Telescope and RoboNet}
\label{subsubsec:liverpool}

The 2\,m Liverpool Telescope (LT) at the Observatorio del Roque de los Muchachos at La Palma, Canary Islands is a fully robotic facility telescope for rapid transient follow-up, with fast slewing, automated enclosure operation, and a suite of instruments enabling optical/NIR  (UBVRI, m$_{\rm lim}$ $\sim$ 22 – 23 in 300 s), spectroscopy ($\lambda$ $\sim$ 4000 – 8000 $A^{\circ}$), and polarimetry \citep{2004SPIE.5489..679S}. Typical GRB response times are $\sim$ 2 -- 5 minutes, with the large aperture enabling deeper photometry and spectroscopy than smaller robotic facilities. Through coordinated programs such as RoboNet (including the Faulkes telescopes in earlier configurations), LT-class facilities have produced dense early-time multi-color light curves that capture the diversity of afterglow behaviors and connect those features to contemporaneous X-ray evolution \citep{2006NCimB.121.1303G}. 

\begin{figure*}[!ht]
    \centering
    \includegraphics[width=0.65\textwidth]{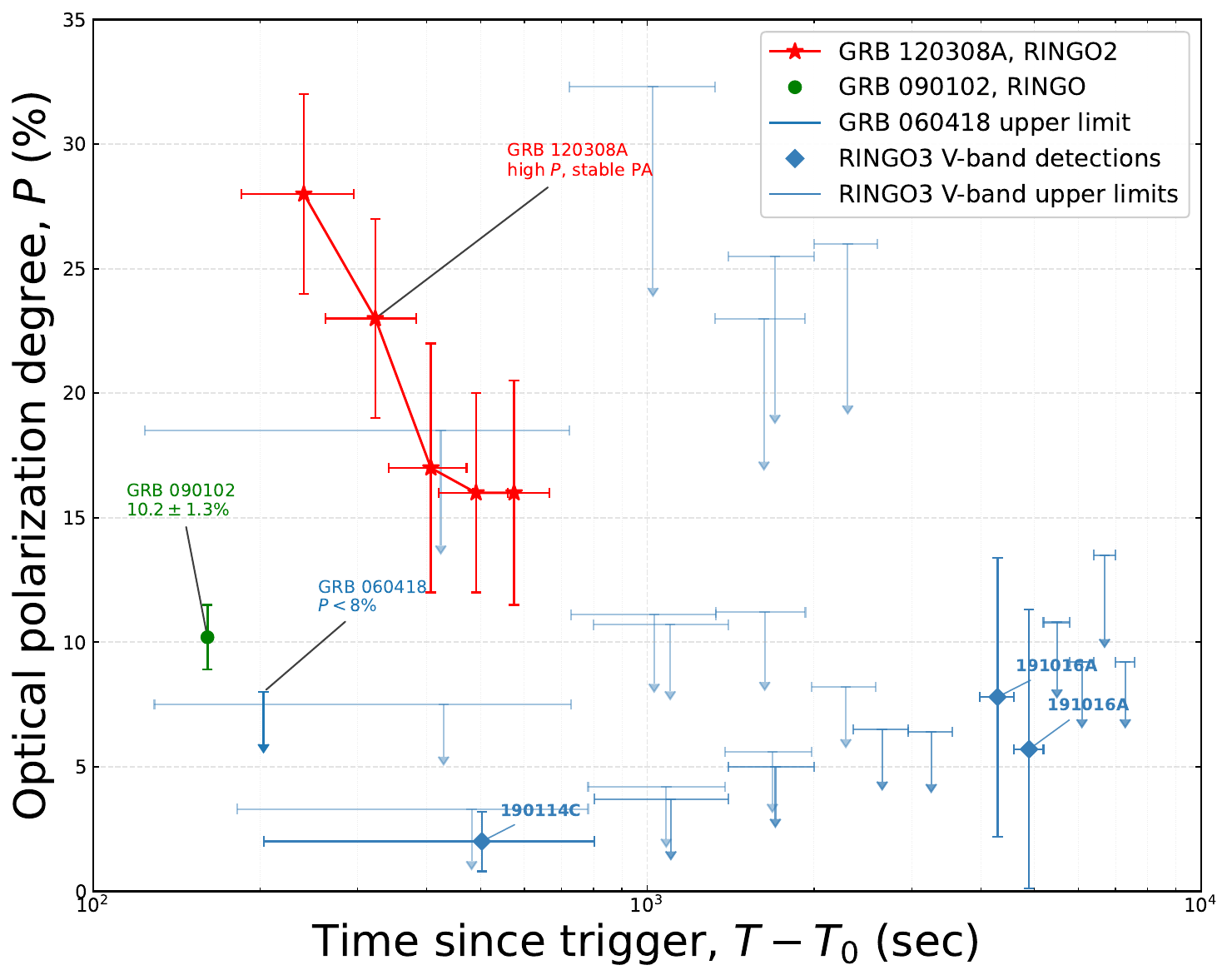}
\caption{Early optical polarimetry of GRB afterglows with the Liverpool Telescope RINGO/RINGO2/RINGO3 instruments. The panel summarizes the landmark LT measurements of GRB~060418 \citep{2007Sci...315.1822M}, GRB~090102 \citep{2009Natur.462..767S}, and GRB~120308A, together with the later RINGO3 sample. The high early polarization and nearly stable position angle of GRB~120308A support a reverse-shock origin with a large-scale ordered magnetic field in the ejecta \citep{2013Natur.504..119M}. In contrast, several later RINGO2/RINGO3 events show low polarization or upper limits, consistent with forward-shock-dominated emission, dust-induced polarization, or weak/absent reverse shocks \citep{2017ApJ...843..143S, 2022MNRAS.516.1584S}. The figure summarizes the LT contribution to establishing rapid robotic polarimetry as a direct probe of ejecta magnetization and magnetic-field geometry.}
\label{fig:LT_RINGO_polarimetry}
\end{figure*}

A particularly distinctive contribution of the Liverpool Telescope to GRB physics has been early-time optical polarimetry (see Figure \ref{fig:LT_RINGO_polarimetry}). The combination of a 2-m fully robotic telescope, automatic response to high-energy triggers, and the dedicated RINGO/RINGO2/RINGO3 polarimeters enabled polarimetric measurements on timescales of minutes, when reverse-shock and forward-shock components can still be separated, directly constraining magnetic field geometry and the ordering of fields in the emitting region \citep[e.g.,][]{2006SPIE.6269E..5MS, 2012SPIE.8446E..2JA}. These polarimetric constraints are uniquely diagnostic for differentiating between magnetized outflows, patchy shells, and shock-generated fields, especially when combined with early SEDs and high-energy light curves. The first demonstration came from GRB~060418, observed with RINGO at $t \simeq 203$~s, for which a $2\sigma$ upper limit of $\Pi < 8\%$ near the fireball deceleration epoch ruled out some models requiring a dominant large-scale ordered magnetic field in the emitting region \citep{2007Sci...315.1822M}. The breakthrough detection followed for GRB~090102: a 60-s RINGO exposure beginning at $t=160.8$~s measured $\Pi = 10.2\pm1.3\%$ during a steep-to-shallow optical transition, naturally interpreted as reverse-shock-dominated emission and providing direct evidence for ordered magnetic fields in the ejecta with a magnetization below the level that would suppress the reverse shock \citep{2009Natur.462..767S}. RINGO2 then provided the most striking time-resolved case, GRB~120308A, for which the early optical afterglow showed very high polarization, $\Pi = 28\pm4\%$ at about four minutes, declining to $\sim16\%$ over the following $\sim10$~min while the polarization position angle remained nearly stable \citep{2013Natur.504..119M}. This behavior strongly favors a reverse shock carrying a large-scale, engine-advected toroidal or helical magnetic field, rather than a purely shock-generated random field. Later, LT polarimetric samples placed these landmark events in a broader statistical context. The RINGO2 sample showed that early polarization detections are preferentially associated with afterglows displaying reverse-shock signatures \citep{2017ApJ...843..143S}, while the RINGO3 seven-year sample responded to 67 GRB alerts, detected 28 optical afterglows, and obtained useful polarimetry for seven of the ten events bright enough for detailed analysis \citep{2022MNRAS.516.1584S}. These sample studies found no clear correlation between polarization and global GRB properties, and many non-detections or low-polarization cases are consistent with forward-shock-dominated emission, dust-induced polarization, limited sensitivity, or the absence of a luminous reverse shock. Thus, the LT legacy is not simply the detection of a few highly polarized afterglows but rather the demonstration that rapid robotic polarimetry can distinguish magnetized reverse-shock ejecta from the more common low-polarization forward-shock phase.

\subsubsection{Gravitational-wave Optical Transient Observer (GOTO)}
\label{subsubsec:goto}
GOTO \citep{2022MNRAS.511.2405S, Dyer24} is a wide-field, modular robotic facility originally optimized for rapid tiling of gravitational-wave localization regions, but its design is also relevant to GRB follow-up when localizations are large (e.g., \fermi/GBM) or when rapid wide-field coverage is required. GOTO consists of 32 40\,cm-class wide-field unit telescopes (UTs), each with a 2.1$^\circ$ $\times$ 2.8$^\circ$ FoV. These are spread across 4 mounts at two nodes: a northern site at Observatorio del Roque de los Muchachos, La Palma, and a southern site at Siding Spring Observatory, Australia. The incorporation of 8 UTs per mount achieves a large instantaneous FoV of 44 square degrees per pointing and fast survey speed, enabling efficient probability-coverage strategies in L-band filter (400 -- 700\,nm) for poorly localized transients. GOTO achieves limiting magnitudes of m$_L$ $\sim$ 20 -- 21 in 4x45 second exposures and can tile thousands of square degrees per night. The system is robotically controlled by a custom `pilot' \citep{Dyer18}, while data are rapidly delivered to users by a bespoke and autonomous photometry pipeline \citep{Lyman26}.

Although primarily designed to search the many thousand square degree localization regions associated with GW triggers \citep[for which it has contributed counterpart upper limits;][]{2020MNRAS.497..726G}, GOTO also responds rapidly to GRB alerts, providing early-time optical constraints where targets are immediately observable and routinely identifying optical counterparts to \fermi-GBM triggers with poorly constrained positional uncertainties \citep{Belkin24, Kumar25}.

\subsubsection{MITSuME}
\label{subsubsec:mitsume}
MITSuME (Multicolor Imaging Telescopes for Survey and Monstrous Explosions) was developed to provide rapid, simultaneous multi-band optical observations of fast transients, with a particular emphasis on the early afterglows of GRBs. The core MITSuME optical system consists of 50-cm robotic telescopes located at Akeno Observatory and Okayama Astrophysical Observatory in Japan. Each telescope is equipped with a three-channel CCD camera that uses dichroic beam
splitters to obtain simultaneous images in the $g'$, $R_{\rm C}$, and $I_{\rm C}$ bands \citep{2008AIPC.1000..543S}. This simultaneous three-color capability is particularly valuable for GRB follow-up, because it allows color evolution to be measured without the ambiguity
introduced by filter-cycling observations of rapidly fading sources.

The main scientific strength of MITSuME is therefore not only rapid response, but also contemporaneous optical spectral information during the first minutes to hours after the burst. Such data can help distinguish intrinsic spectral evolution from apparent color changes caused by non-simultaneous sampling, constrain dust extinction and high-redshift suppression, and separate different early afterglow components. In particular, MITSuME observations have been used to study rapid color evolution and early light-curve transitions associated with reverse-shock emission, forward-shock onset, refreshed shocks, and continued central
engine activity \citep{2010AIPC.1279..466Y}. An associated near-infrared extension of the project was also designed to complement the optical telescopes with broader wavelength coverage, improving constraints on high-redshift and dust-obscured GRB afterglows.

\subsubsection{Nazarbayev University Transient Telescope at Assy-Turgen (NUTTelA-TAO)}
\label{subsubsec:nuttel}
NUTTelA-TAO is a 0.7\,m robotic telescope located at the Assy-Turgen Astrophysical Observatory in Kazakhstan, designed for rapid, autonomous follow-up of high-energy transients, including GRBs. The system is equipped with a 10$'$ $\times$ 10$'$ field of view and supports multi-band optical observations, including simultaneous $g'r'i'$ imaging via the Burst Simultaneous Three-channel Imager \citep[BSTI,][]{2020SPIE11447E..9IG, 2022SPIE12184E..8AG}. This capability enables the construction of early-time spectral energy distributions without temporal interpolation, which is critical during the rapidly evolving phases of GRB afterglows.

A key strength of NUTTelA-TAO is its ability to perform truly simultaneous multi-color observations within the first minutes after a GRB trigger. This capability was demonstrated in the case of GRB~230328B \citep{2023GCN.33539....1K,2026arXiv260604585K}, where NUTTelA-TAO obtained the earliest simultaneous multi-band ($g'r'i'$) optical observations starting at $\sim$41 seconds after the trigger. These data represent one of the earliest multi-color optical detections of a GRB afterglow, enabling direct measurement of the initial spectral shape and its rapid temporal evolution without the systematic uncertainties associated with filter switching. 

NUTTelA-TAO observations have also contributed to detailed studies of early afterglow spectral evolution \citep{2023MNRAS.520.6104K, 2025RMxAC..59..109A}. For GRB~201015A, simultaneous multi-band data obtained within the first hour revealed a clear transition in the optical spectral slope, interpreted as evidence for time-dependent dust extinction along the line of sight. The observed evolution from a redder to a bluer spectrum is consistent with dust destruction by the intense prompt and early afterglow radiation field, providing rare observational support for theoretical predictions of dust sublimation in GRB environments. Such measurements require both rapid response and simultaneous multi-band coverage, highlighting the unique role of instrument systems like the NUTTelA-TAO.

In the broader GRB follow-up ecosystem, NUTTelA-TAO contributes in several key ways: (i) its geographic location provides coverage of a sky region underserved by other robotic facilities, filling a critical longitude gap in global GRB follow-up networks; (ii) delivering simultaneous multi-band photometry that constrains early-time SEDs, color evolution, and rapid spectral variability that are difficult to capture with sequential observations. These capabilities make NUTTelA-TAO particularly valuable for probing the physical conditions of GRB jets and their environments during the earliest observable phases.

\subsubsection{D50 and FRAM telescopes}
\label{subsubsec:D50}

The Czech D50 and FRAM telescopes illustrate how small- and medium-aperture autonomous facilities can contribute to the earliest phases of GRB optical follow-up. Both descend from a long Ond\v{r}ejov tradition of robotic GRB response and are operated through the RTS2 observatory-control and data-reduction system \citep{Kubanek2010,Nekola2010,Strobl2023}. This lineage began with the early Ond\v{r}ejov robotic GRB-monitoring programme led by R.~Hudec \citep{2000AIPC..526..265H}, in particular the Burst Alert Robotic Telescope (BART), for which a new autonomous control system -- later developed into RTS2 -- was implemented from around 1999, leading to first light in 2001 and the first GRB follow-ups in 2002.

The D50 is a 0.5-m autonomous telescope at the Ond\v{r}ejov Observatory of the Astronomical Institute of the Czech Academy of Sciences, built primarily for rapid ground-based follow-up of GRB triggers distributed through real-time alert networks. Since its first light in 2007 it has maintained a near-continuous observational archive and followed more than one hundred \textit{Swift} and INTEGRAL localizations, including GRB~090726 \citep{2010A&A...510A..49S}, the faint GRB~210312B \citep{2025A&A...698A.162J}, and the short GRB~160927A, the first short burst followed from the Czech Republic \citep{Trcka2019}; weighted image coaddition is used to improve sensitivity for faint or rapidly fading afterglows \citep{Jelinek2019}. With a typical slew time of $\sim30$~s, D50 routinely begins observing within the first minute after a trigger. This allowed it to catch GRB~250702F $27.8$~s after the alert, while the brightest MeV pulses were still ongoing; the resulting high-cadence light curve -- capturing a prompt optical flare followed by a broader afterglow flare whose thermal-electron interpretation is discussed above -- makes it the scientifically richest event in the D50 record to date (Figure~\ref{fig:250702F}; \citealt{Jelinek2026}).

FRAM, the F/(Ph)otometric Robotic Atmospheric Monitor, denotes a family of small telescopes originally developed for atmospheric monitoring at the Pierre Auger Observatory and later at the Cherenkov Telescope Array sites \citep{2021JInst..16P6027A}. The first FRAM was deployed in Argentina in 2005, and already in early 2006 its wide-field camera detected the very bright early optical emission of GRB~060117, which reached $R \sim 10$~mag and faded rapidly -- an early-\textit{Swift}-era example of a luminous optical flash \citep{Jelinek2006,Ebr2014}. The 25-cm FRAM-ORM at the Observatorio del Roque de los Muchachos (La Palma, Spain) is the node most actively used for GRB work; together with D50, it observed GRB~210619B from within $\sim30$~s of the trigger, while high-energy activity was still ongoing, following a $\sim10$~mag transient through its smooth fading and color evolution \citep{Oganesyan2023}. These data showed that the early optical emission exceeded the extrapolation of the prompt $\gamma$-ray spectrum, best interpreted as a bright reverse shock transitioning to forward-shock emission in a narrow, magnetized jet expanding into a very low-density medium \citep{Oganesyan2023}. The remaining FRAM units, at the Auger and CTA sites, carry only wide-field cameras for atmospheric calibration and are not used for GRB follow-up.

\subsubsection{GROWTH India Telescope}
\label{subsubsec:GIT}

The GROWTH-India Telescope (GIT) has added an important Indian longitude node to the global robotic GRB follow-up ecosystem. GIT is a 0.7-m fully robotic optical telescope located at the Indian Astronomical Observatory, Hanle, Ladakh, and operated through the Indian Institute of Astrophysics and IIT Bombay. Its wide optical field of view ($\sim0.7$~--~$0.8$ deg), SDSS $ugriz$ filter set, automated target-of-opportunity response, and image-processing pipelines make it well suited for rapid localization, confirmation, multi-band photometry, and constraining upper limits of GRB optical afterglows \citep{Kumar2022GIT}. Although GIT is smaller than facilities such as LT or P60, its longitudinal location is highly valuable for events occurring during Asian night-time and for maintaining near-continuous global coverage in coordination with facilities such as ZTF, P60, DOT, HCT, GMRT, and other GROWTH-network assets \citep{Kasliwal2019GROWTH,Kumar2022GIT}. GIT has contributed both to rapid GRB afterglow discovery/confirmation and to longer-term afterglow monitoring. For example, GIT optical observations formed part of the multi-wavelength campaign on the TeV-detected GRB~190114C, where late optical data helped connect the optical afterglow with the radio/mm evolution and revealed departures from simple standard afterglow closure relations \citep{Misra2021GRB190114C}. GIT also contributed to the monitoring of GRB~210204A, whose optical afterglow displayed extremely delayed flaring activity, interpreted most plausibly as refreshed-shock emission from late-arriving ejecta \citep{Kumar2022GRB210204A}. More broadly, the large number of GIT GCN Circulars reporting detections, confirmations, and constraining upper limits for \swift, \fermi, MAXI, {\it SVOM}, and {\it Einstein Probe} GRBs demonstrates that GIT is now a regular component of the rapid-response GRB follow-up network, particularly for early optical photometry and source confirmation from the Indian subcontinent.

\subsubsection{COLIBRÍ}
\label{subsubsec:colibri}

COLIBR\'I is the French-Mexican Ground Follow-up Telescope (FM-GFT) of the \textit{SVOM} mission. It is a 1.3~m robotic telescope installed at the Observatorio Astronómico Nacional in the Sierra de San Pedro Mártir, Mexico, and is developed through a collaboration involving French institutions (Aix--Marseille Universit\'e, CNRS, CNES, and associated laboratories) and Mexican partners (UNAM and SECIHTI), with additional participation from Arizona State University \citep{2026arXiv260424259B}. Its primary role within the \textit{SVOM} ground segment is rapid multi-band follow-up of GRB afterglows, providing accurate optical/NIR localizations and early photometric information, which is useful for subsequent large-aperture spectroscopy.

The design of COLIBR\'I is driven by the need to respond rapidly to high-energy transient alerts while maintaining sufficient sensitivity and field coverage for \textit{SVOM}/ECLAIRs localizations. The telescope has a design pointing delay of about 20~s after alert reception, a field of view of $\simeq 26'$ in the visible channels and $\simeq 22'$ in the NIR channel, and a real-time processing goal of identifying and localizing optical counterparts within $\sim$5~min \citep{2026arXiv260424259B}. In practice, the measured delay from an on-board \textit{SVOM} trigger to the start of COLIBR\'I observations is typically of order one minute, because this includes the full alert chain, from satellite triggering and alert transmission to the interruption of the ongoing observation and execution of the new observing sequence \citep{2026arXiv260424259B}.

The principal optical instrument, DDRAGO, is a two-channel wide-field imager using blue and red CCD cameras, while the NIR camera CAGIRE is designed to provide $J$- and $H$-band imaging on the same telescope \citep{2026arXiv260424259B,2023ExA....56..645N}. In its final configuration, COLIBR\'I is therefore intended to deliver simultaneous multi-band coverage from the visible to the NIR, enabling rapid identification of optical afterglows, dark bursts, and high-redshift GRB candidates. This capability is particularly important for distinguishing between dust-extinguished afterglows and very high-redshift events, and for providing early photometric-redshift estimates that can trigger larger facilities for spectroscopy.

Science operations began in 2025, and the first year of operation has demonstrated that COLIBR\'I can routinely respond to \textit{SVOM} and other high-energy transient alerts. Its western-hemisphere location provides valuable longitudinal coverage for the global network of robotic GRB follow-up facilities, complementing telescopes in Europe, Asia, and Africa. Beyond \textit{SVOM} GRBs, COLIBR\'I is also designed to contribute to the broader transient-sky program, including fast follow-up of alerts from facilities detecting gravitational-wave events, high-energy neutrinos, fast radio bursts, fast blue optical transients, and other rapidly evolving astrophysical sources \citep{2026arXiv260424259B}.

\subsubsection{Other robotic and rapid-response facilities in the GRB literature}
\label{subsubsec:other_robotic}

Beyond the systems highlighted above, several additional robotic and rapid-response facilities have contributed to early GRB optical studies. Representative examples include the Zadko Telescope \citep{Coward2010, Coward2017}, Skynet/Panchromatic Robotic Optical and Monitoring Polarimetry Telescopes (PROMPT) Network \citep{Reichart2005, Dutton2022}, Pi of the Sky \citep{Burd2005}, Thai Robotic Telescope network, the T80S 0.8\,m robotic telescope and its S-PLUS Transient Extension Program \citep{MendesdeOliveira2019,Santos2024}, RC80 robotic telescope at Piszkesteto Station of Konkoly Observatory\footnote{\url{https://ccdsh.konkoly.hu/wiki/RC80_observing}}, 30\,cm FRAM-Auger telescope in Malargue, Argentina \citep{BenZvi2007,2010AdAst2010E..31P}, 0.5\,m Virgin Islands Robotic Telescope (VIRT) at the University of the Virgin Islands' Etelman Observatory \citep{Gendre2019}, the robotic telescope at IAU station 565, Bassano Bresciano Observatory, Italy, Reionization and Transients InfraRed camera (RATIR\footnote{\url{https://ratir.astroscu.unam.mx/}}), 0.76\,m Katzman Automatic Imaging Telescope (KAIT) at Lick Observatory \citep{Li2003}, 1.3\,m Peters Automated Infrared Imaging Telescope (PAIRITEL) located at the Fred L. Whipple Observatory on Mt. Hopkins, Arizona \citep{2006ASPC..351..751B}, the Watcher 40\,cm robotic telescope located at Boyden Observatory \citep{Melady2009}, Gamma-Ray burst Optical/Near-infrared Detector (GROND) on the 2.2\,m MPG telescope at La Silla \citep{Greiner2024}, the 0.5\,m COATLI robotic telescope
\citep{Watson2016COATLI}, and the wide-field Deca-Degree Optical Transient Imager (DDOTI, \citealt{Watson2016DDOTI}). These robotic facilities collectively enable multi-band, multi-epoch observations of GRB optical counterparts from the prompt phase through late-time afterglow evolution, providing the empirical foundation for testing and refining emission models and constraining physical parameters. They also probe complementary regions of discovery space: (i) very wide-field instruments can capture prompt optical flashes before precise positions are available, (ii) robotic 1 -- 2\,m telescopes dedicated to rapid ToO programs and dense monitoring, and (iii) emerging survey/tiling facilities that can rapidly cover large probability regions and provide candidate vetting for deeper pointed follow-up. In practice, the most complete physical picture of early optical behavior is obtained by combining these heterogeneous assets, with wide-field instruments constraining prompt emission over large sky areas, fast-slewing small/medium apertures deliver sub-minute light curves and color information, and larger robotic telescopes provide early spectroscopy/polarimetry and sustained monitoring.

\begin{figure*}[!ht]
    \centering
    \includegraphics[width=0.99\textwidth]{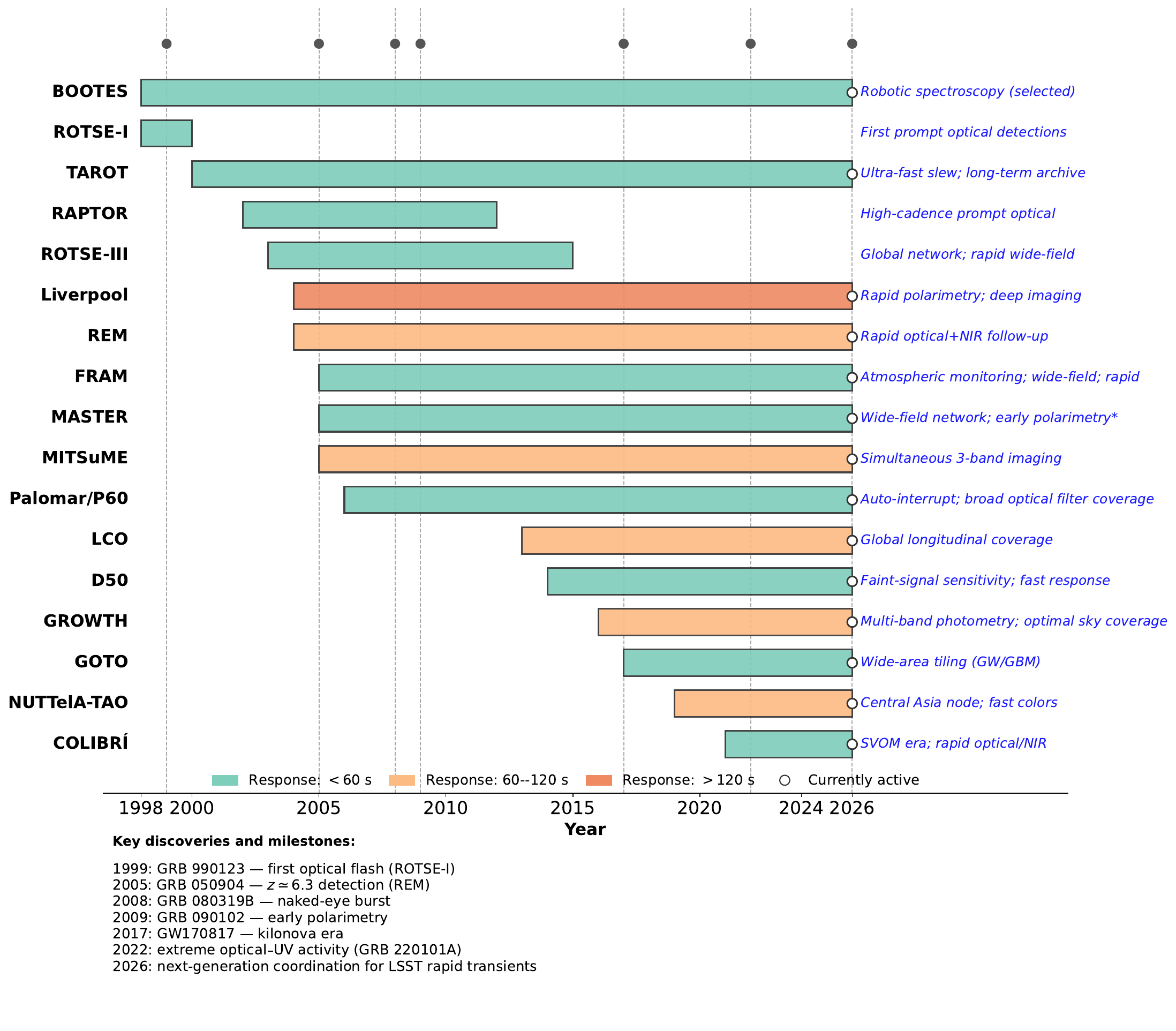}
\caption{Evolution of key robotic optical facilities and their milestones in early GRB follow-up (1998--2026): Horizontal bars represent operational periods for major robotic telescopes and networks; bar colors denote typical response time classes ($<60$\,s, $60$--$120$\,s, and $>120$\,s post-trigger), illustrating the progressive improvement from single-site, rapid-response wide-field instruments to coordinated global networks. Filled circles mark facilities operational through 2026. Vertical dashed lines indicate significant observational milestones, including the discovery of prompt optical emission, high-redshift detections, early-time polarimetry capabilities, and the advent of multi-messenger astronomy. This timeline demonstrates how advances in autonomous triggering, rapid scheduling algorithms, and global network coordination have progressively enabled observations during the most diagnostic phases of optical emission from seconds to hours post-burst.}
\label{fig:robotic}
\end{figure*}

Figure~\ref{fig:robotic} highlights the evolution of robotic GRB follow-up by integrating operational timelines, response-time performance, and key observational milestones. The first generation of systems prioritized minimal latency and wide-field coverage, enabling the initial prompt optical detections that established the contemporaneity of optical and high-energy $\gamma$-ray emission. Subsequent facilities extended this capability through enhanced longitudinal coverage via coordinated networks, rapid multi-band photometry spanning optical through NIR wavelengths, and, in select cases, early-time polarimetry and spectroscopy. These advances enabled physically discriminating observations of reverse-shock emission, forward-shock onset dynamics, and magnetic field structure in the outflow. By the mid-2010s, coinciding with the emergence of multi-messenger astronomy, globally coordinated robotic networks have increasingly coupled rapid discovery with deeper and more versatile follow-up capabilities. This evolution underscores the continued necessity of a heterogeneous infrastructure combining fast wide-field survey instruments with larger-aperture robotic telescopes to fully capture the diverse phenomenology of early optical emission in GRBs.

\subsection{Real-time data processing pipelines for robotic telescopes}
\label{data analysis}

\begin{figure*}[!ht]
    \centering
    \includegraphics[width=0.7\linewidth]{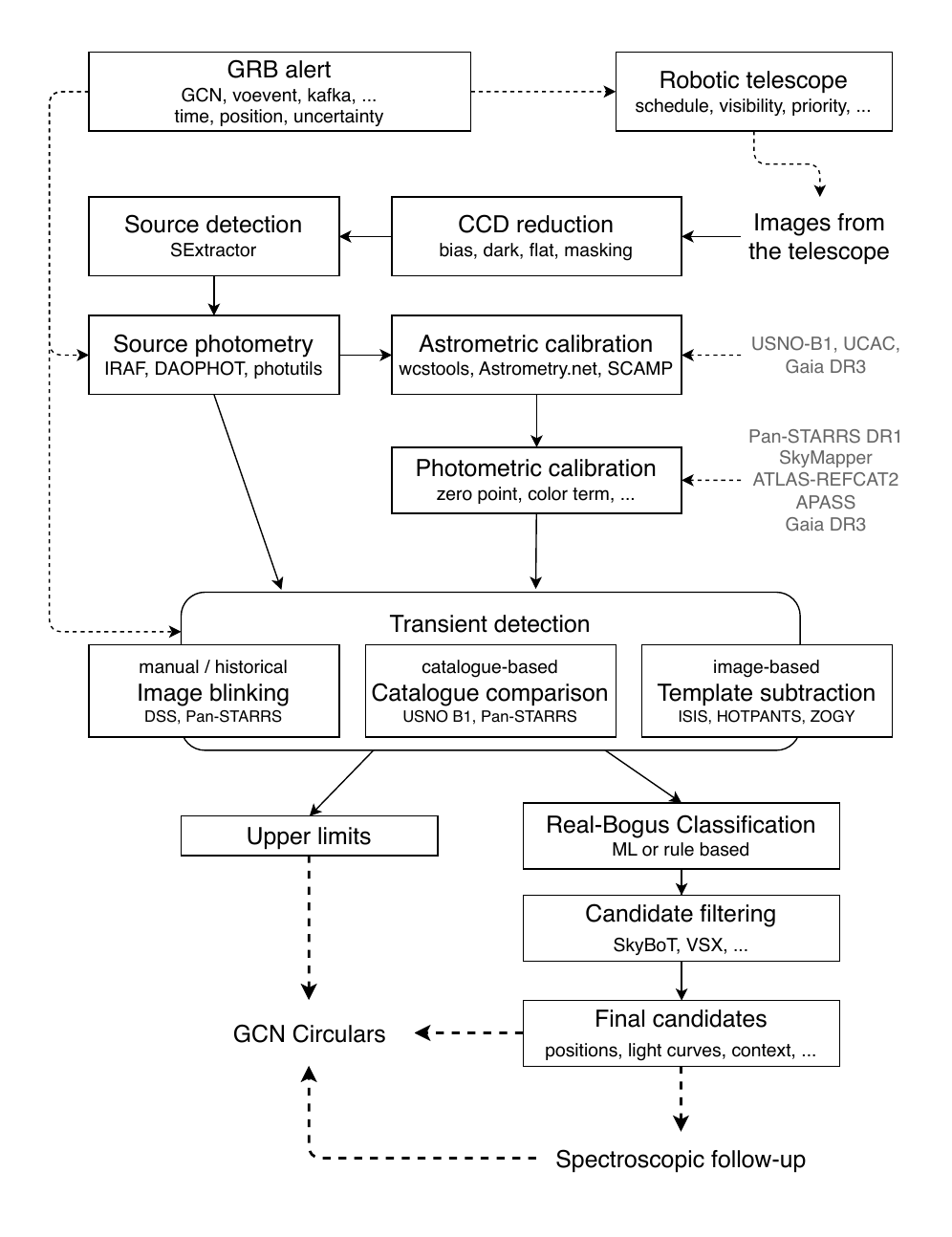}
\caption{Schematic diagram of a typical GRB data processing pipeline, outlining major steps performed by it. The exact path through the pipeline depends on the localization accuracy and scientific goal. For well-localized GRBs, the pipeline may only perform forced photometry or derive an upper limit at the reported position, whereas poorly localized triggers require a wider transient-search workflow involving catalog comparison, image subtraction, and candidate filtering.
}
\label{fig:pipeline}
\end{figure*}

Real-time data-processing pipelines for robotic telescopes have historically served to minimize the delay between image acquisition and the reliable detection, characterization, and reporting of transient candidates. This is particularly important for GRB follow-up, where optical afterglows may fade rapidly and where even early non-detections provide valuable constraints. Figure~\ref{fig:pipeline} shows the basic structure of such a pipeline. The raw data are first corrected for instrumental signatures through bias and dark subtraction, flat-fielding, and, when necessary, masking of bad pixels, saturated regions, and cosmic-ray hits. The calibrated images are then searched for sources, astrometrically registered, photometrically calibrated, and analyzed for transient candidates. Depending on the accuracy of the GRB localization, this final stage may vary from measuring a source or an upper limit at a known X-ray position to performing a wide-field transient search over a tiled probability region.

Over time, many software packages and standalone utilities have been adopted for individual stages of this workflow. SExtractor \citep{sextractor} became a standard tool for rapid source detection and basic object characterization, while more accurate aperture or PSF photometry, especially in crowded fields or for late-time observations affected by host-galaxy contamination, has often been performed using aperture- or PSF-photometry tools such as DAOPHOT \citep{daophot}, often through IRAF-based workflows \citep{iraf}, or through facility-specific wrappers around these packages. For astrometric calibration, early systems commonly used catalog matching based on geometric pattern recognition, such as triangle-matching algorithms \citep{triangles}, often requiring a reasonably good initial pointing estimate. The development of Astrometry.net \citep{astrometrynet} was an important step toward more robust automation, since it allowed blind plate solving without reliable \textit{a priori} pointing information. In large-format and survey-style reductions, astrometric refinement and distortion modeling have also frequently been performed with tools such as SCAMP \citep{scamp}, often in combination with resampling and co-addition utilities such as SWarp \citep{swarp}.

For transient identification, early visual or catalog-based comparison with archival images was gradually complemented, and in many cases replaced, by image-subtraction techniques. Packages such as ISIS \citep{isis}, HOTPANTS \citep{hotpants}, and later ZOGY \citep{zogy} made it possible to detect faint or host-contaminated transients by matching the PSF and background of science and reference images and searching the residual frames.

This evolution was enabled by the growing availability of deep and homogeneous digital sky surveys covering large regions of the sky, including the Sloan Digital Sky Survey \citep{sdss}, Pan-STARRS \citep{ps1,ps1surveys}, SkyMapper \citep{skymapper}, and later Gaia-based catalogs. These surveys provided both reference images and catalogs for transient searches and dense grids of accurately measured stars (secondary standards) for astrometric and photometric calibration, enabling rapid zero-point determination and thus allowing measurements from different epochs, filters, and telescope apertures to be placed on a common scale. At the same time, the software basis of astronomical pipelines shifted from heterogeneous collections of IRAF tasks, DAOPHOT, SExtractor, shell scripts, and observatory-specific codes toward more modular Python-based systems. Packages such as Astropy \citep{astropy}, ccdproc \citep{ccdproc}, astroquery \citep{astroquery}, photutils \citep{photutils}, and reproject \citep{reproject} have made Python a primary implementation language for modern pipelines rather than merely a wrapper around older reduction software.

The most recent stage of this development is the convergence of GRB follow-up pipelines with general time-domain survey pipelines. While well-localized \swift-like GRBs may still require only rapid photometry or limiting-magnitude estimation at a known position, poorly localized \fermi/GBM and multi-messenger triggers require wide-field tiling, image subtraction, rejection of moving objects and known variables, candidate ranking, and rapid coordination of further observations. Machine-learning methods, including real-bogus classifiers, image-stamp classifiers, and light-curve classifiers, have therefore become increasingly important. This marks a transition from simple GRB detection workflows to full transient detection and characterization pipelines, such as those now used by wide-field robotic facilities and networks such as GOTO \citep{Lyman26}.

\section{Summary and Conclusions}
\label{sec:conclusion}

Early optical observations have transformed GRB studies from a primarily high-energy trigger and localization field into a time-domain laboratory for relativistic jet physics. The first seconds to hours after the burst contain several short-lived emission components that are either absent or strongly blended at later epochs: prompt optical emission associated with internal dissipation, reverse-shock flashes produced as the ejecta decelerate, the onset of the forward-shock afterglow, optical flares, and rebrightenings linked to late energy release or refreshed shocks. Robotic telescopes are uniquely suited to this phase because they can respond automatically to space-based alerts, begin observations while the burst is still active or shortly after it ends, and obtain high-cadence photometry before the diagnostic early components fade.

A major conclusion of this review is that the most powerful constraints come from separating multiple early components rather than from detecting a single early optical point. High-cadence light curves establish whether an early peak is temporally connected to the prompt phase, the reverse shock, or the forward-shock onset. Multi-band photometry distinguishes chromatic internal activity from achromatic external-shock evolution and constrains extinction, spectral slopes, and the passage of synchrotron break frequencies. Simultaneous optical, X-ray, and $\gamma$-ray coverage tests whether optical emission tracks the high-energy prompt light curve or evolves independently. Early polarimetry adds an additional, highly diagnostic dimension: the moderate-to-high polarization detected in events such as GRB~090102 and GRB~120308A indicates that at least some reverse-shock-dominated afterglows carry ordered, large-scale magnetic fields advected from the central engine, while the lower polarization commonly measured at later times is more consistent with forward-shock emission from shock-generated or partially tangled fields \citep{2009Natur.462..767S,2013Natur.504..119M,2022MNRAS.516.1584S}. Thus, early optical polarimetry directly links observed afterglow behavior to jet composition and magnetic-field geometry.

The development of robotic facilities has been central to these advances. Wide-field, fast-response instruments such as ROTSE, TAROT, RAPTOR, Pi of the Sky, MASTER, and BOOTES established the importance of sub-minute optical coverage and provided early detections or constraining upper limits for large GRB samples. Larger-aperture robotic telescopes and networks, including the Liverpool Telescope, REM, LCO, GOTO, and COLIBR\'I, have expanded this capability by adding deeper imaging, multi-band optical/NIR coverage, rapid spectroscopy, polarimetry, and coordinated longitudinal monitoring. The field has therefore evolved from isolated early detections to a heterogeneous global infrastructure in which different facilities play complementary roles: wide-field systems are essential for rapid localisation and early limits, while larger robotic telescopes provide the sensitivity and instrumentation needed for physical interpretation.

Despite this progress, several open problems remain. The relative contribution of internal dissipation, reverse shocks, and forward shocks to prompt and early optical emission is still uncertain for many bursts. Reverse-shock signatures are detected only in a subset of well-observed events, leaving open whether their absence is mainly caused by high magnetization, low circumburst density, early peak times missed by observations, or observational sensitivity limits. The degree to which optical flares share a common origin with X-ray flares also remains unclear, especially in cases where the two bands show different temporal or spectral evolution. Population-level studies are still limited by heterogeneous cadence, filter coverage, depth, response time, and publication bias toward bright or unusual events. A major task for the field is therefore to build uniform early-time samples with well-characterized upper limits, rapid color information, and consistent modeling across prompt, optical, X-ray, radio, and high-energy data.

In conclusion, rapid robotic optical follow-up is indispensable for understanding GRB physics. It provides the only practical route to observing the prompt-to-afterglow transition, measuring reverse-shock and forward-shock onset signatures, identifying early flares, constraining initial Lorentz factors and ejecta magnetization, and probing magnetic-field geometry through polarimetry. The key lesson from more than two decades of robotic GRB observations is that the earliest optical light is not merely the beginning of the standard afterglow; it is a physically rich phase in which the central engine, jet composition, shock formation, and circumburst environment are all imprinted on rapidly evolving observables. Continued coordination between high-energy satellites, real-time alert systems, robotic optical/NIR facilities, and larger spectroscopic follow-up telescopes will remain essential for converting fleeting early detections into robust physical constraints.

\section{Future Prospects}
\label{sec:future}

The next decade of early GRB follow-up will be driven by the coupling of rapid high-energy triggers, wide-field optical/UV surveys, and deeper robotic follow-up facilities. The launch and operation of SVOM opens a particularly important phase for prompt-to-afterglow studies: its wide-field high-energy triggering, autonomous spacecraft repointing, onboard optical capability, and dedicated ground segment are designed to deliver rapid multi-wavelength constraints on GRBs from the prompt phase through the first minutes of afterglow evolution \citep{2022IJMPD..3130008A}. In parallel, Einstein Probe extends the discovery space to soft X-ray transients and X-ray-rich GRBs, including high-redshift or relativistic events whose early emission may be missed by harder $\gamma$-ray instruments \citep{2022hxga.book...86Y,2025SCPMA..6839501Y}. These missions will increase the demand for robotic optical/NIR systems capable of responding on timescales of seconds, obtaining simultaneous multi-band photometry, and triggering larger telescopes for spectroscopy before the afterglow fades. Facilities such as COLIBR\'I, BOOTES, GOTO, LCO, MASTER, and future 2 -- 4 m class robotic telescopes will be essential for separating prompt optical emission, reverse-shock flashes, forward-shock onset, and early flares through high-cadence light curves and color evolution. A key future priority is to make early polarimetry and rapid spectroscopy more routine: polarimetry within the first few minutes can directly test whether the ejecta carry ordered magnetic fields, while low-resolution spectroscopy or fast multi-band photometry can identify high-redshift, dust-obscured, or intrinsically faint afterglows in time for deeper observations with 8 -- 10 m telescopes and, eventually, ELT-class facilities. Wide-field time-domain surveys will provide a complementary discovery channel. The Rubin Observatory alert stream will generate millions of real-time alerts per night, enabling searches for orphan afterglows, late afterglow rebrightenings, GRB-supernovae, and kilonova-like counterparts, although rapid classification and broker filtering will be required to isolate rare GRB-related events from the much larger transient population \citep{Ivezic2019}. In the UV, ULTRASAT will provide wide-field, high-cadence near-UV coverage, offering a powerful probe of early hot components, shock breakout/cocoon emission, and blue kilonova emission associated with compact-object mergers \citep{Shvartzvald2024}. Finally, the future of early GRB observations will be increasingly multi-messenger: gravitational-wave and neutrino alerts will motivate rapid tiling of large localization regions, while TeV detections with current and next-generation Cherenkov facilities will require simultaneous optical, X-ray, radio, and high-energy coverage to determine whether inverse-Compton, hadronic, or structured-jet processes dominate the highest-energy emission. In this landscape, robotic telescopes will remain indispensable, not simply as follow-up instruments, but as real-time decision-making nodes that classify events, measure early colors and polarization, identify the most valuable targets, and trigger deeper observations before the diagnostic early emission disappears.

\backmatter





\bmhead{Acknowledgements} We thank the teams operating robotic telescope facilities worldwide for their dedication to rapid transient follow-up. This research has made use of data obtained from the High Energy Astrophysics Science Archive Research Center (HEASARC) and the GRB Coordinates Network (GCN). RG is thankful to Dr. Daniel A. Perley, Dr. Anna Y. Q. Ho, Dr. Alan Watson, and Dr. Mike Moss for the fruitful comments on the draft. RG was sponsored by the National Aeronautics and Space Administration (NASA) through a contract with ORAU. The views and conclusions contained in this document are those of the authors and should not be interpreted as representing the official policies, either expressed or implied, of the National Aeronautics and Space Administration (NASA) or the U.S. Government. The U.S. Government is authorized to reproduce and distribute reprints for Government purposes notwithstanding any copyright notation herein.
This work was co-funded by the EU and supported by the Czech Ministry of Education, Youth and Sports through the project CZ.02.01.01/00/22\_008/0004596 (SenDISo). 
AJCT acknowledges support from the Spanish Ministry project 
PID2023-151905OB-I00 and Junta de Andaluc\'ia grant P20\_010168.
TK acknowledges support from the Science Committee of the Ministry of Science and Higher Education of the Republic of Kazakhstan (Grant No. AP26102915).
AK,KZh studied under the state assignment of Lomonosov Moscow State University. MASTER research is partially carried out using the equipment of the shared research facilities of HPC computing resources at Lomonosov MSU.
M.G.D. acknowledges the support of the JSPS Grant-in-Aid for Scientific Research (KAKENHI) (A), Grant Number JP25H00675.

\section*{Declarations}

\bmhead{Conflict of interest/Competing interests}  Not applicable.

\bibliography{GRBs}

\end{document}